\documentclass[traditabstract]{aa}  

\usepackage{graphicx}
\usepackage{natbib}
\usepackage{lscape}
\usepackage{longtable}
\usepackage{txfonts}

\usepackage{float}

\usepackage{xcolor}

\newcommand{\rey}{\mathcal{R}}

\newcommand{\degree}{\ensuremath{^\circ}}

\def\nodata {...}

\def\arcsec{\hbox{$^{\prime\prime}$}}
\newcommand{\rdust}{{$R_{\rm disk,dust}$}}
\newcommand{\reff}{{$R_{\rm eff,dust}$}}
\newcommand{\rgas}{{$R_{\rm disk,gas}$}}

\begin{document}

   \title{Observational constraints on dust disk sizes in tidally truncated protoplanetary disks in multiple systems in the Taurus region}

\titlerunning{Binaries in Taurus}
\authorrunning{Manara et al.}


   \author{C.F. Manara \inst{1}\fnmsep\thanks{ESO Fellow}, M. Tazzari\inst{2}, F. Long\inst{3,4}, G.J. Herczeg\inst{3}, G. Lodato\inst{5}, A.A. Rota\inst{5}, P. Cazzoletti\inst{6}, G. van der Plas\inst{7}, \\P. Pinilla\inst{8}, G. Dipierro\inst{9}, S. Edwards\inst{10}, D. Harsono\inst{11}, D. Johnstone\inst{12,13}, Y. Liu\inst{6,14}, F. Menard\inst{7}, B. Nisini\inst{15}, \\E. Ragusa\inst{9,5}, Y. Boehler\inst{7}, \and S. Cabrit\inst{16,7}
          }

   \institute{European Southern Observatory, Karl-Schwarzschild-Strasse 2, 85748 Garching bei M\"unchen, Germany\\
              \email{cmanara@eso.org}
              \and
  Institute of Astronomy, University of Cambridge, Madingley Road, CB3 0HA, Cambridge, UK
  \and
  Kavli Institute for Astronomy and Astrophysics, Peking University, Beijing 100871, China
\and
Department of Astronomy, School of Physics, Peking University,  Beijing 100871, China
\and
Dipartimento di Fisica, Universita Degli Studi di Milano, Via Celoria, 16, I-20133 Milano, Italy
  \and
Max-Planck-Institut f\"ur Extraterrestrische Physik, Giessenbachstrasse 1, 85748, Garching bei M\"unchen, Germany
\and
Univ. Grenoble Alpes, CNRS, IPAG, F-38000 Grenoble, France
\and
Max-Planck-Institut f\"{u}r Astronomie, K\"{o}nigstuhl 17, 69117, Heidelberg, Germany
  \and
  Department of Physics and Astronomy, University of Leicester, Leicester LE1 7RH, UK
\and
Five College Astronomy Department, Smith College, Northampton, MA 01063, USA
\and
Leiden Observatory, Leiden University, P.O. box 9513, 2300 RA Leiden, The Netherlands
  \and
  NRC Herzberg Astronomy and Astrophysics, 5071 West Saanich Road, Victoria, BC, V9E 2E7, Canada
\and
Department of Physics and Astronomy, University of Victoria, Victoria, BC, V8P 5C2, Canada
\and
Purple Mountain Observatory, Chinese Academy of Sciences, 2 West Beijing Road, Nanjing 210008, China
\and
  INAF–Osservatorio Astronomico di Roma, via di Frascati 33, 00040 Monte Porzio Catone, Italy
  \and
  Sorbonne Universit\'{e}, Observatoire de Paris, Universit\'{e} PSL, CNRS, LERMA, F-75014 Paris, France
}
             
   \date{Received May 27, 2019; accepted July 6, 2019}

 
  \abstract
  {The impact of stellar multiplicity on the evolution of planet-forming disks is still the subject of debate. 
  Here we present and analyze disk structures around ten multiple stellar systems that were included in an unbiased, high spatial resolution survey performed with ALMA of 32 protoplanetary disks in the Taurus star-forming region. At the unprecedented spatial resolution of $\sim$0.12\arcsec \ we detect and spatially resolve the disks around all primary stars, and  those around eight secondary and one tertiary star. The dust radii of disks around multiple stellar systems are smaller than those around single stars in the same stellar mass range and in the same region. The disks in multiple stellar systems  also show a steeper decay of the millimeter continuum emission at the outer radius than disks around single stars, suggestive of the impact of tidal truncation on the shape of the disks in multiple systems. 
  However, the observed ratio between the dust disk radii and the observed separation of the stars in the multiple systems is consistent with analytic predictions of the effect of tidal truncation only if the eccentricities of the binaries are rather high (typically $>$0.5) or if the observed dust radii are a factor of two smaller than the gas radii, as is typical for isolated systems. Similar high-resolution studies targeting the gaseous emission from disks in multiple stellar systems are required to resolve this question.
 }


   \maketitle
%

\section{Introduction}

A physical theory to explain the origin of the observed populations of exoplanets relies on stringent observational constraints on the properties of protoplanetary disks, the place where planets form and evolve. 
In this context we must consider that a large fraction of stars are born in multiple stellar systems \citep[e.g.,][]{monin07},  and that exoplanets are detected around multiple stellar systems \citep[e.g.,][]{hatzes16}. 
However, stellar multiplicity may have a negative effect on the formation of planetary systems \citep[e.g.,][]{kraus16}. 

The initial conditions of planet-forming disks in multiple systems are likely set at the protostellar phase, with distinct pathways depending on whether the fragmentation occurs within the envelope or in a gravitationally unstable disk \citep[e.g.,][]{tobin16a,tobin16b}.  
At later stages of protostellar and disk evolution, dynamical interactions between disks in multiple stellar systems have a severe impact on their evolution \citep[e.g.,][]{CP93,bate18,RC18}. In particular, the sizes of the gaseous component of disks surrounding stars in multiple systems are expected to be truncated to sizes that are a fraction of the distance between the two components, with a dependence on the eccentricity of the orbit, the stellar mass ratio, the viscosity and temperature of the disks, and their co-planarity \citep[e.g.,][]{PP77,AL94,lubow15,ML15}. Although the dynamics of these systems is clearly understood from the theoretical viewpoint, observations are still lagging behind in confirming these theories, mainly due to the lack of resolved measurements of disk radii in multiple stellar systems.

Similarly to the case of isolated disks, the physical processes regulating the evolution of disks in multiple stellar systems are still a matter of debate. Theoretical work assuming X-ray driven photoevaporation as the driver of disk evolution has demonstrated that a different morphology is expected in disks in close binaries, for example a rarer appearance of transition disks \citep{RC18}.

Finally, the observed ratio of gas to dust radii in isolated disks is mainly regulated, on the one hand, by the effects of optical depth of the CO emission and, on the other hand, by the growth and drift of dust grains \citep[e.g.,][]{dutrey98,BA14,facchini17,trapman19}. Typically, the observed ratio in isolated and large disks is $\sim$1.5-3 \citep{ansdell18}, in line with the theoretical expectations. However, no observational information on this ratio in disks in multiple systems are available to date, except for the RW Aur system, which appears to have a gas radius larger by a factor $\sim$2 than the dust radius \citep{rodriguez18}. Further data is needed to constrain models of dust grain evolution in truncated disks.

The highest resolution observations obtained with the Submillimeter Array (SMA) and the advent of Atacama Large Millimeter/submillimeter Array (ALMA) are starting to provide constraints on the theory of disk evolution in multiple stellar systems. In particular, disks in multiple stellar systems
are on average fainter in the (sub)millimeter at any given stellar mass than those in single
systems, and that the disk around the primary component,
the most massive star, is usually brighter than the disk around the
secondary component \citep{harris12,AJ14,cox17,AJ19}. This result seems to hold in the young Taurus and $\rho$-Ophiucus regions, while disks around singles or binaries have a similar submillimeter brightness in the older Upper Scorpius region \citep{barenfeld19}. 
Stringent constraints on theory of disk truncation and evolution in multiple systems can only be obtained by resolving the spatial extent of individual disks in multiple stellar systems. The spatial resolution $\sim$0.2\arcsec \ - 0.5\arcsec, or higher, of previous observations \citep[e.g.,][]{AJ19} was not sufficient for the majority of the targets. 
   \begin{figure*}
   \centering
  \includegraphics[page=1,width=\textwidth,trim=7cm 5cm 8cm 5cm, clip]{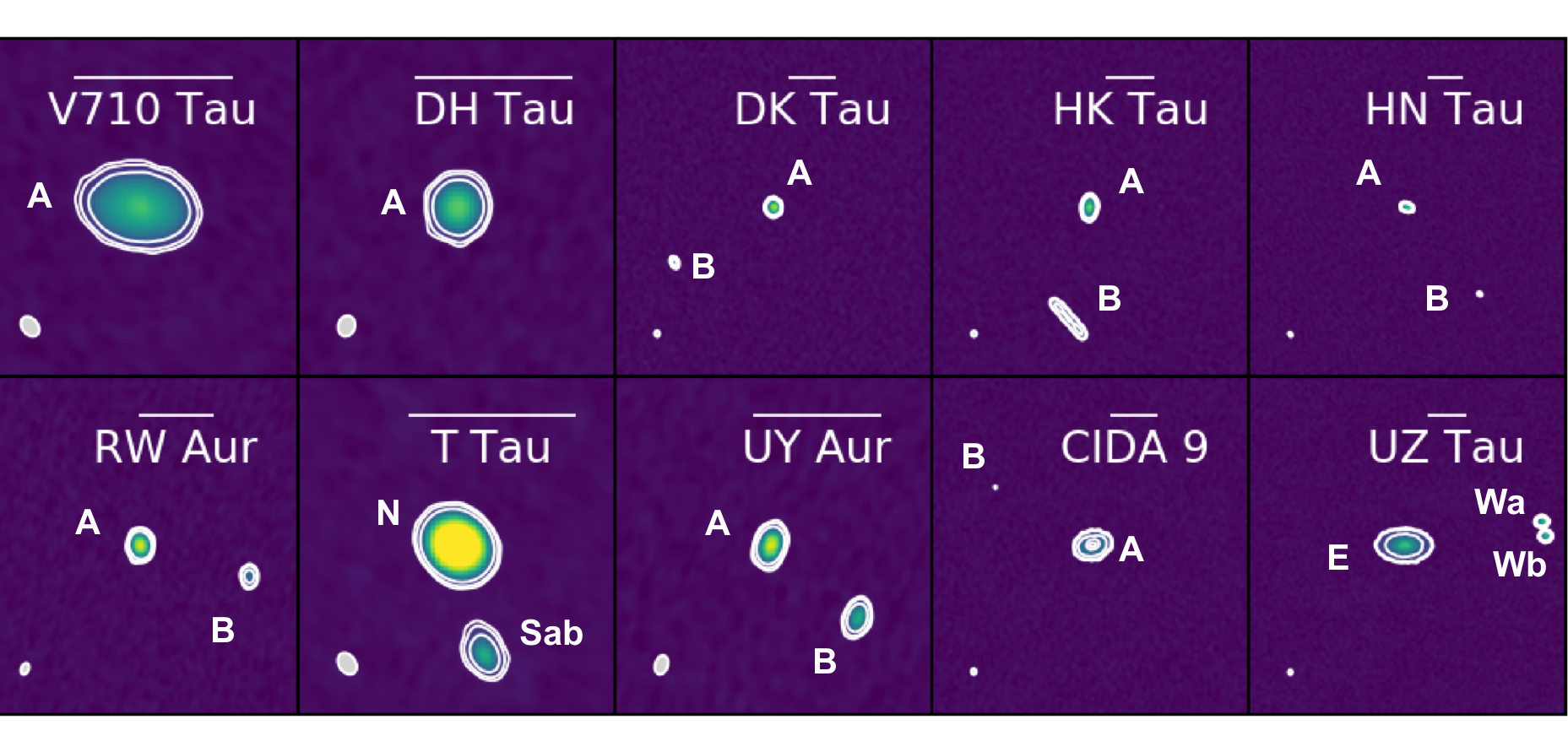}
   \caption{Continuum images of the disks around multiple stars in the Taurus star-forming region studied here. All bars above the names are 1\arcsec \ long, which is $\sim$140 au at the distance of Taurus. Beams are shown in the bottom left. Contours show 5, 10, 30 times
   the rms of the image. The components of the systems are labeled. The label for undetected secondary component is not shown. }
              \label{fig::binaries}%
    \end{figure*}

Here we present the first homogeneous analysis of the disks in ten multiple stellar systems taken with ALMA in the 1.3 mm dust continuum at the unprecedented spatial resolution of $\sim$0.12\arcsec, about $\sim$15-20 au at the distance of Taurus. These systems were included in our snapshot survey of 32 targets in the Taurus star-forming region \citep[][]{long19}. 
This is to date the largest sample of multiple systems in a single star-forming region observed at millimeter wavelengths with a spatial resolution better than 0.2\arcsec.

The paper is organized as follows. The sample, observations, and data reduction are discussed in Sect.~\ref{sect::data}, while the analysis of the data is presented in Sect.~\ref{sect::analysis}. The main results are presented in Sect.~\ref{sect::results}, and the comparison with analytical models of tidal truncation is then discussed in Sect.~\ref{sect::trunc_models}. Finally, we discuss our results in Sect.~\ref{sect::discussion} and draw our conclusions in Sect.~\ref{sect::concl}.


\section{Sample and observations}\label{sect::data}

\begin{figure*}
   \centering
  \includegraphics[page=2,width=\textwidth,trim=7cm 5cm 8cm 5cm, clip]{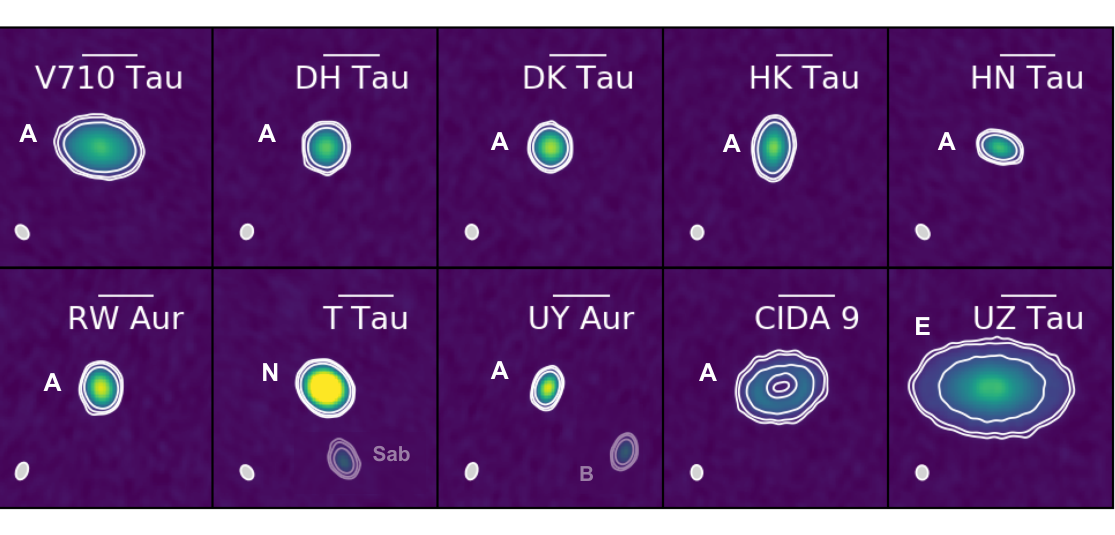}
   \caption{Continuum images of the primary component of multiples. All tiles are 2\arcsec$\times$2\arcsec. Bars are 0.5\arcsec \ long. Beams are shown in the bottom left. Contours show 5, 10, 30 times the rms of the image. The secondary component, when closer than 1\arcsec, is shaded out. }
              \label{fig::binaries_prim}%
    \end{figure*}

\begin{figure*}
   \centering
  \includegraphics[page=3,width=\textwidth,trim=7cm 5cm 8cm 5cm, clip]{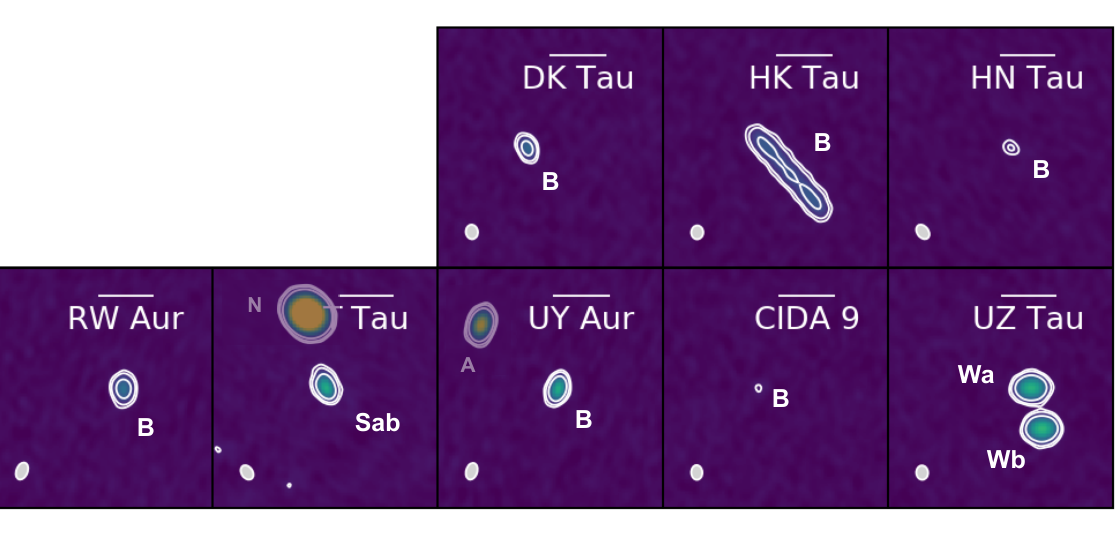}
   \caption{Continuum images of the secondary components of multiples. All tiles are 2\arcsec$\times$2\arcsec. Bars are 0.5\arcsec \ long. Beams are shown in the bottom left. Contours show 5, 10, 30 times the rms of the image. The primary component, when closer than 1\arcsec, is shaded out. }
              \label{fig::binaries_sec}%
    \end{figure*}

The survey of disk structures in Taurus (program 2016.1.00164.S, PI Herczeg) covered with  $\sim$0.14\arcsec$\times$ 0.11\arcsec \ resolution ALMA Band 6 observations 32 targets located in the Taurus star-forming region. All targets had spectral types earlier than M3 and were selected to avoid biases related to either disk brightness or inference of substructures from previous ALMA observations or spectral energy distribution modeling. Binaries with separations between 0.1\arcsec \ and 0.5\arcsec \ and targets with high extinction ($A_V>3$ mag) were also excluded. The most significant bias is the exclusion of disks with previous high-resolution ALMA images (spatial resolution better than 0.25\arcsec) in Taurus.  A more complete description of the sample selection, including the sources that were excluded, is provided in \citet{long19}. 
The combination of the high resolution and sensitivity of the survey has allowed us to detect continuum emission of primary disks in ten  wide binaries, the secondary disks in eight cases, and tertiary disks in one case (Figures~\ref{fig::binaries}, \ref{fig::binaries_prim}, and~\ref{fig::binaries_sec}, respectively).

The analysis of the whole sample is presented in two companion papers \citep{long18,long19}, along with a detailed study of one specific highly structured disk, MWC480 \citep{liu18}, and an analysis of the putative planet population inferred from substructures \citep{lodato19}. The survey found 12 disks with prominent dust gaps and rings \citep{long18} and 12 smooth disks around single stars \citep{long19}. Of the ten multiple systems, eight have smooth disks around the primary stars while two of the primaries show disks with substructure \citep[CIDA~9A and UZ~Tau~E,][]{long18}. The focus of this paper is on the characteristics of  the primary and the secondary disks in  multiple systems in comparison to those from  single systems with smooth disks in the snapshot survey \citep{long19}.

\begin{table*} 
\caption{\label{tab::binaries} Information from the literature on the targets}
\centering 
\begin{tabular}{l*{12}{c}}
\hline
Name of the &   Separation &    PA &     $d$ & SpT1 &   SpT2  & M1 &    M2 &       $q$ & $\mu$ &   Cont det        & $^{13}$CO det? \\
System & [\arcsec]      & [\degree]     & [pc]  &  &    & [$M_\odot$]   & [$M_\odot$]     &       & &     & \\
\hline
T Tau & 0.68 &  179.5 & 144 & K0 &      \nodata & 2.19$^{+0.38}_{-0.24}$ &       2.65$^{+0.10}_{-0.11}$  & 1.21 & 0.55   & NS    & ? \\
T Tau S & 0.09 &        4.9 &   144 & K0 &      \nodata & 2.12$\pm$0.10 &       0.53$\pm$0.06   & 0.25 & 0.20     & S     & ? \\
UY Aur  & 0.89  & 227.1 & 155 & K7      &M2.5   & 0.65$^{+0.17}_{-0.13}$        &  0.32$\pm$0.2  & 0.49 & 0.33    &  AB   & ? \\
RW Aur & 1.49 & 254.6 & 163 & K0 & K6.5 & 1.20$^{+0.18}_{-0.13}$ & 0.81$\pm$0.2 & 0.67 & 0.41 & AB & A \\
DK Tau & 2.38 & 117.6 & 128 & K8.5 & M1.5 & 0.60$^{+0.16}_{-0.13}$ & 0.44$\pm$0.2 & 0.73 & 0.42 & AB & A \\
HK Tau  & 2.32 & 170.4 & 133 & M1.5 & M2 & 0.44$^{+0.14}_{-0.11}$ & 0.37$\pm$0.2 & 0.84 & 0.46 & AB & AB \\
CIDA 9$^\dagger$ & 2.35 & 50 & 171 & M2 & M4.5 & 0.43$^{+0.15}_{-0.10}$ & 0.19$\pm$0.1 & 0.44 & 0.31 & AB & A \\
DH Tau & 2.34 & 130 & 135 & M2.5 &  M7.5 & 0.37$^{+0.13}_{-0.10}$ &  0.04$\pm$0.2 &  0.11 &  0.10 & A & \nodata \\
V710 Tau$^*$ & 3.22 & 176.2 & 142 & M2 & M3.5 & 0.42$^{+13}_{-0.11}$ & 0.25$\pm$0.1 & 0.60 & 0.37 & A & A \\
HN Tau & 3.16 & 219.1 & 136 & K3 & M5 & 1.53$\pm$0.15 & 0.16$\pm$0.1 & 0.10 & 0.09 & AB & A \\
UZ Tau & 3.52 & 273.1 & 131 & M2 & M3 & 1.23$\pm$0.07
& 0.58$\pm$0.2 & 0.47 & 0.32 &  EWaWb & E \\
UZ Tau W & 0.375 & 190 & 131 & M3 & M3 & 0.30$\pm$0.04 & 0.28$\pm$0.2 & 0.93 & 0.48 &        WaWb & \nodata \\
\hline
\end{tabular}
\tablefoot{When both disks are detected, separations are measured as the distance between the fitted center of the two disks. Otherwise, the value is taken from the literature \citep{WG01,KH09,kohler16}. Distances are obtained by inverting the Gaia parallax when the uncertainty on the parallax is less than 10\% of the measured parallax. See Sect.~\ref{sect::sample} for more information. The values $q$ and $\mu$ are derived as described in Sect.~\ref{sect::sample}. The last column reports whether $^{13}$CO emission is detected. A question mark flags when the detection is contaminated by cloud emission. For the two triple systems (T~Tau, UZ~Tau), the first line reports the information for the primary and the center of mass of the secondary, while the second line reports the information on the secondary and tertiary stars. $^\dagger$ The disk around the primary component is a well-resolved transition disk \citep[e.g.,][]{long18}. $^*$ The disk around component A (north) is detected (see Appendix~\ref{app::v710tau}). } 
\end{table*}

%

\subsection{Sample}\label{sect::sample}

The ten multiple stellar systems discussed in this work cover a wide range of system parameters. The  projected separations of the individual components in the systems ($a_p$) range from 0.7\arcsec \ to 3.5\arcsec \ ($\sim$100 au to 500 au at 140 pc, the typical distance to these targets), the mass ratios ($q=M_2/M_1$) from $\sim$1 to $\sim$0.1, and the mass parameters ($\mu=M_2 / (M_1+M_2)$ ) from $\sim$0.1 to $\sim$0.6 (Table~\ref{tab::binaries}). Two systems are triples: T~Tau, composed of a star in the north and two close-by stars in the south \citep[e.g.,][see also Appendix~\ref{app::ttau}]{kohler16}, and UZ~Tau, composed of a star to the east and two on the west side of the system \citep[e.g.,][]{WG01}.  While it has been suggested that UY~Aur~B could also be a binary \citep{tang14}, it is considered  one object here.

The stellar properties are obtained as in \citet{long19}. All the targets in the sample have been extensively studied with spectroscopy. We use the spectral type and the luminosity of the target, when
available, derived by \citet{HH14}. The stellar luminosities are then rescaled to the distance obtained from the parallaxes measured by the Gaia satellite \citep{gaia} and released in Data Release 2 \citep{gaiadr2}. Parallaxes to both components of the multiple systems are measured with relative uncertainty on the parallax smaller than 10\% and good-quality astrometric fit for most of the objects in the sample. For RW Aur, we adopt the parallax to RW Aur B because the {\it Gaia} DR2 astrometric fit to RW Aur A is poor.  For all other objects the distance is obtained from the weighted average of the parallax to the system members. 
As listed in Table~\ref{tab::binaries}, the final adopted values for the distances range from 131 pc to 171 pc. 
We then assign an effective temperature to the targets using the relation between spectral type and effective temperature by \citet{HH14}, and combine this with the distance corrected stellar luminosity to infer the stellar masses using the evolutionary models by \citet{baraffe15} and the nonmagnetic models by \citet{feiden16}, as in \citet{pascucci16}.  The combination of these two sets of evolutionary tracks covers the full range of effective temperatures of the stars in our sample. 

Only in three cases are the stellar masses  not derived as just described. 
T Tau S has a very high extinction and therefore lacks sufficient spectroscopic data for this analysis.  However, the masses of both T Tau Sa and Sb have been accurately measured from orbital dynamics \citep{schaefer14,kohler16}. The dynamical mass estimate is also assumed for UZ Tau E \citep{simon00} and HN Tau A \citep{simon17}, as also adopted by \citet{long19}.

   \begin{figure}
   \centering
  \includegraphics[width=0.5\textwidth]{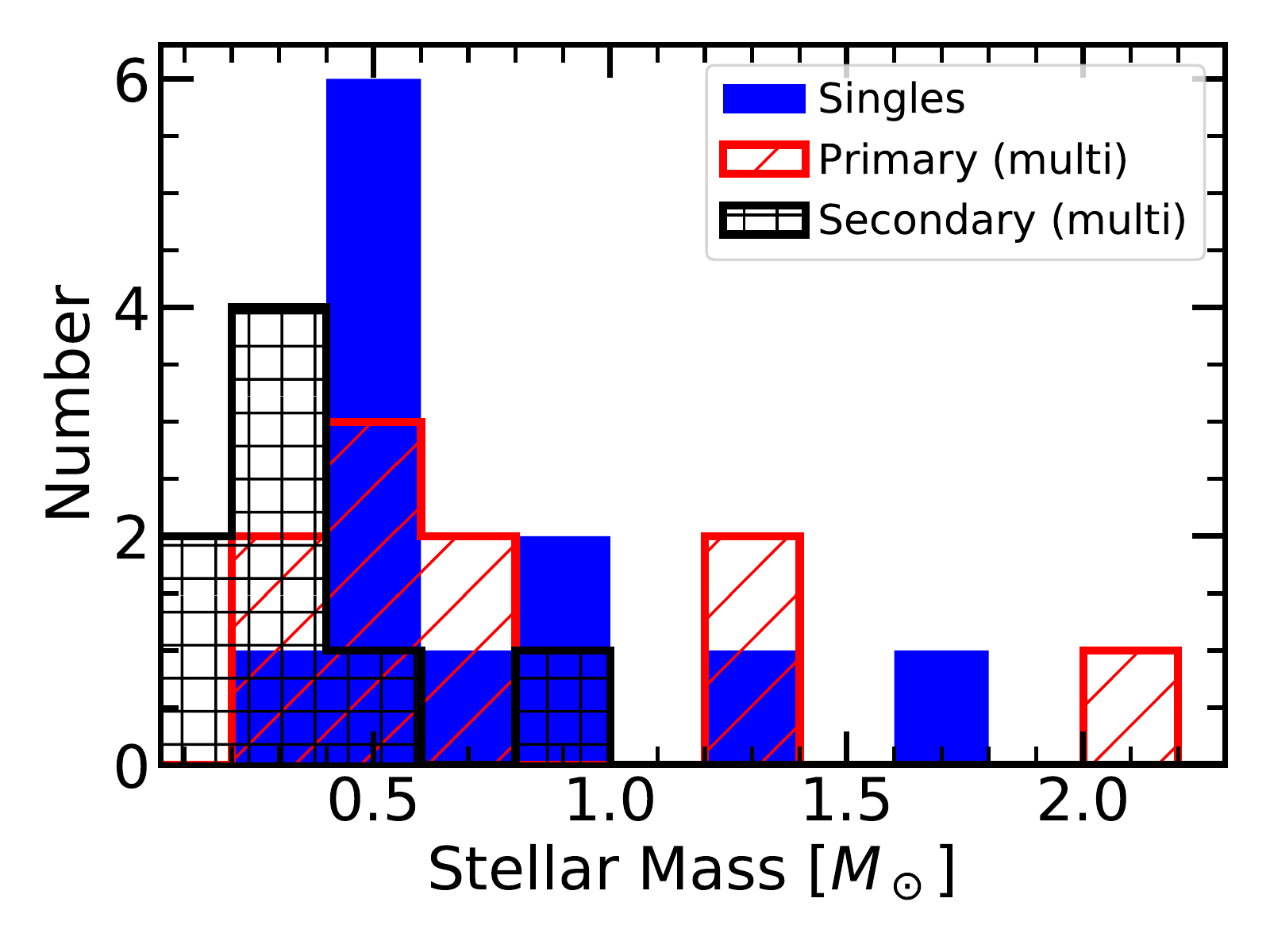}
   \caption{Stellar masses for the primary stars of the multiple systems analyzed here and for the stars with smooth disks analyzed by \citet{long19}.
   }
              \label{fig::masses}%
    \end{figure}

Figure~\ref{fig::masses} shows the distribution of stellar masses for the primary stars and secondary stars in the multiple systems analyzed here compared with the distribution of stellar masses for single stars surrounded by smooth disks analyzed in the companion paper by \citet{long19}. The two samples of primaries and singles cover the same range of stellar masses, and the secondaries are also compatible, although slightly skewed to lower stellar masses. 

\subsection{Observations and data reduction}

Our sample was observed with ALMA in Band 6 in 2017 August--September. The continuum spectral windows were centered at 218 and 233 GHz, each with a bandwidth of 1.875 GHz, for an averaged frequency of 225.5 GHz (corresponding to 1.3 mm). Each target was observed for $\sim 8-10$ min. The observing conditions and calibrators for individual targets can be found in Table 2 of \citet{long19}, where the details of the data reduction and calibration are described. In short, phase and amplitude self-calibrations were applied to our targets to maximize the image signal-to-noise ratio after the standard calibration procedure. The continuum images were then created with the \textsc{CASA} task \textit{tclean}, using Briggs weighting with a robust parameter of 0.5. These images have a typical beam size of 0.12\arcsec \ and continuum rms of 50 $\mu$Jy beam$^{-1}$. The images of the continuum emission from our targets in shown in Figs.~\ref{fig::binaries},~\ref{fig::binaries_prim}, and \ref{fig::binaries_sec}.

  \begin{figure*}
  \centering
 \includegraphics[width=\textwidth,page=2]{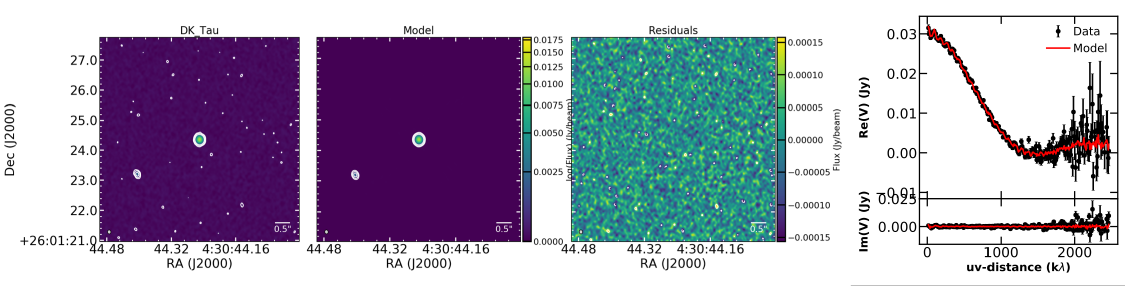}
  \caption{Example of the best fit of our data obtained as described in Sect.~\ref{sect::analysis}. Here we show the image of the data, model, and residuals for the DK Tau system, together with the visibilities of the data. }
             \label{fig::DKTau_example}%
    \end{figure*}

%



\section{Analysis}\label{sect::analysis}

 \begin{table*} 
\caption{\label{tab::fits_res} Coordinates, radii, fluxes, inclinations, and PA derived from the fit of the continuum emission for the detected disks.}
\renewcommand{\arraystretch}{1.5}
\centering 
\begin{tabular}{l | cc| c| c | c | c | c | c | c }
\hline
Name & RA & Dec &       \reff    & \reff        & \rdust & \rdust & $F_{\rm tot}$ & inc & PA\\
& [h:m:ss] & [d:m:ss] & [arcsec]        &  [au] & [arcsec]      & [au] & [mJy]   & [\degree] & [\degree]  \\
\hline
T Tau N      & 04:21:59.4 & +19:32:06.18  & 0.1111$_{-0.0001}^{+0.0001}$        & 16.0$_{-0.01}^{+0.01}$  &  0.1434$_{-0.0002}^{+0.0002}$ &  20.6$_{-0.03}^{+0.03}$ & 179.72$_{-0.18}^{+0.19}$ & 28.3$^{+0.2}_{-0.2}$ & 87.5$^{+0.3}_{-0.3}$ \\
UY Aur A     & 04:51:47.4 & +30:47:13.09  & 0.0332$_{-0.0004}^{+0.0021}$        & 5.1$_{-0.06}^{+0.32}$   &  0.0432$_{-0.0042}^{+0.0106}$ &  6.7$_{-0.65}^{+1.64}$   & 20.09$_{-0.57}^{+0.82}$ & 23.5$^{+8.6}_{-9.4}$ & -53.6$^{+10.1}_{-10.7}$\\
RW Aur A     & 05:07:49.6 & +30:24:04.70  & 0.1009$_{-0.0008}^{+0.0007}$        & 16.5$_{-0.13}^{+0.11}$  &  0.1317$_{-0.0008}^{+0.0014}$ &   21.5$_{-0.13}^{+0.23}$  & 35.61$_{-0.19}^{+0.18}$ & 55.0$^{+0.5}_{-0.4}$ & 41.1$^{+0.6}_{-0.6}$\\
DK Tau A     & 04:30:44.2 & +26:01:24.35 & 0.0916$_{-0.0010}^{+0.0007}$ & 11.8$_{-0.13}^{+0.09}$  &  0.1168$_{-0.0010}^{+0.0014}$ &   15.0$_{-0.13}^{+0.18}$ & 30.08$_{-0.18}^{+0.18}$ & 12.9$^{+2.5}_{-2.8}$ & 4.5$^{+9.9}_{-9.7}$\\
HK Tau A     & 04:31:50.6 & +24:24:17.37  & 0.1565$_{-0.0013}^{+0.0009}$        & 20.9$_{-0.17}^{+0.12}$  &  0.2157$_{-0.0019}^{+0.0030}$ &   28.7$_{-0.25}^{+0.40}$ & 33.15$_{-0.23}^{+0.23}$ & 56.9$^{+0.5}_{-0.5}$ & -5.1$^{+0.5}_{-0.5}$\\
CIDA 9A      & 05:05:22.8 & +25:31:30.50  & 0.2827$_{-0.0016}^{+0.0013}$        & 48.4$_{-0.27}^{+0.22}$  &  0.3598$_{-0.0027}^{+0.0020}$ &   61.6$_{-0.46}^{+0.34}$ & 36.8$_{-0.1}^{+0.1}$ & 46.4$^{+0.5}_{-0.4}$ & -76.5$^{+0.6}_{-0.6}$\\
DH Tau A     & 04:29:41.6 & +26:32:57.76  & 0.1053$_{-0.0009}^{+0.0006}$        & 14.2$_{-0.12}^{+0.08}$  &  0.1456$_{-0.0031}^{+0.0030}$ &   19.7$_{-0.42}^{+0.40}$  & 26.68$_{-0.18}^{+0.17}$ & 16.9$^{+2.0}_{-2.2}$ & 18.9$^{+7.4}_{-7.3}$ \\
V710 Tau A   & 04:31:57.8 & +18:21:37.64  & 0.2379$_{-0.0007}^{+0.0008}$        & 33.8$_{-0.10}^{+0.11}$  &  0.3174$_{-0.0021}^{+0.0023}$ &   45.0$_{-0.30}^{+0.33}$  & 55.20$_{-0.20}^{+0.19}$ & 48.9$^{+0.3}_{-0.3}$ & 84.3$^{+0.4}_{-0.4}$\\
HN Tau A     & 04:33:39.4 & +17:51:51.98  & 0.1037$_{-0.0019}^{+0.0018}$        & 14.1$_{-0.26}^{+0.24}$  &  0.1363$_{-0.0023}^{+0.0036}$ &   18.5$_{-0.31}^{+0.49}$  & 12.30$_{-0.32}^{+0.37}$ & 69.8$^{+1.4}_{-1.3}$ & 85.3$^{+0.7}_{-0.7}$\\
UZ Tau E     & 04:32:43.1 & +25:52:30.63  & 0.4424$_{-0.0022}^{+0.0011}$        &  57.9$_{-0.29}^{+0.14}$ &  0.6588$_{-0.0032}^{+0.0020}$ &   86.3$_{-0.42}^{+0.26}$  & 131.9$_{-0.2}^{+0.1}$ & 55.2$^{+0.2}_{-0.2}$ & 89.4$^{+0.2}_{-0.2}$\\

\hline

\multicolumn{4}{c}{Secondary}\\
\hline
T Tau S         & 04:21:59.4 & +19:32:05.52 & 0.2615$_{-0.2167}^{+0.3476}$      & 37.7$_{-31.2}^{+50.1}$          & 1.7805$_{-1.6699}^{+0.8360}$ &  256.5$_{-240.5}^{+120.4}$ & 9.72$_{-0.44}^{+0.48}$ & 61.6$^{+8.8}_{-4.8}$ & 7.9$^{+3.7}_{-3.5}$\\
UY Aur B    & 04:51:47.3 & +30:47:12.53 & 0.0118$_{-0.0164}^{+0.0137}$  & 1.9$_{-2.5}^{+2.1}$             & 0.0427$_{-0.0240}^{+0.0527}$ &  6.7$_{-3.7}^{+8.2}$ & 5.78$_{-0.47}^{+0.81}$ & 25.6$^{+35.3}_{-30.6}$ & -35.0$^{+186.8}_{-73.1}$\\
RW Aur B    & 05:07:49.5 & +30:24:04.29 & 0.0716$_{-0.0082}^{+0.0069}$  &  10.9$_{-1.3}^{+1.1}$   & 0.0894$_{-0.0096}^{+0.0167}$ &  13.4$_{-1.6}^{+2.7}$ & 4.11$_{-0.89}^{+1.40}$ & 74.6$^{+3.8}_{-8.2}$ & 41.0$^{+3.6}_{-3.7}$\\
DK Tau B    & 04:30:44.4 & +26:01:23.20 & 0.0571$_{-0.0142}^{+0.0114}$  &  7.9$_{-1.8}^{+1.5}$            & 0.0679$_{-0.0179}^{+0.0209}$ &  9.5$_{-2.3}^{+2.7}$ & 2.45$_{-0.58}^{+1.89}$ & 78.0$^{+6.1}_{-11.0}$ & 28.0$^{+5.2}_{-5.4}$\\
HK Tau B    & 04:31:50.6 & +24:24:15.09 & 0.4337$_{-0.0027}^{+0.0018}$  &  57.7$_{-0.4}^{+0.2}$   & 0.5112$_{-0.0113}^{+0.0079}$ &  68.0$_{-1.5}^{+1.1}$ & 15.85$_{-0.28}^{+0.28}$ & 83.2$^{+0.2}_{-0.2}$ & 41.2$^{+0.2}_{-0.2}$\\
CIDA 9B     & 05:05:22.9 & +25:31:31.70 & Point Source & \nodata & \nodata & \nodata & 0.32$_{-0.03}^{+0.03}$ &\nodata & \nodata \\
HN Tau B    & 04:33:39.2 & +17:51:49.6 & 0.0009$_{-0.0001}^{+0.0097}$   &  0.2$_{-0.01}^{+1.3}$   & 0.0047$_{-0.0068}^{+0.1076}$ & 0.7$_{-0.9}^{+14.6}$ & 0.23$_{-0.02}^{+0.04}$ & 49.5$^{+52.3}_{-62.5}$ & -2.7$^{+120.7}_{-112.7}$\\
UZ Tau Wa   & 04:32:42.8 & +25:52:31.2 & 0.0991$_{-0.0022}^{+0.0021}$   &  13.0$_{-0.3}^{+0.3}$ & 0.1267$_{-0.0024}^{+0.0099}$ & 16.6$_{-0.3}^{+1.3}$ & 14.42$^{-0.24}_{+0.26}$ & 61.2$^{+1.1}_{-1.0}$ & 91.5$^{+0.8}_{-0.9}$\\
UZ Tau Wb   & 04:32:42.8 & +25:52:30.9 & 0.0982$_{-0.0020}^{+0.0020}$   &  12.8$_{-0.3}^{+0.3}$ & 0.1238$_{-0.0020}^{+0.0100}$ & 16.2$_{-0.3}^{+1.3}$ & 15.57$_{-0.22}^{+0.24}$ & 59.9$^{+0.9}_{-0.9}$ & 92.9$^{+0.8}_{-0.8}$\\

\hline
\end{tabular}
\tablefoot{RA and Dec indicate the coordinates of the center of the disks determined by our fits. }
\end{table*}

Our main goal was to obtain intensity profiles of the dust emission of the individual circumstellar disks in order to measure their dust radii.  

Following \citet{tazzari17}, \citet{tripathi17}, and \citet{trapman19} among others, we defined the disk radius (\rdust) as the radius containing 95\% of the 1.33 mm continuum flux, and the effective radius (\reff) as that containing 68\% of the continuum flux: 

\begin{align}
\label{eq:effective.radii}
\int_{0}^{R_{\rm disk,dust}} 2 \pi \cdot I(r) \cdot r \cdot dr / F_{\rm tot} = 0.95, \\ 
\int_{0}^{R_{\rm eff,dust}} 2 \pi \cdot I(r) \cdot r \cdot dr / F_{\rm tot} = 0.68,
\end{align}
where $F_{\rm tot}$ is the total continuum flux and $I(r)$ is the intensity profile of the emission as a function of disk radius. To estimate these two quantities we performed a fit of the observed continuum visibilities ($V_\mathrm{obs}$). 
To account for bandwidth smearing effects for disks that are located away from the phase center of the observations, we averaged the visibilities in each of the continuum spectral windows to one channel and we normalized the $u, v$ coordinates of each visibility point using its exact observing wavelength. 

We used \texttt{Galario} \citep{tazzari18} to compute the model visibilities ($V_\mathrm{mod}$) of a given axisymmetric brightness profile $I(r)$. 
We explored the parameter space using the Markov chain Monte Carlo  (MCMC) ensemble sampler \texttt{emcee} \citep{emcee} and we adopted a Gaussian likelihood:
\begin{equation}
\mathcal{L}\propto \exp\left(-\frac{1}{2}\,\chi^2\right)=\exp\left(-\frac{1}{2}\sum_{j=1}^{N}|V_\mathrm{mod}-V_\mathrm{obs}|^2w_j\right)\,,
\end{equation}
where $N$ is the total number of visibility points and $w_j$ is the weight corresponding to the $j$-th visibility. 
To fit the continuum visibilities of a multiple system made of $M$ components, we computed the total model visibilities as the sum of the visibilities of the individual components, namely
\begin{equation}
    V_\mathrm{mod}=\sum_{i=1}^M V_{\mathrm{mod}\,i}\,,
\end{equation}
where $V_{\mathrm{mod}\,i}$ is a function of the brightness profile parameters, of the offset with respect to the phase center ($\Delta$RA, $\Delta$Dec), of the north to east disk position angle (PA), and of the disk inclination ($i$).
Therefore, the total number of free parameters scales with the number of detected components in a multiple system since each disk is modeled  with an independent brightness profile and geometry. 

The functional form adopted to fit the individual disks is a power law  with exponential cutoff function, where the exponent of the power law ($\gamma1$) and that of the exponential cutoff ($\gamma2$) are independent:
\begin{equation}\label{eq::powexp}
I(r) = I_0 ~ r^{-\gamma1}  \exp\left(-\frac{r}{R_{\rm c}}\right)^{\gamma2}. 
\end{equation}
In this equation, $I_0$ is such that
\begin{equation}
I_0  = F_{\rm tot} / \int_0^\infty 2\pi ~ r ~ r^{-\gamma1} ~ \exp\left(-\frac{r}{R_{\rm c}}\right)^{\gamma2} ~ dr.
\end{equation}
A functional form is preferred to a 
simple
Gaussian,  to the Nuker profile \citep[e.g.,][]{tripathi17}, or to a self-similar solution \citep[e.g.,][]{tazzari17} as it provides a better description of the cutoff of the outer disk  than  a Gaussian profile, it has one less parameter than the Nuker profile, and the power-law exponent is not related to the exponential cutoff  as it is  in the similarity solution. We verified that the values of \reff \ obtained with this functional form are compatible with those obtained using a Gaussian function, and that both the \reff \ and \rdust \ values are compatible with the results obtained using a Nuker profile. The same functional form is used in the companion paper by \citet{long19} to describe the intensity profile of smooth disks around single stars, and this choice enables us to make a direct comparison with the rest of the disks in Taurus.

Only for two targets (CIDA~9 and UZ Tau E) did we need to adopt a different functional form to describe the brightness profile given the presence of large-scale rings in their profiles.  Following \citet{long18}, we used a single Gaussian ring profile for CIDA~9A, and a point-source for the secondary unresolved disk, while we made use of three concentric Gaussian rings to fit the profile of UZ~Tau~E. The difference with respect to the fit of \citet{long18} is that here we fitted all the disks in the system, while \citet{long18} considered only the primary disk. The results for the primary are similar to those obtained by \citet{long18}. 
Recently, \citet{czekala19} published ALMA observations of the dust continuum, CO, $^{13}$CO, and C$^{18}$O of UZ Tau E, demonstrating that this system is nearly coplanar by finding similar values as reported here for the disk inclination.

The number of walkers and steps needed to achieve the convergence of the chain varied depending on the number of disk components: typically we used 150 walkers and $\sim$20000 steps for the cases where only one disk is detected (8 parameters), 200 walkers and more than 25000 steps for the cases where two disks are detected (16 parameters), and 400 walkers and 31000 steps for UZ Tau where three disks are detected (29 parameters). These steps are sufficient to reach convergence, and the last 3000 - 10000 steps are used to sample the posterior distribution.

The adopted final parameters of the models are taken as the median of the posterior probability function of each parameter. The adopted brightness profiles and the best fit parameters are discussed in Appendix~\ref{app::bestfit} and listed in Tables~\ref{tab::best_fit_pars}-\ref{tab::best_fit_pars2} (inclinations and position angles are also given in Table~\ref{tab::fits_res}). The uncertainties are estimated as the central interval between the 16th and 84th percentiles. These uncertainties are likely underestimated. The values of the normalized $\chi^2$ obtained with the fit are usually $\sim$3.5, with the only exception of T~Tau, whose fit yields  a value of $\chi^2$ of 6.4, possibly due to the presence of substructures in both detected disks  (see Appendix~\ref{app::ttau} for discussion on this target). The reason why a very good fit with small residuals (see, e.g., Fig~\ref{fig::DKTau_example})  yields  values of $\chi^2 > 1$  is that the weights in \texttt{CASA} are underestimated by a factor typically very close to this value of $\chi^2$. We apply this correction factor and report the correct uncertainties in the tables.

Once the best parameters for the model of each individual disk are found, we can derive \rdust \ and \reff.  
The values for \reff \ and \rdust \ derived for the primary and secondary components of the multiple systems are given in Table~\ref{tab::fits_res} together with the fitted coordinates and the flux density of the individual disks ($F_{\rm tot}$), the latter obtained by taking into account the inclination of the model disk, reported together with the position angles of the disks in the same table. The uncertainties on these values are obtained by calculating the radii and the flux with the values obtained with 1000 different chains, and deriving the 16th and 84th percentiles of the distribution. The dust radii measured for all the disks around the primaries have small uncertainties, while large uncertainties suggest that the dust radii estimates for T~Tau~S, UY~Aur~B, and HN~Tau~B are not well constrained.

   \begin{figure}
   \centering
  \includegraphics[width=0.45\textwidth]{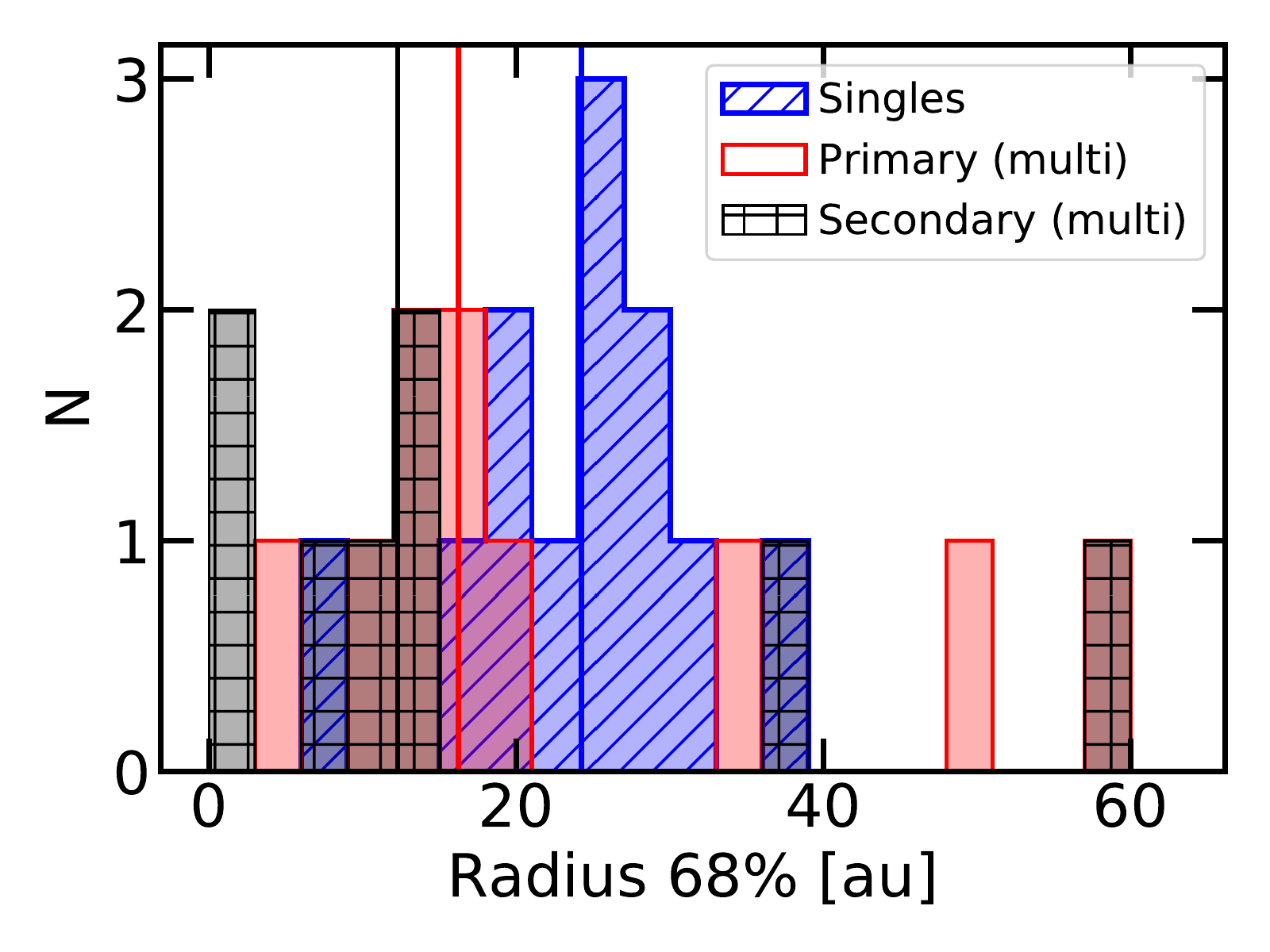}
  \includegraphics[width=0.45\textwidth]{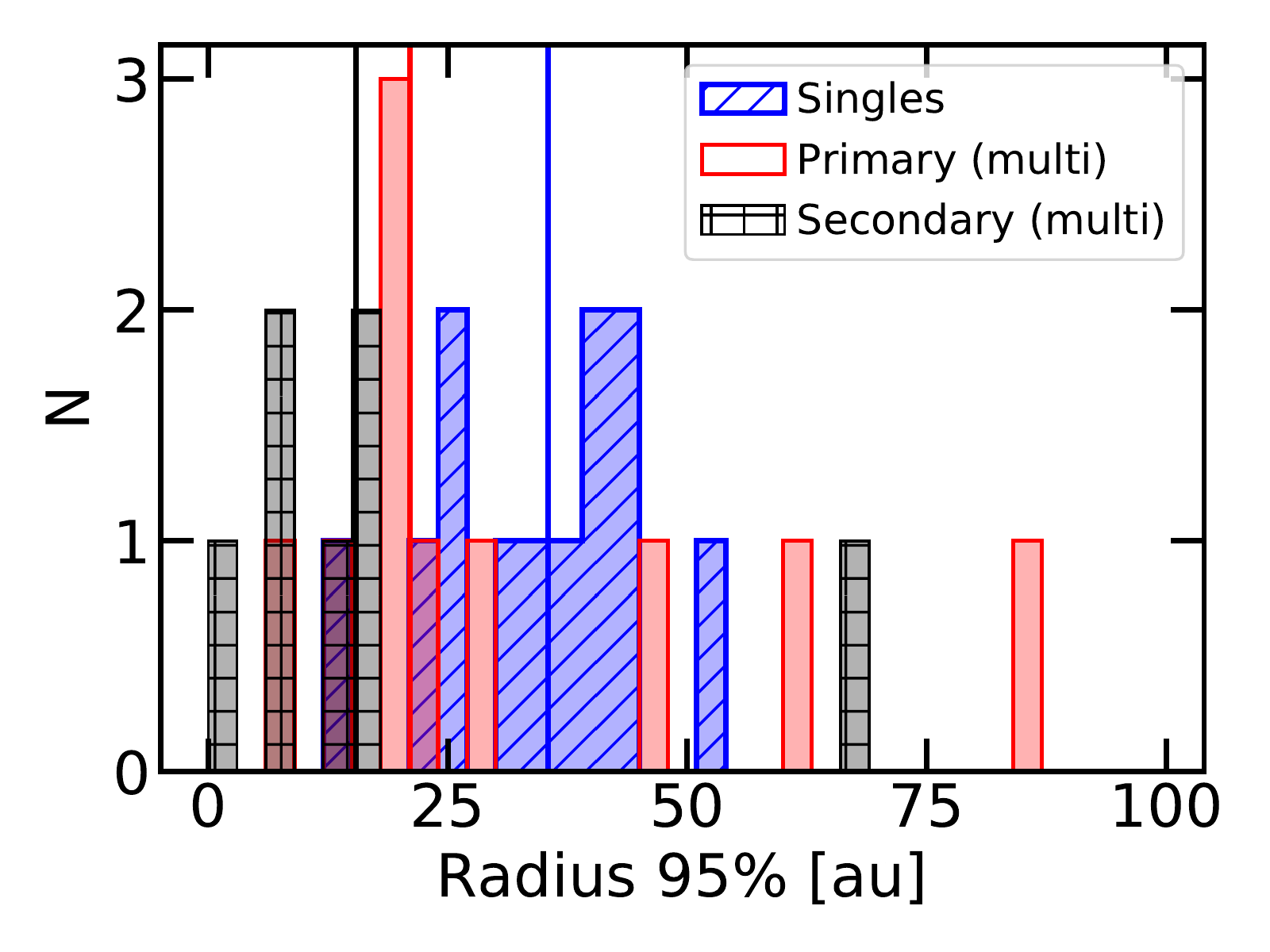}
   \caption{Histogram showing the disk sizes for disks around smooth single stars \citep[from][]{long19} and disks around the stars in multiple stellar systems. The black, red, and blue lines show the median of the distributions for smooth single disks, primaries, and secondaries, respectively. The radius of T~Tau~S, which is an unresolved binary (see Appendix~\ref{app::ttau}), is outside the plotted range.
    The two disks around primary stars with \reff$>$40 au and \rdust$>$60 au show clear substructures in the dust distribution \citep[CIDA~9A, UZ~Tau~E,][]{long18}. HK~Tau~B, an edge-on disk, also has  \rdust$>$60 au. See text for discussion. }     \label{fig::hist_rad_sin_bin_au}%
    \end{figure}

   \begin{figure}
   \centering
  \includegraphics[width=0.45\textwidth]{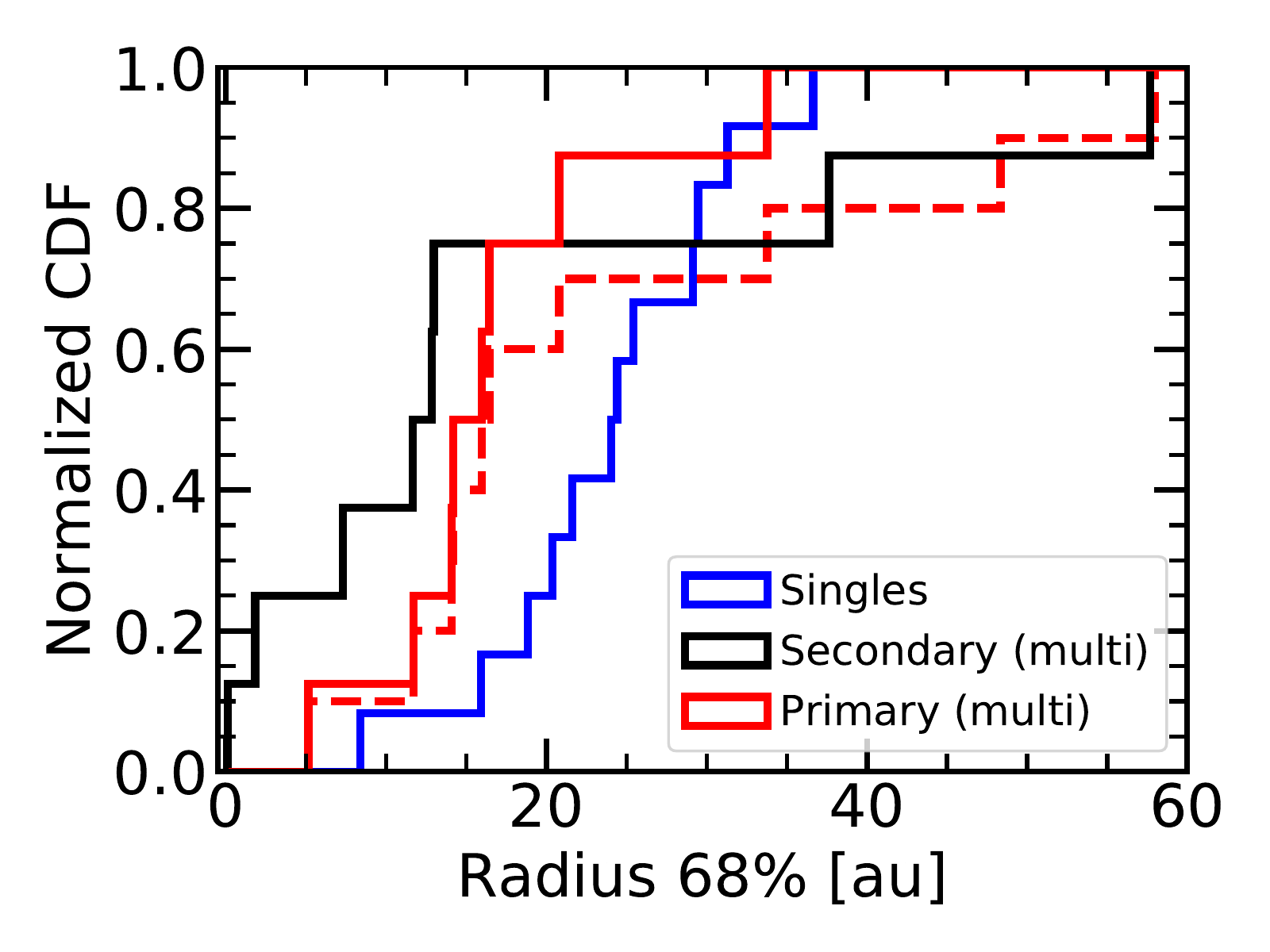}
  \includegraphics[width=0.45\textwidth]{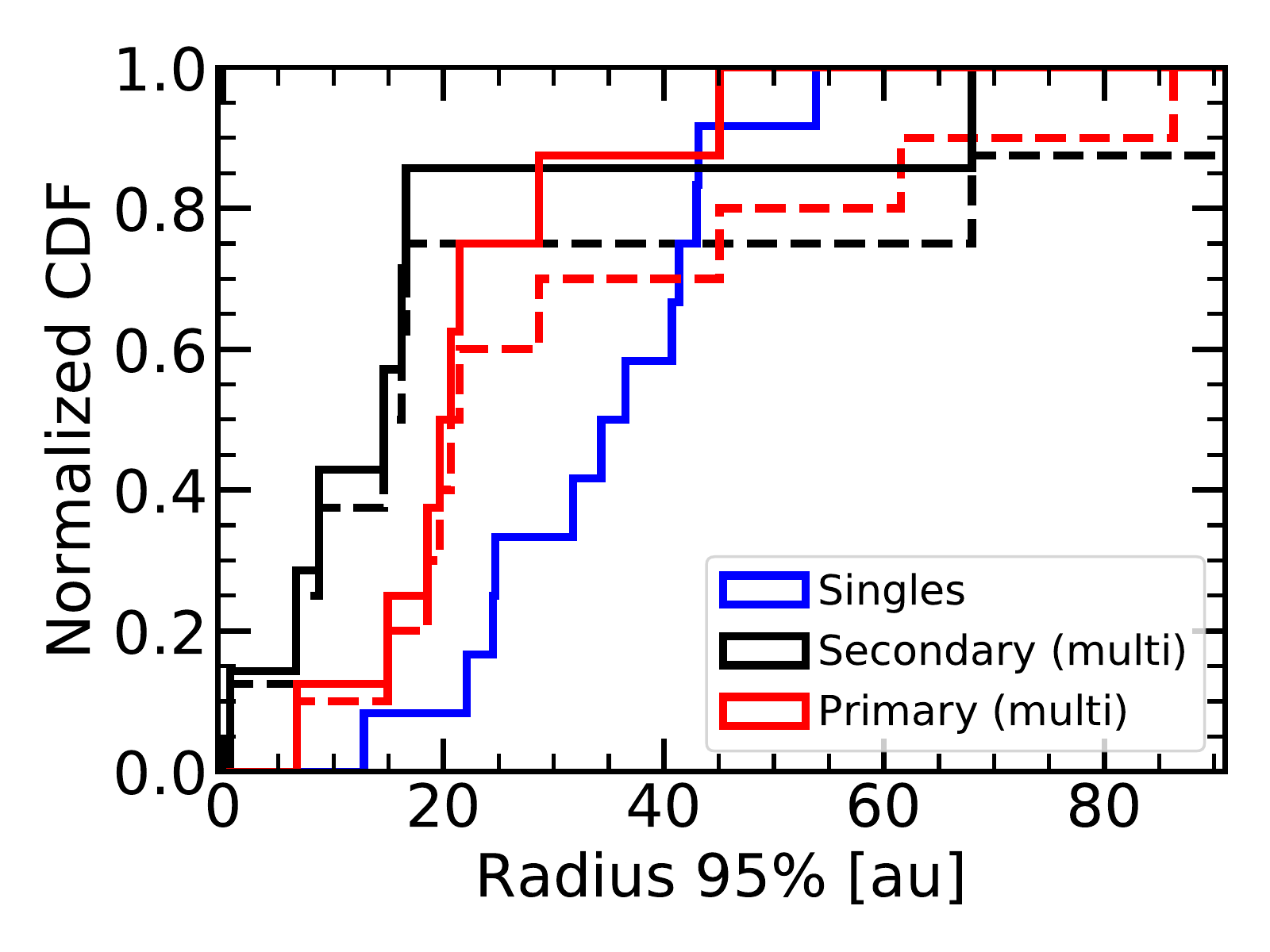}
   \caption{Normalized cumulative distribution function for the disk sizes for disks around smooth single stars \citep[blue, from][]{long19} and disks around the stars in multiple stellar systems (red for primary component, black for secondary). The red dashed line shows the distribution for the disks around primary stars in multiple systems including CIDA 9 A and UZ Tau E, who are known to have ring-like structures \citep{long18}, and the black dashed line shows the distribution including T~Tau~S, which is an unresolved binary (see Appendix~\ref{app::ttau}).}
              \label{fig::rad_sin_bin_au}%
    \end{figure}

\section{Results}\label{sect::results}

In a companion paper, \citet{long19} have used our same analysis strategy to determine the intensity profile of the dust emission in the 12 smooth single disks observed in the ALMA survey of disk structures in Taurus (see Sect.~\ref{sect::data}). These targets have similar stellar properties to those of the disks in multiple systems presented here (see Fig.~\ref{fig::masses} and Sect.~\ref{sect::sample}) and are located in the same star-forming region. Their measurements are thus the ideal sample for  comparing the properties of disks in isolated systems against those in disks located in multiple systems. 

Here we exclude the disks showing prominent gaps and rings structures discussed by \citet{long18}. However, CIDA~9A and UZ~Tau~E are members of multiple stellar systems, but  they show rings in their disks at the same time. We will note when these two targets are included in our analysis and when they are not.

   \begin{figure}
   \centering
  \includegraphics[width=0.45\textwidth]{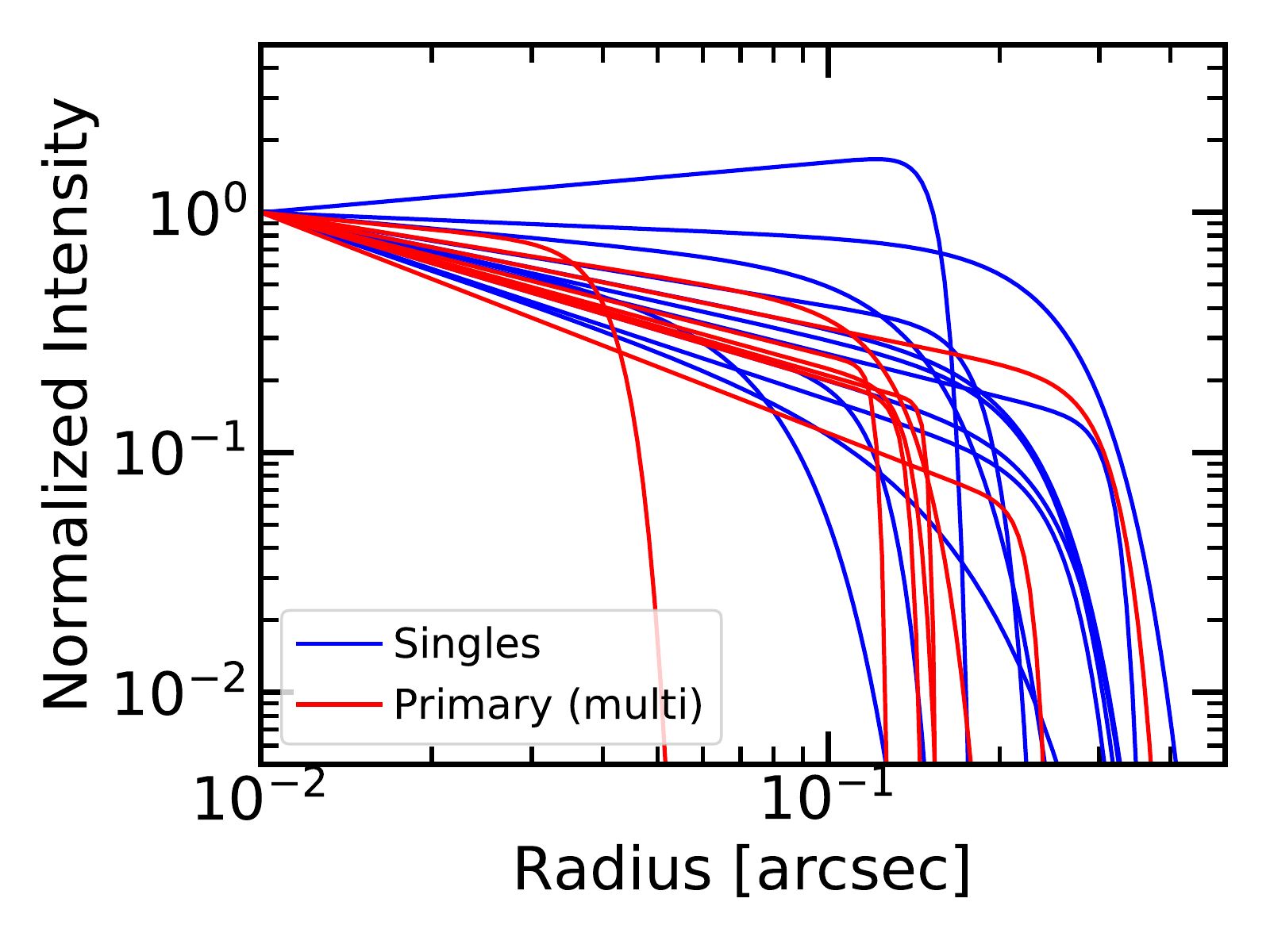}
   \caption{Normalized intensity profile for disks around smooth single objects \citep{long19} and for disks around the primary star in multiple systems, with the exclusion of CIDA~9A and UZ~Tau~E. 
   }
              \label{fig::profiles_bin_singl}%
    \end{figure}

\subsection{Disk sizes in multiple systems}\label{sect::radii}

We first compare the dust sizes of the disks in multiple systems versus the isolated disks to test whether the presence of companions correlates with disks being smaller than if the disks evolve in isolation. 
The dust radii are measured from the radius that encircles 68\% and 95\% of the millimeter  emission.  The dust radii  of disks around single stars, primary stars in multiple systems, and detected secondaries or tertiary stars in multiple systems are shown in Figs.~\ref{fig::hist_rad_sin_bin_au} and~\ref{fig::rad_sin_bin_au}. In both the histograms and the cumulative distributions, the median value of the dust radii distribution (\rdust=21 au and \rdust=15 au, respectively) is smaller in multiple systems than in single systems (\rdust=34 au).   We perform also the Kolmogorov--Smirnov  (K-S) two-sided test for the null hypothesis that the samples of dust radii around single stars and around primary stars in multiple systems are drawn from the same continuous distribution. The results exclude this hypothesis with p-values$<10^{-5}$, thus confirming that the distributions of dust radii of disks around single stars or in multiple systems are statistically different. This same result is found  when excluding the two disks around primary stars showing substructures (CIDA~9A and UZ~Tau~E) and when comparing the dust radii of disks around primary and secondary stars. 

While the bulk of the distribution of disk dust radii is statistically significantly smaller for disks around primary stars in multiple systems with respect to disks around single objects, a few outliers are present in the former group. In particular, the disks around CIDA~9A and UZ~Tau~E have dust radii \rdust$>$60 au, which makes them larger than any smooth disk around single objects in our sample. These two disks also show large-scale substructures \citep{long18}, which could be related to dust traps helping to keep the millimeter dust at larger radii \citep[e.g.,][]{pinilla12}. When comparing their dust radii with those of disks around single objects showing prominent substructures, both disks are below the median values of the distribution, implying that these disks are smaller than typical disks with substructures around single stars. 

This analysis shows that the dust radii of the disks around secondary or tertiary components of multiple systems are statistically smaller than  the sizes of the disks around primary stars in multiple systems and of disks around single stars. Two main outliers are however present:  
T~Tau~S and HK~Tau~B. The fit to T Tau S is poor, possibly because the system is a close binary (see Sect.~\ref{app::ttau}), and  the large uncertainties on the value of \rdust \ do not allow us to make any conclusion. HK Tau B is a disk observed edge-on, with optical depth effects that might increase the observed size of the emission of the disk at 1.3 mm.

   \begin{figure}
   \centering
  \includegraphics[width=0.45\textwidth]{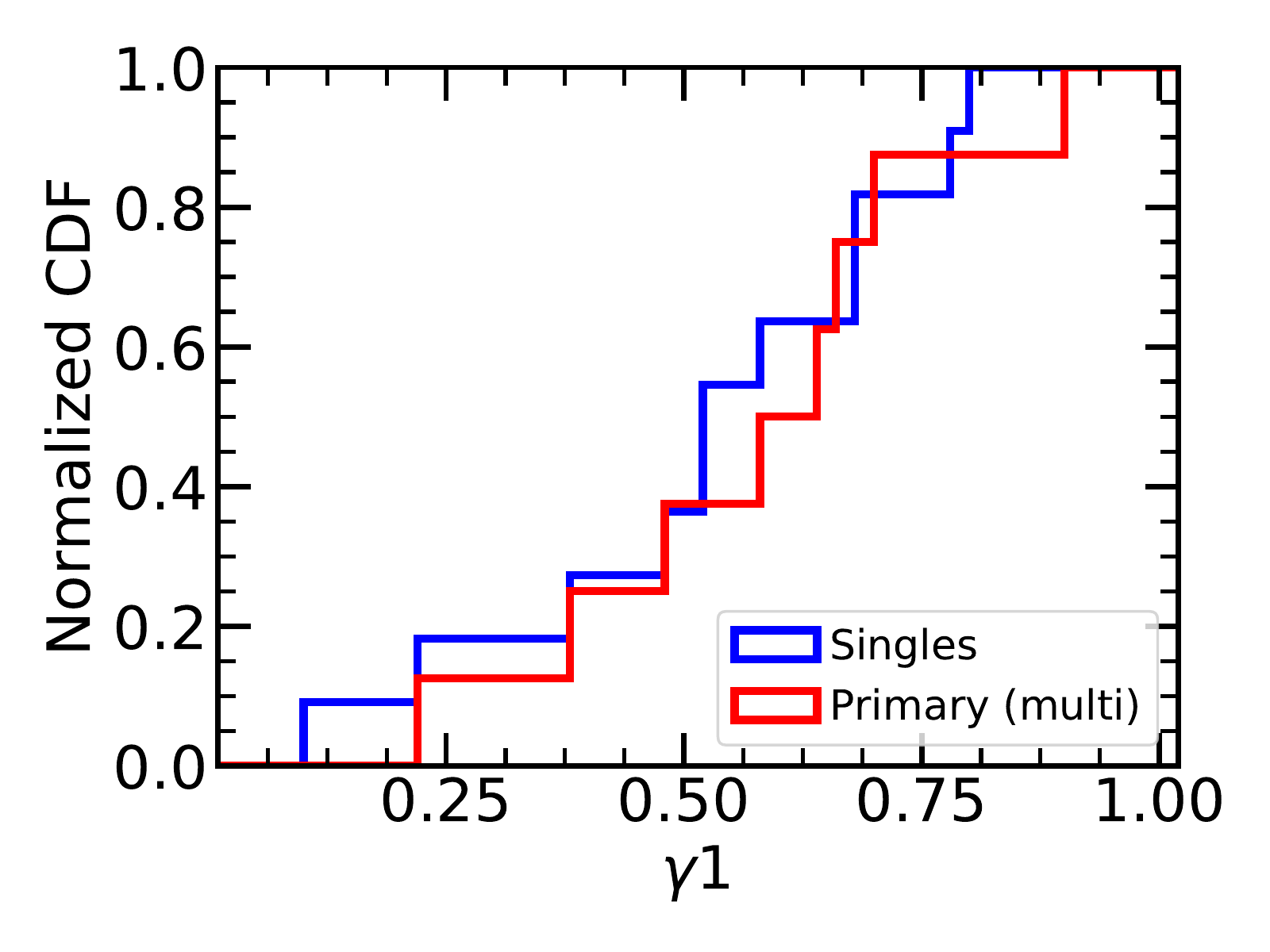}
  \includegraphics[width=0.45\textwidth]{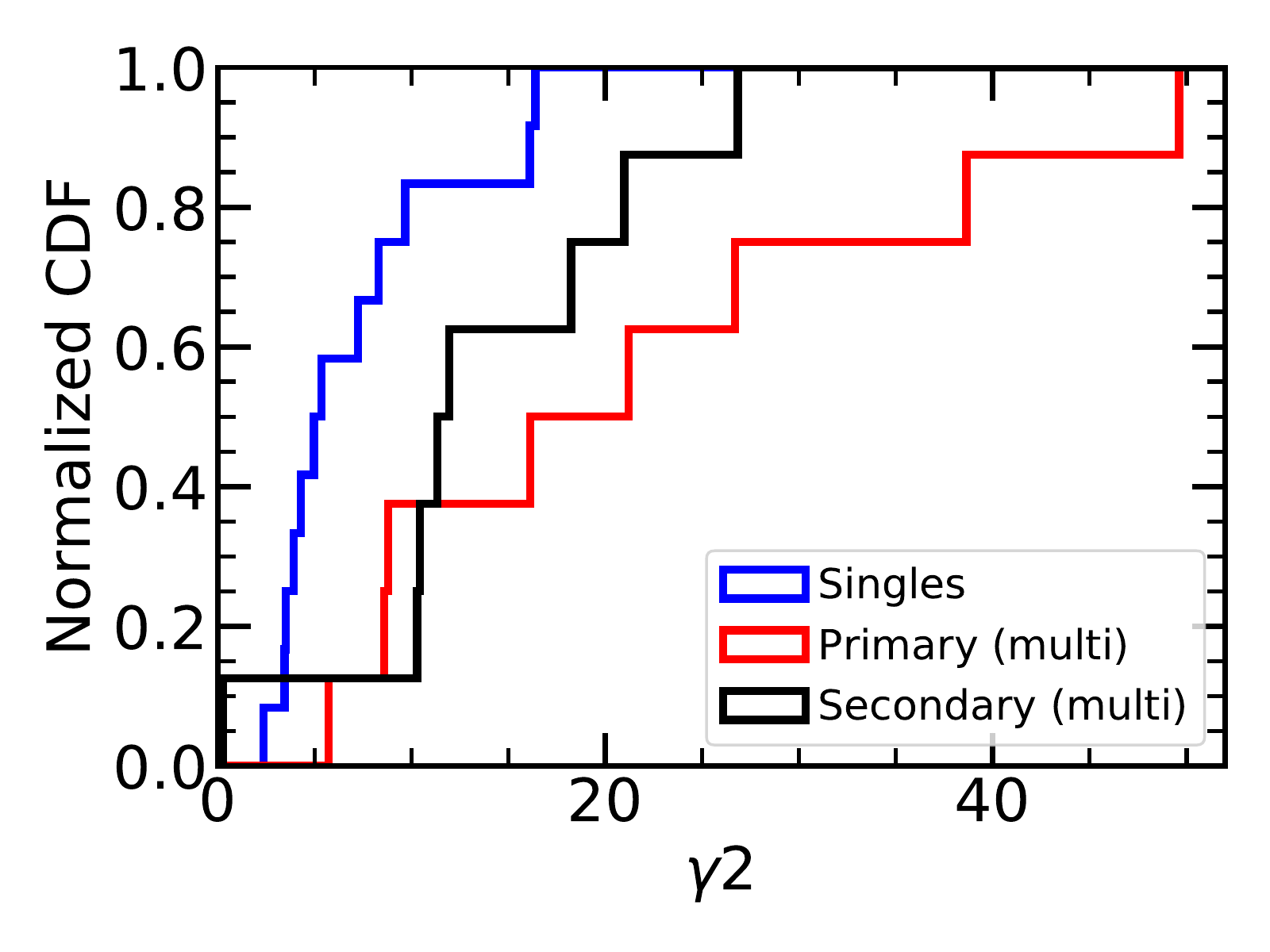}
   \caption{Normalized cumulative distribution function for the parameters of Eq.~\ref{eq::powexp} $\gamma$1 (top) and $\gamma$2 (bottom) for smooth disks around single stars \citep{long19} and smooth disks around the primary star in multiple stellar systems. 
   }
              \label{fig::gamma2_sin_bin}%
    \end{figure}

\subsection{Disk surface brightness profiles}\label{sect::profiles}

In this section, we compare the the overall shape of the brightness profiles of the disks in multiple and single stellar systems to evaluate whether a difference is observed, possibly due to the effect of tidal truncation by the companions. It should be immediately noted that our observations probe the dust emission profile, and not the gas emission. The latter is directly affected by dynamical interactions, and this effect on the gas surface density can impact the drift and growth of dust particles in the disk, and thus the shape of the dust brightness profile. 
In order to be able to perform the comparison, we consider in this subsection only the targets that have been fitted with a power law plus an exponential cutoff profile (Eq.~\ref{eq::powexp}), which means all the smooth single disks of \citet{long19} and the multiple disks analyzed here with the exclusion of CIDA~9 and UZ~Tau~E.

The normalized intensity profiles as a function of radius for all the disks around primary stars and single stars are shown in Fig.~\ref{fig::profiles_bin_singl}. Given that the peak S/N of our data is always higher than 175 for the primary components of the systems studied here and for the single disks \citep{long19}, our data allow us to probe the entire brightness dynamic range shown in this figure. The target showing the most compact profile is UY~Aur~A, while the other disks in multiple systems show  profiles that resemble that of smooth single disks in the inner regions (i.e., with a much more abrupt exponential cutoff at outer radii).

To better quantify these behaviors, we compare in Fig.~\ref{fig::gamma2_sin_bin} both the exponent of the power-law part of the intensity profile, $\gamma1$, which traces the inner part of the disk, and that of the exponential cutoff for the smooth single disks and for the disks in multiple systems, $\gamma2$. The distributions of the power-law exponent are statistically indistinguishable for the disks in single or multiple systems, with a p-value of 1. Instead, the values of $\gamma2$, the exponent of the exponential cutoff, are statistically smaller for smooth disks around single stars with respect to the ones for disks in multiple stellar systems, with p-values$<10^{-5}$ when performing a K-S  test of the two distributions.  

This result suggests that the disks around multiple stellar systems present a statistically significant sharper outer edge in the dust emission than the disks in isolated systems. Even at the high resolution of our observations, characterizing the steep brightness drop at the disk's outer edge is a challenging task, and we thus caution on the uncertainties of the inferred $\gamma2$ values (Table~\ref{tab::fits_res}).

\subsection{Relative inclinations of disks in multiple systems}\label{sect::incl}
The amount of alignment of the plane of rotation of disks in a multiple systems can be used as a constraint to star formation models \citep[e.g.,][]{bate18}. Although dynamical evolution can alter the relative inclination of the disks one with the other and with respect to the orbital plane, it is instructive to explore the observed values of the relative inclinations for the disks in our sample to be used to constrain disk evolution models in multiple systems.

\begin{figure}
   \centering
  \includegraphics[width=0.45\textwidth]{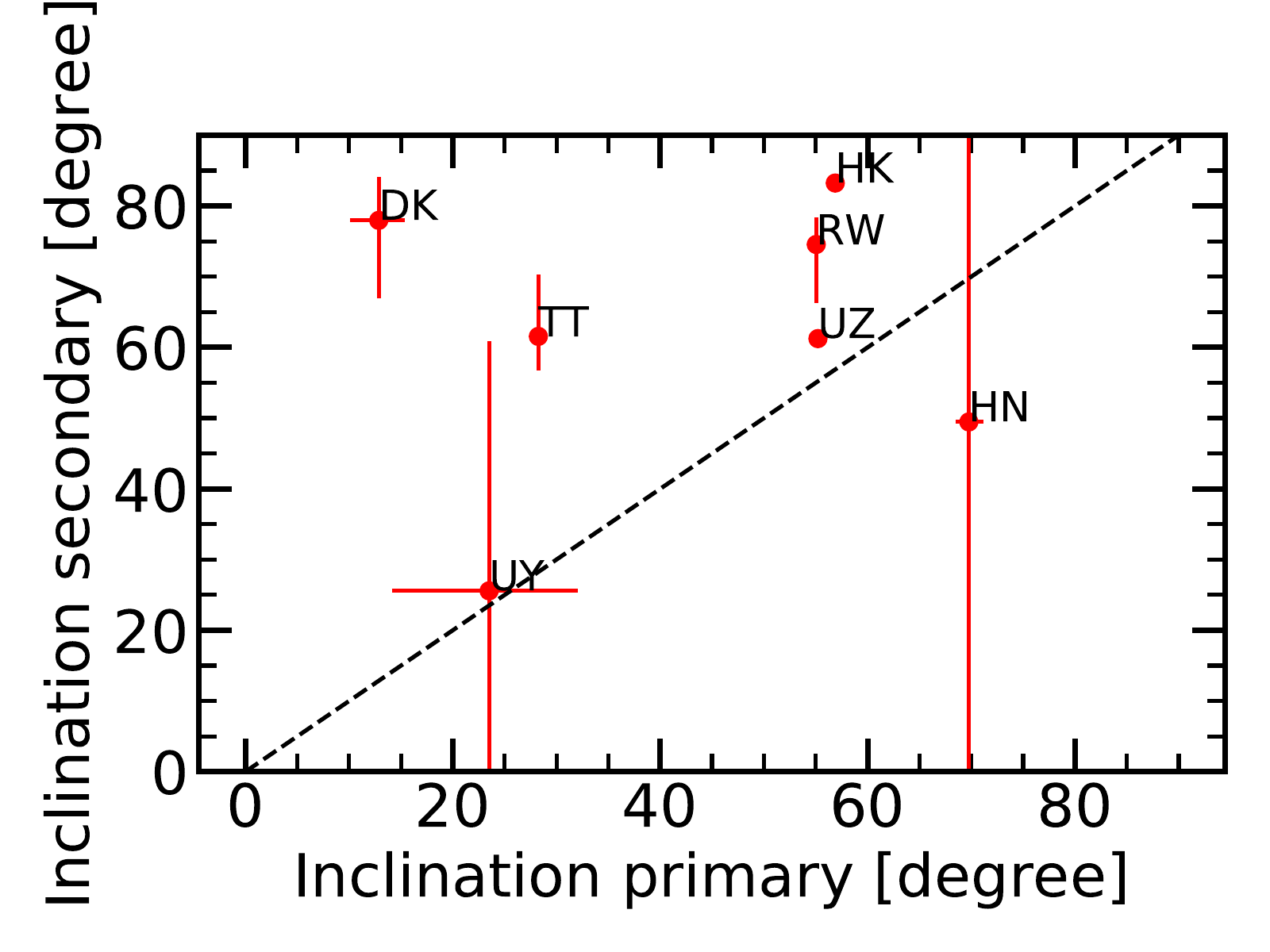}
\includegraphics[width=0.45\textwidth]{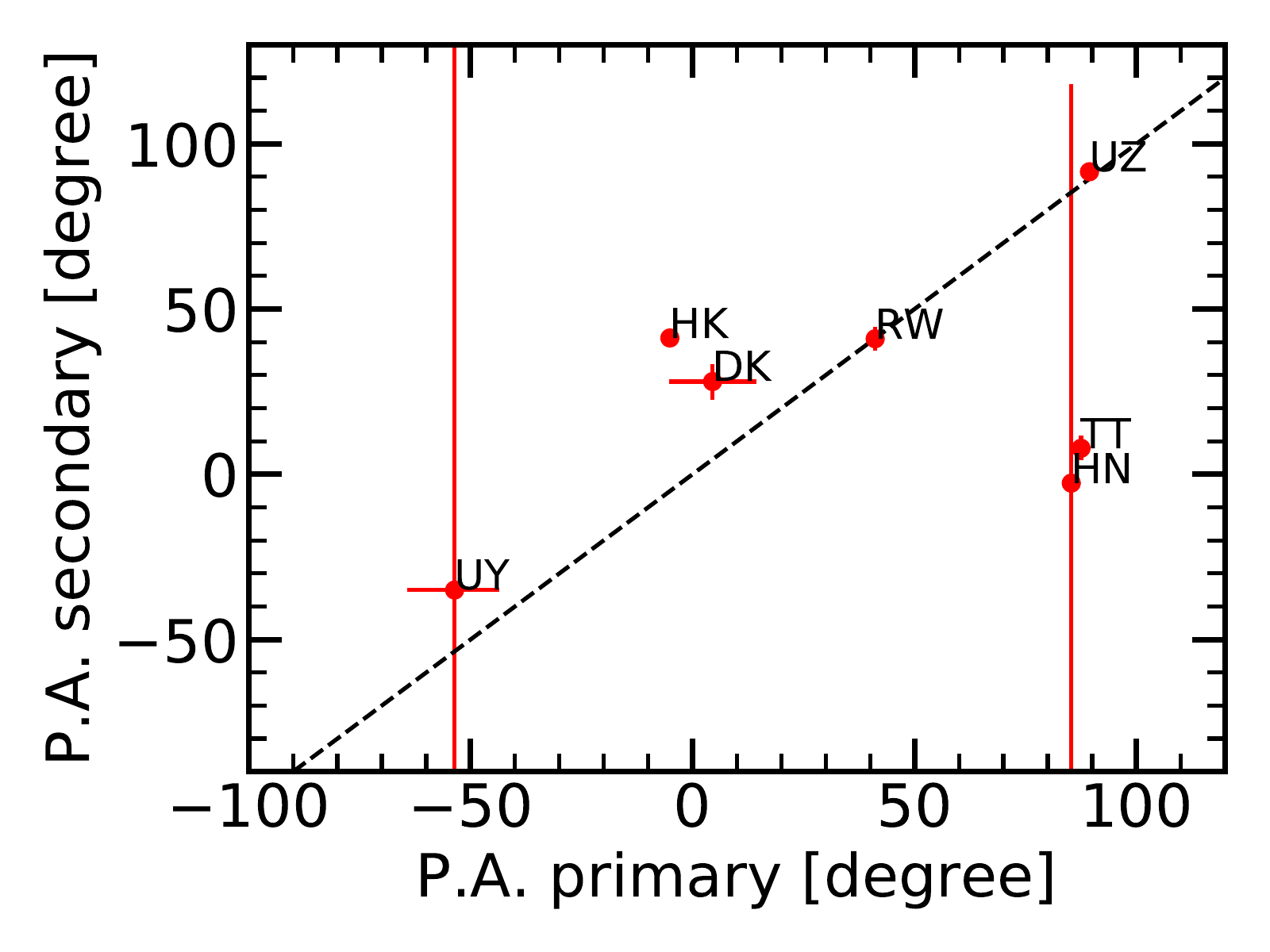}
\caption{Comparison between the inclination (top) and position angles (bottom) in degrees for the disks around the primary and  the secondary in each multiple stellar system where both disks are detected and resolved. The first two letters of the names are shown as labels. The point labeled UZ shows the value of the eastern component (primary) vs that of the Wa component (secondary). Both disks in the western component have a very similar inclination of 60\degree. }
              \label{fig::incl}%
    \end{figure}

In Fig.~\ref{fig::incl} we investigate the alignment of the disks around the two components of the binaries using the information on disk inclinations and position angles reported in Table~\ref{tab::fits_res}. In the case of the UZ~Tau triple system, we show the comparison of the inclination of the eastern component (primary) and the western a (Wa) component. Both disks in the western pair have a very similar inclination of 60\degree. The fit to HN~Tau~B, a disk around a very low-mass secondary star, has a large error bar because the disk is unresolved (or only marginally resolved) in our observations. In general, the secondary component is found to be on a more inclined plane with respect to the primary, and  the position angles of the disks are  usually different.  Only the disks in the RW~Aur and UZ~Tau systems have similar position angles. 

\begin{figure}
   \centering
  \includegraphics[width=0.45\textwidth]{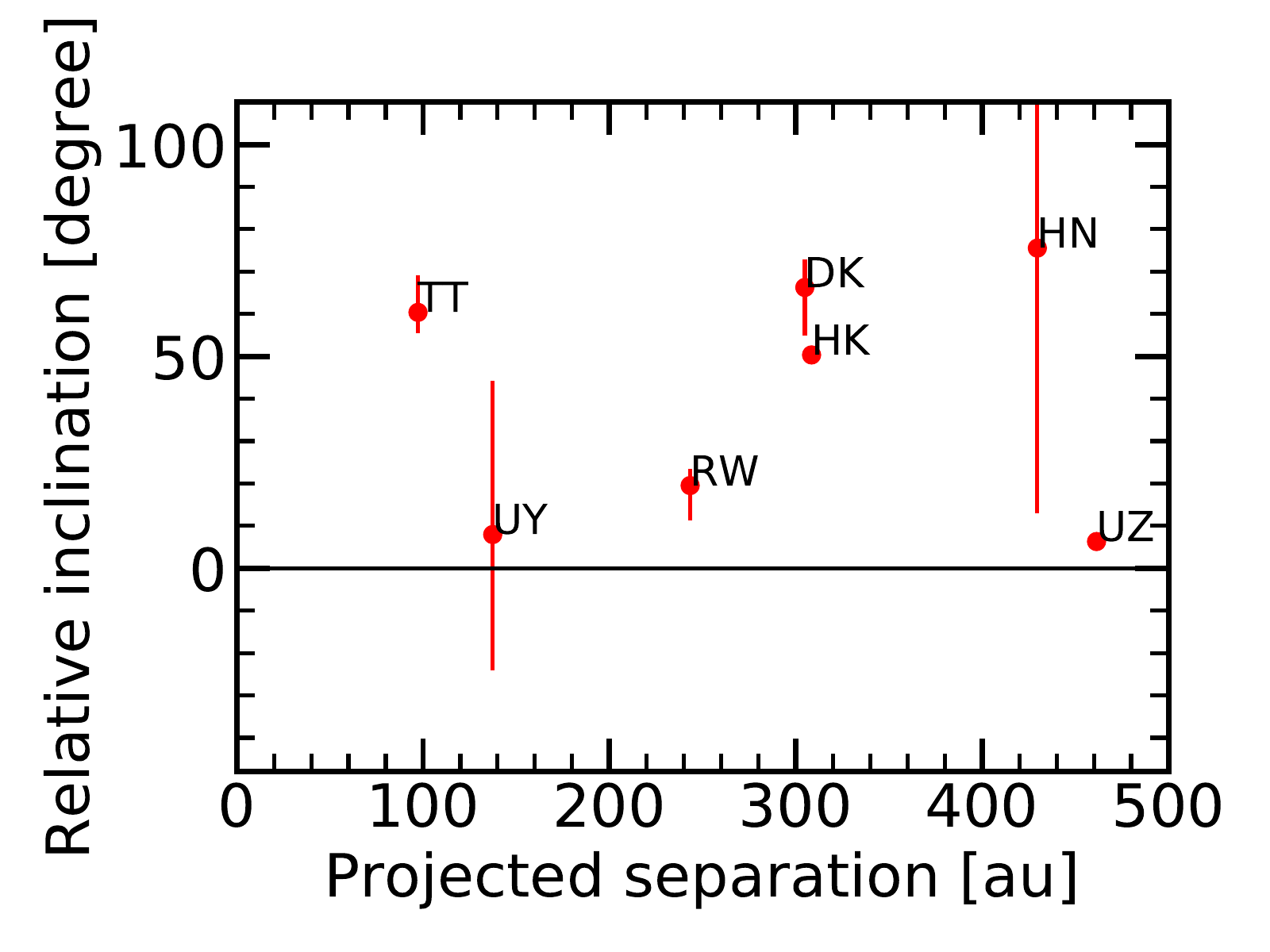}
   \caption{Relative inclination in degrees of the two disks in each multiple stellar systems as a function of their separation. The first two letters of the names are shown as labels. The point labeled UZ shows the value of the eastern component (primary) vs that of the Wa component (secondary). Both disks in the western component have a very similar inclination of 60\degree. }
              \label{fig::incl_rel}%
    \end{figure}

We  computed the relative inclination of the disks in each system using the spherical law of cosines ($\cos(\Delta i) = \cos(i_1) \cdot \cos(i_2) + \sin(i_1) \cdot \sin(i_2) \cos(\Omega_1-\Omega_2)$, with $\Omega$ the position angle), and show the results as a function of the projected separation in Fig.~\ref{fig::incl_rel}.
Overall, the only systems with relative inclination close to 0\degree\ are UZ~Tau and UY~Aur, even though the latter has a large uncertainty in the derived inclination for the secondary, and for RW~Aur ($\sim$20\degree). Instead, the relative inclination of the disks around the primary and secondary components is  large for the other four systems with inclinations measured for both components. In our sample of binaries with projected separations larger than $\sim$100 au there is no clear trend of a dependence of the relative inclination with the projected separation. 

A previous attempt to measure the relative disk inclination in binaries was performed by means of polarimetric imaging in $K$ band by \citet{jensen04}. While they observed alignment within $\lesssim$20\degree \ in their targets, the uncertainties in their method does not allow us  to rule out that the relative inclinations can be higher. Their finding that most disks in binary systems are not coplanar is reinforced here by our resolved observations of the disk around the individual components.

 %

 %

   \begin{figure}
   \centering
  \includegraphics[width=0.45\textwidth]{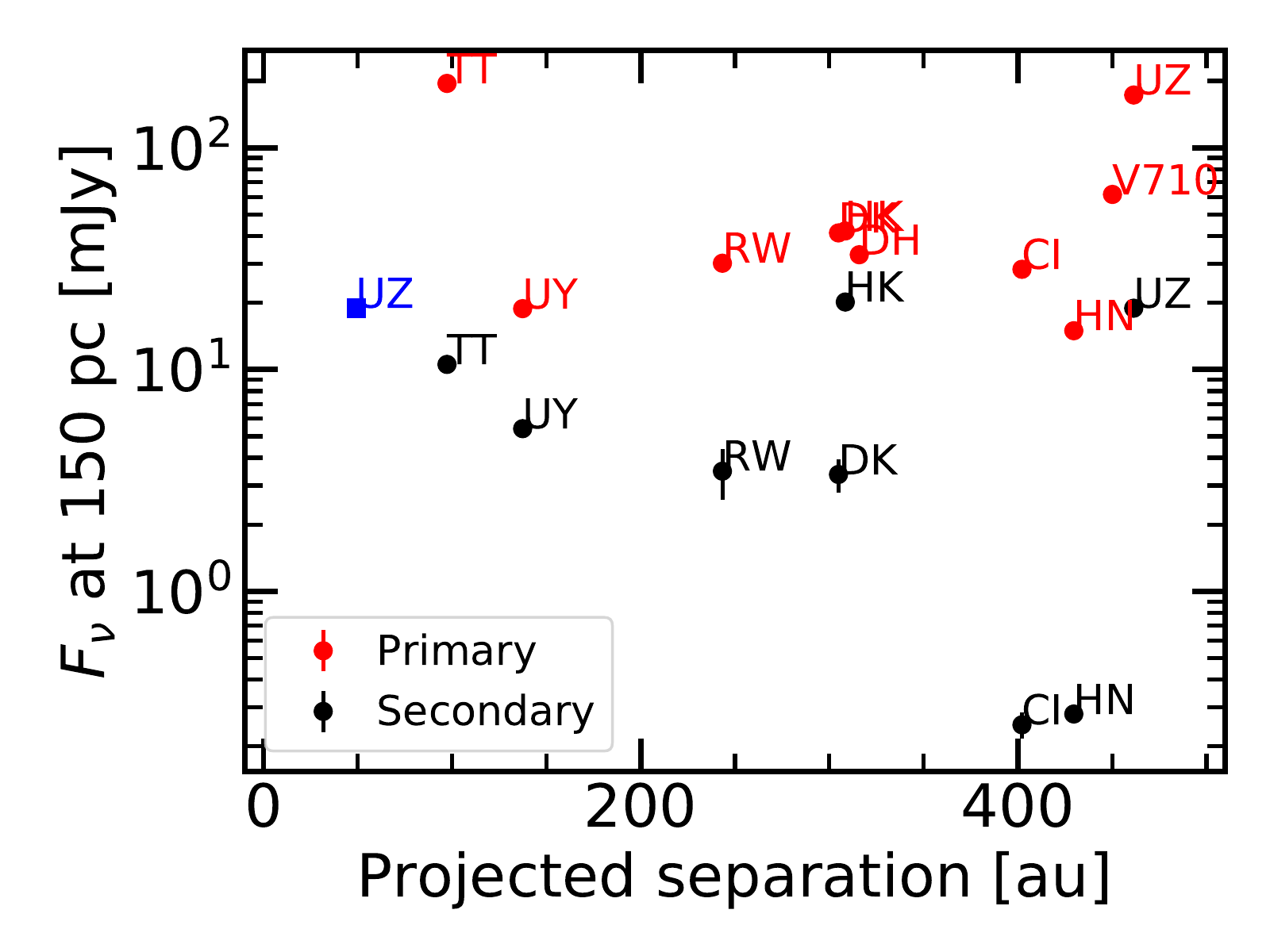}
\includegraphics[width=0.45\textwidth]{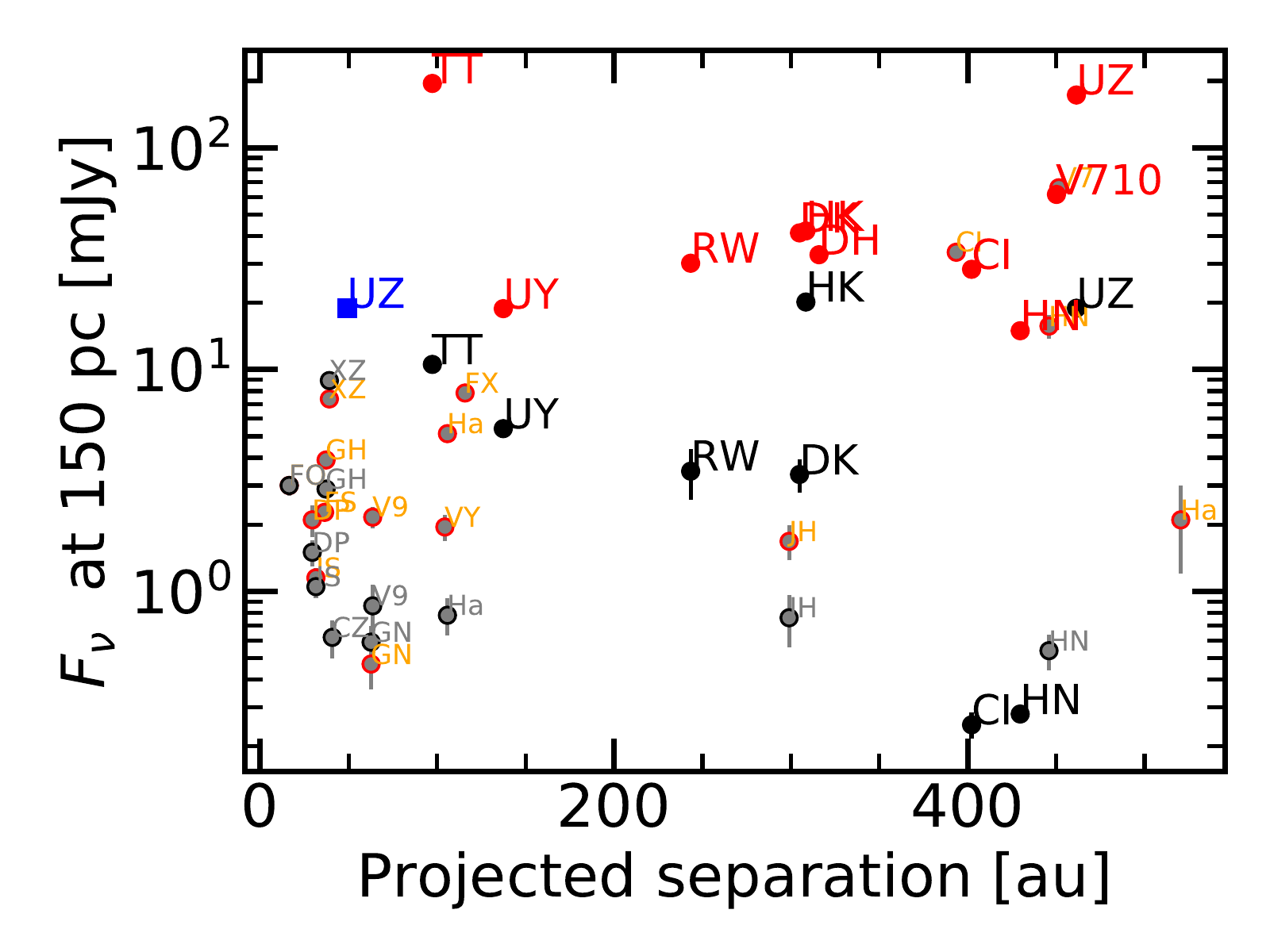}
   \caption{Observed separation of the components in the multiple systems vs the measured flux densities of the individual disks. Only the first two letters of the names are shown, e.g., CI is for CIDA 9. 
The blue point for UZ Tau W represents the values for the flux of the western components of the UZ Tau system, assuming only the distance between Wa and Wb. The top panel shows only the objects studied here, while the bottom panel includes the targets studied by \citet{AJ19}, reported with gray symbols with red and black edges for the primary and secondary disks, respectively.}
              \label{fig::sep_flux}%
    \end{figure}

 %

   \begin{figure}
   \centering
  \includegraphics[width=0.45\textwidth]{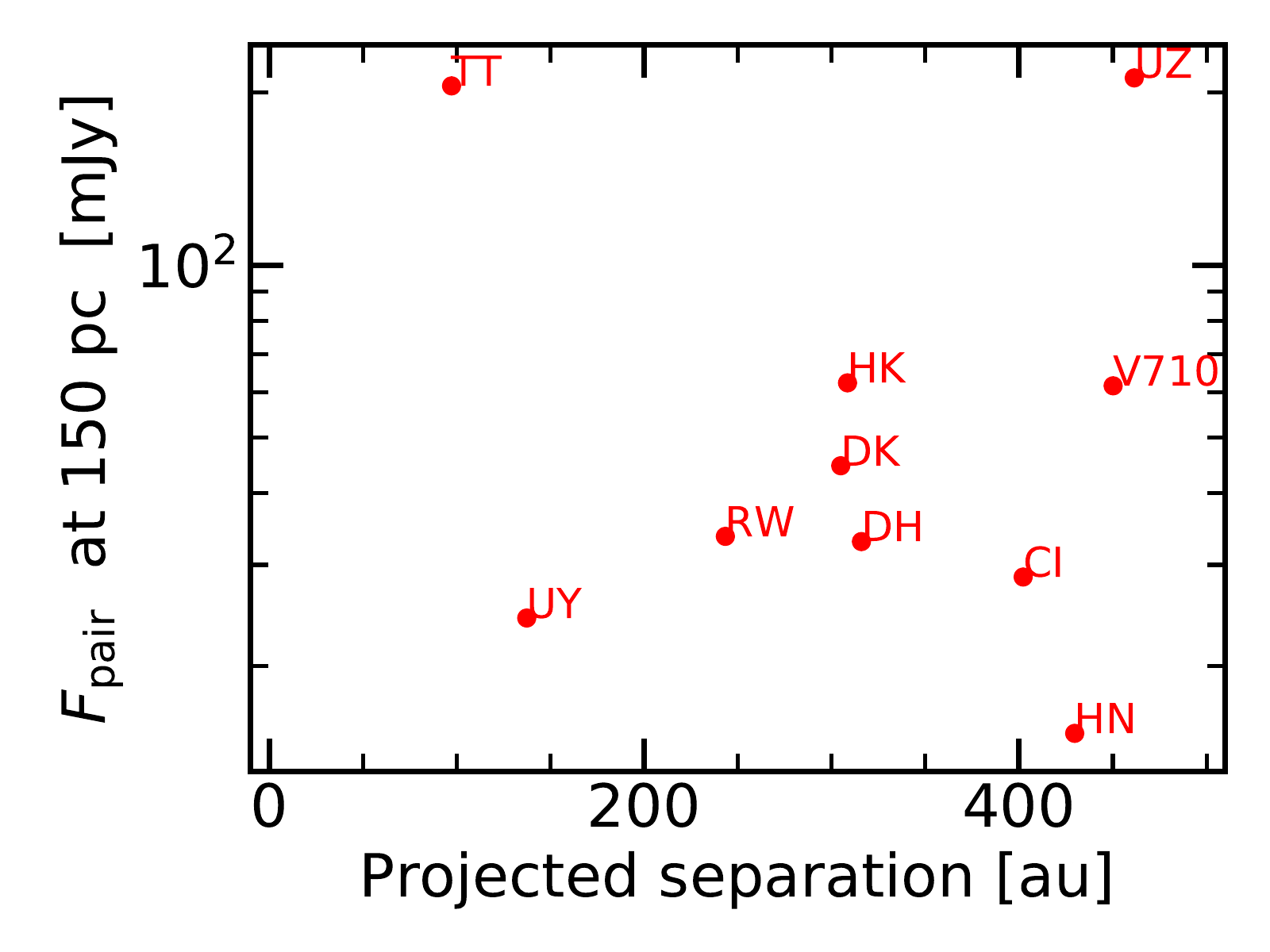}
  \caption{Observed separation of the components in the multiple systems vs the sum of the rescaled measured flux densities of the individual disks. Only the first two letters of the names are shown as labels (e.g., CI is for CIDA 9). }
        \label{fig::sep_flux_sum}%
    \end{figure}

\subsection{Disk flux versus separation}
Previous work has shown that more compact binary systems show a smaller total millimeter continuum flux density than wider binary systems \citep[e.g.,][]{harris12,AJ19}. Since the disk evolution is expected to be faster in tighter binary systems undergoing close-encounters and strong tidal interactions \citep[e.g.,][]{CP93}, these systems should have lower disk mass, and therefore weaker millimeter continuum flux density, than single disks and binaries on wider orbits. 

Although our sample size is smaller than previous samples, which were comprising 20--40 objects \citep[e.g.,][]{harris12,AJ19}, we also explore this relation in our data, which have the advantage of resolving the individual disks and detecting eight to ten secondary disks in the systems. The rescaled flux density of the individual components of the multiple systems is shown in Fig.~\ref{fig::sep_flux} and the combined rescaled flux of the pairs in Fig.~\ref{fig::sep_flux_sum}, in both cases as a function of the projected separation. We do not find a clear correlation between the two components, possibly due to the low number statistics of our sample and the relatively large projected separations of the systems studied here. When combining our data with those by \citet{AJ19}, there is a hint of a tentative trend of increasing millimeter flux for the primary disk with separations from $\lesssim$50 au to $>$400 au (bottom panel of Fig.~\ref{fig::sep_flux}). We note from Fig.~\ref{fig::sep_flux} that the flux in the secondary component is always found to be smaller than that in the primary disk in our data, as previously reported \citep[e.g.,][]{AJ19}.


\section{Quantifying the effects of tidal truncation}\label{sect::trunc_models}
The effects of tidal truncation due to the presence of another disk in a multiple system can be estimated analytically \citep[e.g.,][]{PP77,AL94}. Our data are ideal tests of these analytic predictions since we are able to resolve the individual disks in the systems. In the following, we carry out   simple empirical estimates of the effect of tidal truncation on the measured dust disk radii and  a detailed comparison with analytic predictions.

\begin{table} 
\caption{\label{tab::rta} Ratios of measured dust disk sizes vs observed separation }
\centering 
\begin{tabular}{l | cc }
\hline
Name &  \reff/$a_{\rm p}$ & \rdust/$a_{\rm p}$ \\
\hline
\multicolumn{3}{c}{Primary disks}\\
\hline
T Tau N & 0.1644$^{+0.0001}_{-0.0001}$ & 0.2123$^{+0.0003}_{-0.0003}$ \\
UY Aur A & 0.0375$^{+0.0024}_{-0.0005}$ & 0.0487$^{+0.0120}_{-0.0047}$ \\
RW Aur A & 0.0676$^{+0.0005}_{-0.0005}$ & 0.0882$^{+0.0009}_{-0.0005}$ \\
DK Tau A & 0.0385$^{+0.0003}_{-0.0004}$ & 0.0490$^{+0.0006}_{-0.0004}$ \\
HK Tau A & 0.0675$^{+0.0004}_{-0.0006}$ & 0.0930$^{+0.0013}_{-0.0008}$ \\
CIDA 9 A & 0.1202$^{+0.0006}_{-0.0007}$ & 0.1530$^{+0.0009}_{-0.0011}$ \\
DH Tau A & 0.0450$^{+0.0003}_{-0.0004}$ & 0.0622$^{+0.0013}_{-0.0013}$ \\
V710 Tau A & 0.0750$^{+0.0003}_{-0.0002}$ & 0.1001$^{+0.0007}_{-0.0007}$ \\
HN Tau A & 0.0328$^{+0.0006}_{-0.0006}$ & 0.0431$^{+0.0011}_{-0.0007}$ \\
UZ Tau E & 0.1256$^{+0.0003}_{-0.0006}$ & 0.1870$^{+0.0006}_{-0.0009}$ \\
\hline
\multicolumn{3}{c}{Secondary disks}\\
\hline
T Tau S & 0.3871$^{+0.5145}_{-0.3208}$ & 2.6355$^{+1.2374}_{-2.4718}$ \\ 
UY Aur B & 0.0133$^{+0.0155}_{-0.0185}$ & 0.0482$^{+0.0595}_{-0.0271}$ \\ 
RW Aur B & 0.0480$^{+0.0046}_{-0.0055}$ & 0.0599$^{+0.0112}_{-0.0064}$ \\ 
DK Tau B & 0.0240$^{+0.0048}_{-0.0060}$ & 0.0285$^{+0.0088}_{-0.0075}$ \\ 
HK Tau B & 0.1870$^{+0.0008}_{-0.0012}$ & 0.2204$^{+0.0034}_{-0.0049}$ \\ 
HN Tau B & 0.0003$^{+0.0031}_{-0.0000}$ & 0.0015$^{+0.0341}_{-0.0022}$ \\ 
UZ Tau Wa$^\dagger$ & 0.0281$^{+0.0006}_{-0.0006}$ & 0.0360$^{+0.0028}_{-0.0007}$ \\ 
UZ Tau Wb$^\dagger$ & 0.0279$^{+0.0006}_{-0.0006}$ & 0.0351$^{+0.0028}_{-0.0006}$ \\ 
\hline
\end{tabular}
\tablefoot{$^\dagger$Considering the separation to Wb. }
\end{table}

\subsection{Disk radii versus separation}
Table~\ref{tab::rta} reports the ratio between \reff \ and \rdust \ measured for the disk around the primary stars and the observed separation between the two components of the multiple systems. These separations are calculated from the fitted positions of the centers of the disks in our data or are obtained from the literature if only one component is detected (V710~Tau and DH~Tau). 

\begin{figure}
   \centering
  \includegraphics[width=0.5\textwidth]{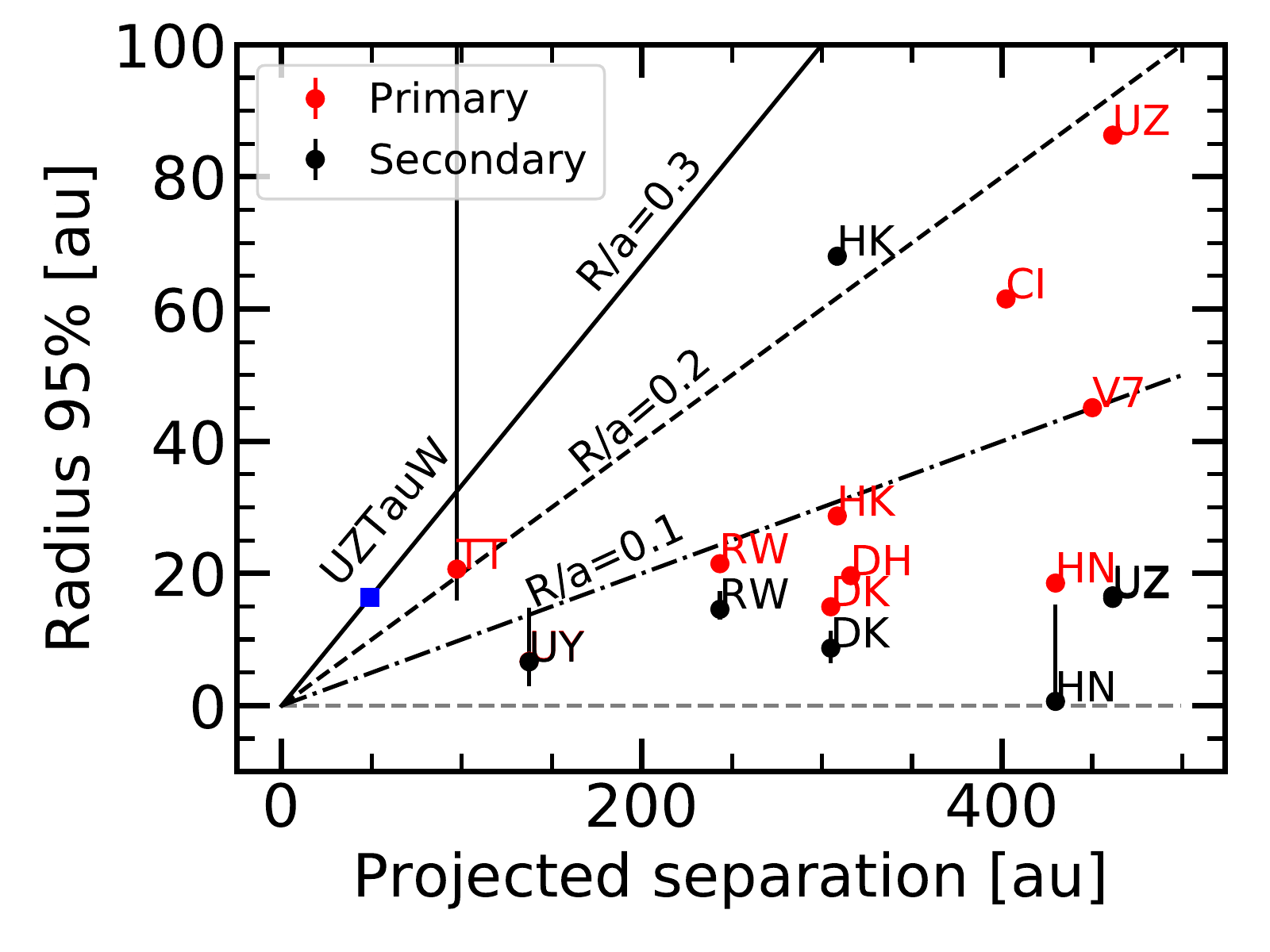}
   \caption{Observed separation of the components in the multiple systems vs the measured radii of the individual disks. Only the first two letters of the names are used  to label the points (e.g., CI is for CIDA 9). The lines represent the ratios of radius to observed separation of 0.3 (solid line), 0.2 (dashed line), and 0.1 (dot-dashed line). 
TTau S is outside of the plot to the top, but the fit is uncertain. UZ Tau W represents the values for the radii of the western components of the UZ Tau system, assuming only the distance between Wa and Wb. }
              \label{fig::sep_rad}%
    \end{figure}

As reported in Table~\ref{tab::rta} and also shown  in Fig.~\ref{fig::sep_rad}, the ratio of the disk radius to projected separation $a_{\rm p}$ is always lower than 0.3 in our sample, with the only exception of the very uncertain radius measured for T~Tau~S. The typical ratio \rdust/$a_{\rm p}$ is  $\lesssim$0.1, and only the western components of UZ~Tau show a value of \rdust/$a_{\rm p}>0.2$. As discussed in the literature  \citep[e.g.,][]{PP77,AL94,RC18}, it can be analytically computed that tidal torques dominate over viscous ones outside a truncation radius, which for a circular orbit is $R_{\rm t}\sim 0.3 \cdot a$, where $a$ is the semimajor axis of the binary orbit, with a dependence on the mass ratio $q$ (see Fig.~\ref{figexpofitpp}).

Since statistically it is more probable to observe stars at the apocenter of the binary orbit, as a very first approximation we assume $a_{\rm p}\sim a$, and therefore that the measured values of \rdust/$a_{\rm p}$ point to dust disk radii that are smaller than  would be expected from tidal truncation models, in line with the results of \citet{cox17}, suggesting that either the binaries are on very eccentric orbits or that dust radii are smaller than gas radii by factors $\gtrsim$2-3, probably due to a more effective drift of the dust probed by our 1.3 mm observations. 

To further verify these possibilities, in the next subsection we present the results of the detailed comparison of  the measured disk radii with analytic models of tidal truncation.

\subsection{Comparison with analytic predictions of tidal truncation}\label{sect::models_analytic}

As described in Appendix~\ref{app::models}, starting from the work of \citet{AL94} a fit to the expected truncation radius can be derived as a function of the semimajor  axis of the orbit ($a$) and the eccentricity ($e$) under the assumption that the disks and the binary orbit are co-planar (Eq. ~\ref{eqfuncfit}). Using Eq.~\ref{eqfuncfit}, and given  that the exact analytic expression for the ratio between the semimajor axis and the projected separation is
\begin{equation}\label{eqseparatio}
F=\frac{a}{a_p}=\bigg[\frac{1-e^2}{1+e\cdot \mbox{cos }\nu}\sqrt{1-\mbox{sin}^2(\omega + \nu)\mbox{ sin}^2 i}\bigg]^{-1}\rm,
\end{equation}
where $e$ is the eccentricity, $\nu$  the true anomaly, $\omega$  the longitude of periastron, and \textit{i}  the inclination of the plane of the orbit with respect to the line of sight, it is possible to obtain the following equation for the ratio of the truncation radius to the projected separation:
\begin{multline}\label{eq::trunc}
\frac{R_{\rm trunc}}{a_{\rm p}} = \frac{0.49 \cdot q^{2/3}_i}{0.6\cdot q_i^{2/3} + \ln(1+q_i^{1/3})}\left(b\cdot e^c + 0.88\mu^{0.01}\right) \cdot \\
\cdot \left[ \frac{1-e^2}{1+e\cdot \cos \nu}\sqrt{1-\sin^2(\omega+\nu)\sin^2i} \right]^{-1},
\end{multline}
where $q_i$ is the mass ratio (either $q_1 = M_1/M_2$ or $q_2 = q = M_2/M_1$), and $b$ and $c$ are the parameters derived in Appendix~\ref{app::models} and tabulated in Table~\ref{tabfitvalues}, which depend on the disk viscosity or equivalently on the Reynolds number, $\rey$.

\begin{figure}
   \centering
  \includegraphics[width=0.45\textwidth]{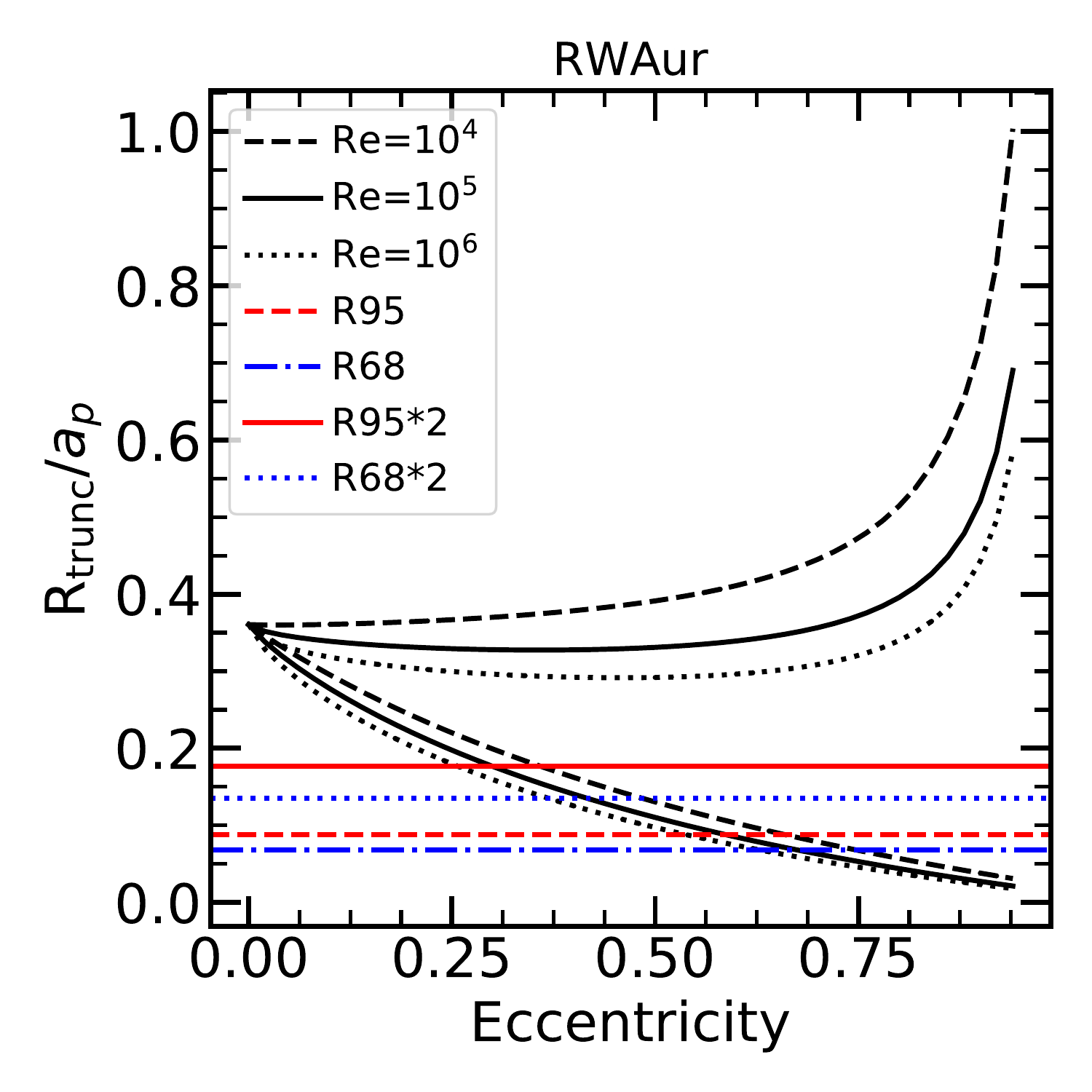}
   \caption{Ratio of the truncation radius to the projected separation of the orbit as a function of eccentricity assuming the parameters of the RW~Aur system and orbital inclination $i$=0\degree. The two sets of black lines are the expectations from analytic models of tidal truncation (Eq.~\ref{eq::trunc}) each one calculated for three different values of $\rey$ (see legend). The set of three lines at the bottom is the estimate for an object observed at apoastron, the top ones at periastron. The red dashed and blue dot-dashed lines report the measured values of \rdust/$a_{\rm p}$ and \reff/$a_{\rm p}$ for RW~Aur~A. respectively, while the red solid and blue dotted lines are a factor of two higher, corresponding to the assumed ratio of the gas to dust radius in the disk.  }
              \label{fig::RWAur_trunc_mod_gas_example}%
    \end{figure}

These values of $R_{\rm trunc}/a_{\rm p}$ have a minimum when the object is at apoastron ($\omega=0$, $\nu=\pi$), meaning when $a_{\rm p} = a\cdot (1+e)$, and a maximum at periastron ($\omega=\nu=0$), when $a_{\rm p} = a\cdot (1-e)$, at a given orbital inclination. The minimum ratio is found at $i$=0\degree \ and this ratio increases for higher orbital inclinations. Under the conservative  assumption of face-on orbits ($i$=0\degree), the two set of lines
plotted in the following plots    (see Fig.~\ref{fig::RWAur_trunc_mod_gas_example} for the specific case of RW Aur; similar curves for the other sources in our sample are presented in Appendix~\ref{app::models_comp}) are described by the equations:
\begin{multline}\label{eq::trunc}
\frac{R_{\rm trunc}}{a_{\rm p}} = \frac{0.49 \cdot q^{2/3}_i}{0.6\cdot q_i^{2/3} + \ln(1+q_i^{1/3})}\left(b\cdot e^c + 0.88\mu^{0.01}\right) \cdot (1+e)^{-1}\\
\frac{R_{\rm trunc}}{a_{\rm p}} = \frac{0.49 \cdot q^{2/3}_i}{0.6\cdot q_i^{2/3} + \ln(1+q_i^{1/3})}\left(b\cdot e^c + 0.88\mu^{0.01}\right) \cdot (1-e)^{-1},
\end{multline}
where the former refers to the truncation radius for an object located at apoastron and the latter at periastron. Each line is plotted for three different values of the Reynolds number ($\rey$), as discussed in Appendix~\ref{app::models}.

The analysis of the data presented in Sect.~\ref{sect::analysis} allows us to derive the dust radii for our targets (\reff, \rdust). At the same time, the projected separation ($a_{\rm p}$) at the time of the observation is measured from the fit. The ratio of the dust radius to the projected separation is reported in Table~\ref{tab::rta}. In the following analysis, we use this information to provide constraints on the truncation models without any prior knowledge of the orbital parameters, in particular the inclination of the orbital plane with respect to the observer, and the orbital eccentricity. Assuming an inclination for the orbital plane it is possible to derive the eccentricity compatible with the case where the object is located at the apoastron of the orbit (see the bottom curves in Fig.~\ref{fig::RWAur_trunc_mod_gas_example} for the case of RW Aur and an orbital inclination equal to 0\degree). This is  a lower limit to the real eccentricity of the orbit, since any other location of the target along the orbit  implies that the theoretical predictions  get closer to the upper lines, which refers to the case when the target is at periastron. Similarly, increasing the orientation of the orbital plane  moves all the theoretical predictions to higher values of $R_{\rm trunc}/a_{\rm p}$; therefore, the eccentricity derived assuming $i$=0\degree \ is  again  the lower limit of the eccentricity. Finally, assuming a noncoplanar disk inclination with respect to the orbital plane  reduces the tidal torque on the disk by a factor of $\sim$cos$^8$($\theta/2$) for a misalignment angle $\theta$ \citep[e.g.,][]{ML15,lubow15}, again leading to a higher $R_{\rm trunc}/a_{\rm p}$.

In general,  there are three considerations that suggest why the eccentricity in these systems is expected to  be small. First, values of $e\sim$1 would imply that at any passage at periastron the effects of tidal interaction on the disks would be massive, leading to a severe truncation that would significantly shorten the disk lifetime and lead to a rapid disk dissipation \citep[e.g.,][]{CP93}. This effect of a close highly eccentric passage of the secondary is observed for example in the RW~Aur system \citep[e.g.,][]{cabrit06,dai15}. Second, the orbital eccentricity may be uniformly random at formation, and decay with time \citep[e.g.,][]{bate18}. Therefore, it is expected that  eccentricities are typically $e<0.5$ or lower for multiple systems in the Taurus region. Third, observed distribution of eccentricities for main-sequence binary systems of low-mass stars show that the median eccentricity is $\sim$0.3 \citep[e.g.,][]{DK13}, and  fewer than 10\% of the systems have $e>0.6$.

We thus explore here the eccentricities that would be derived using  the measured dust disk radii as a
value for the truncation radius, either \reff \ or \rdust, and using an estimate of the gas disk radii that is  two times larger than the dust radii. The latter factor was chosen based on the median value obtained through the observation of disks in the Lupus star-forming region with ALMA \citep{ansdell18}. Although most of the disks analyzed by \citet{ansdell18} are single, this is to date the largest sample of resolved disks observed  in the dust and in the gas emission. A similar value for this factor between the gas and dust disk radius is also observed in the RW~Aur system \citep{rodriguez18} and in the older HD100453 system \citep{vdP19}, but it could in principle be different in binaries, in general. Since this ratio is driven by several effects, including CO optical depth and growth and drift of dust grains \citep[e.g.,][]{dutrey98,BA14,facchini17,trapman19}, the effect on this ratio of an external truncation is still uncertain.
It is also worth noting that this factor can be even larger than 5 in some extreme cases \citep{facchini19}. 

We consider two different cases for the orbital inclination: (1) the conservative case where the binary orbit has $i=0\degree$ and (2) the case where the binary orbit is assumed to be coplanar with the primary disk. We note that for case (1) we still use the theoretical truncation radius obtained for coplanar disks, which again is a conservative choice. 

\begin{figure}
   \centering
  \includegraphics[width=0.45\textwidth]{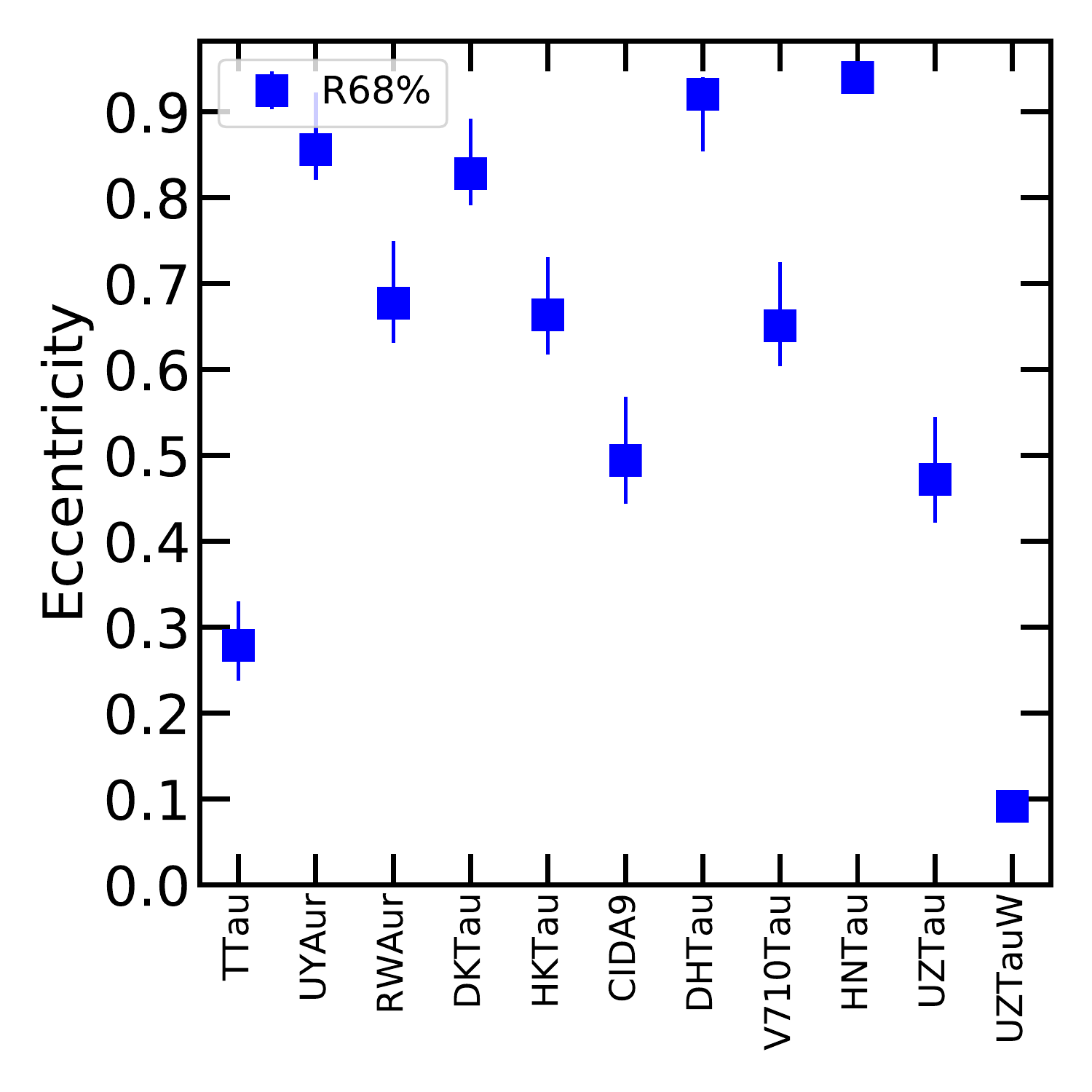}
  \includegraphics[width=0.45\textwidth]{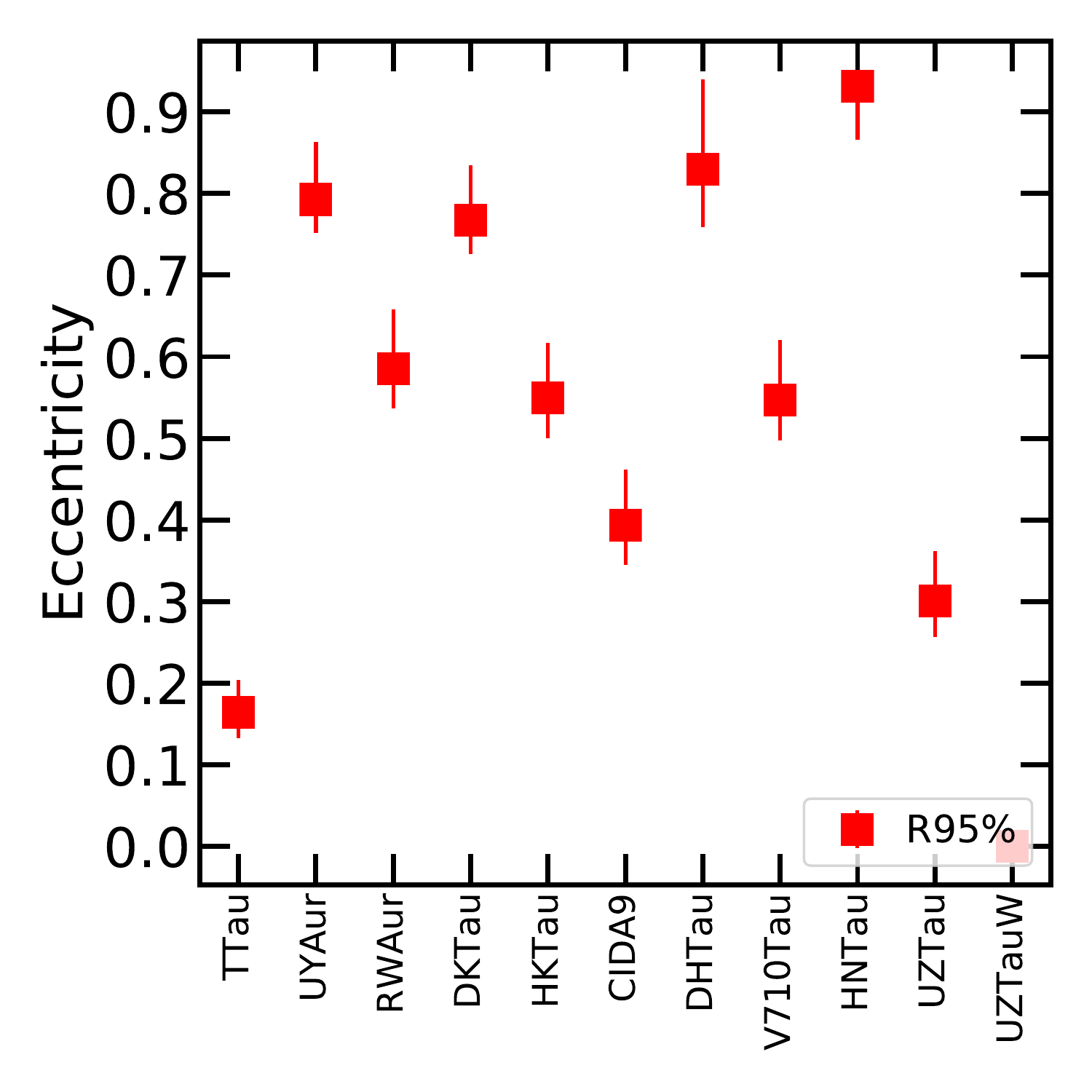}
   \caption{Minimum eccentricities obtained comparing the measured dust disk radii with the theoretical prediction of truncation assuming face-on orbital planes. The error bars are dominated by the difference in the models with different Reynolds numbers. The upper panel refers to the definition of dust radii as \reff, while the lower panel refers to the definition of dust radii as \rdust.}
              \label{fig::ecc_trunc_mod}%
    \end{figure}

\begin{figure}
   \centering
  \includegraphics[width=0.45\textwidth]{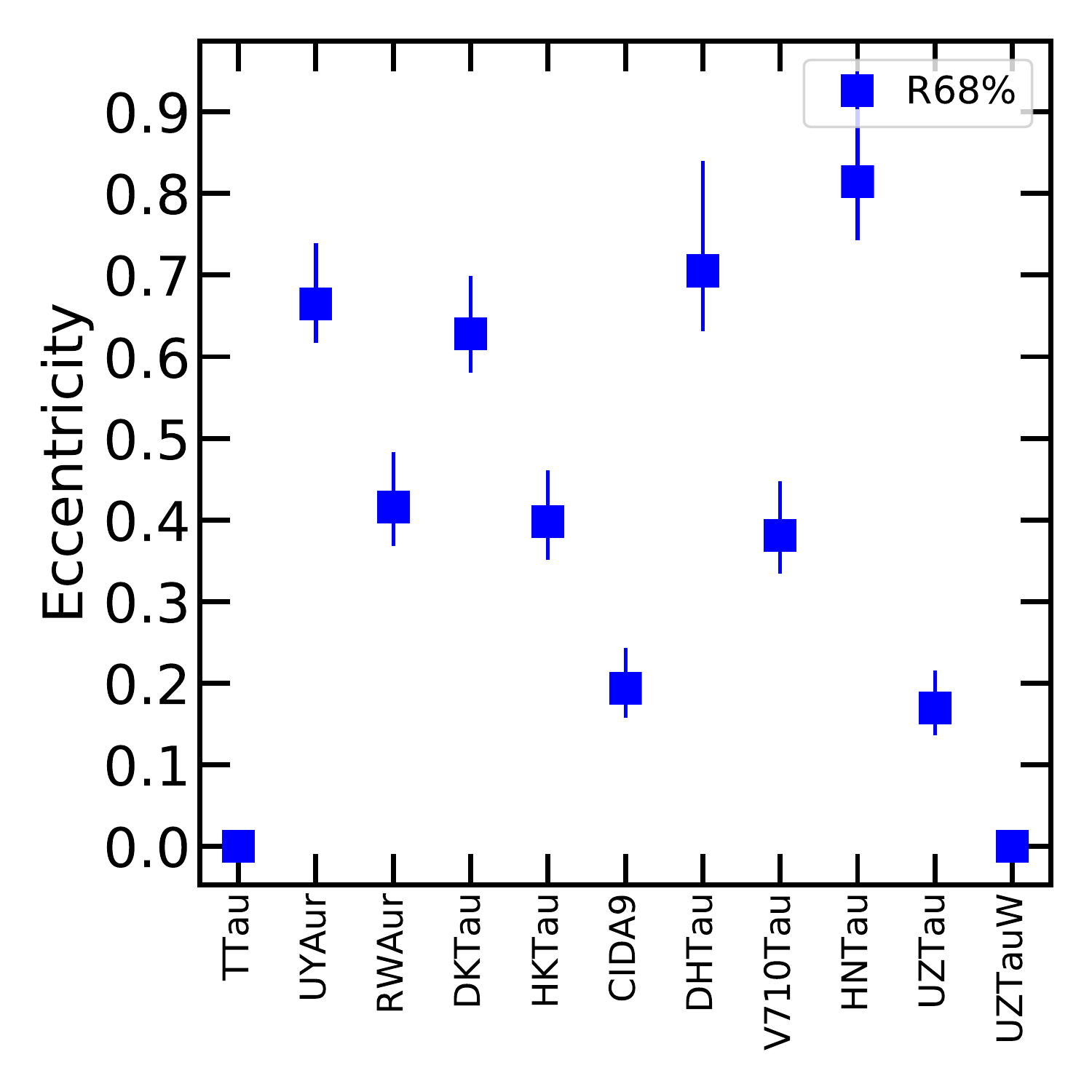}
  \includegraphics[width=0.45\textwidth]{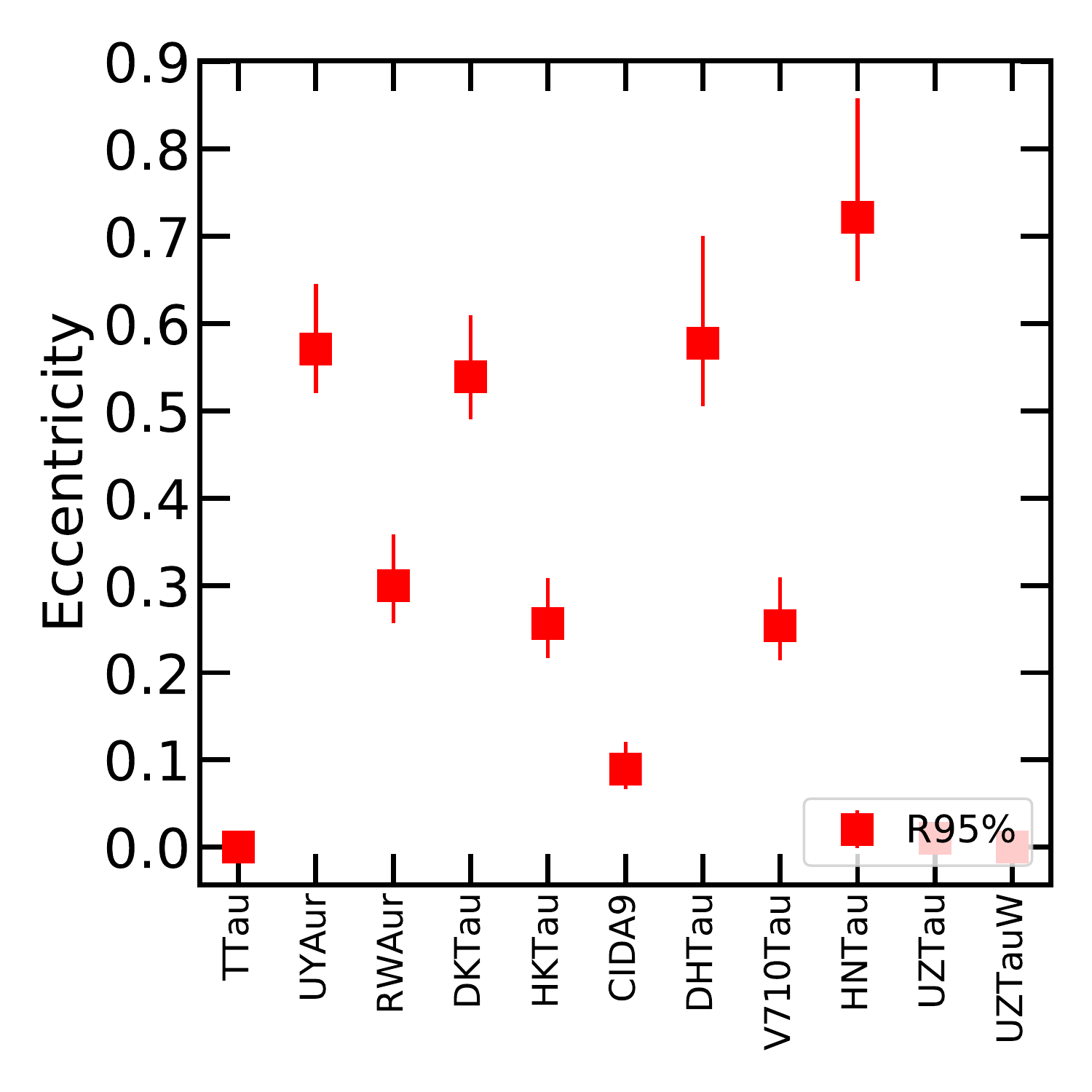}
   \caption{Minimum eccentricities obtained assuming truncation radii equal to twice the measured dust disk radii and comparing them with the theoretical prediction of truncation assuming face-on orbital planes. The error bars are dominated by the difference in the models with different Reynolds numbers. The upper panel refers to the definition of dust radii as \reff, while the lower panel refers to the definition of dust radii as \rdust.}
              \label{fig::ecc_trunc_mod_gas}%
    \end{figure}

{\bf Case 1: Face-on binary ($i=0$\degree)}. We show in Figs.~\ref{fig::DHTau_trunc_mod_gas}-\ref{fig::UZTau_trunc_mod_gas} the comparison between the theoretical predictions and the measured values of \rdust/$a_{\rm p}$ for all the sources in our sample. Both \rdust/$a_{\rm p}$ and \rgas/$a_{\rm p}$ are found to always be compatible with the expected values if the target is currently located at apoastron or between apoastron and periastron. This is expected, as this is the position along the orbit where objects spend the largest amount of time. However, the inferred minimum values of eccentricity are in general quite high ($e>0.5$ in 9/11 cases) assuming the truncation radius to be equal to the dust disk radius. This is shown in Fig.~\ref{fig::ecc_trunc_mod}, where the dust disk radius is estimated either as \reff~(upper panel) or \rdust~(lower panel). We note that the estimated minimum eccentricities do not vary much when changing the definition of the dust radius. A more reasonable distribution of eccentricities is instead found if we assume that the truncation radii equal to twice the dust disk radius, as shown in  Fig.~\ref{fig::ecc_trunc_mod_gas}, where again the upper and lower panels refer to the two choices for the dust radius. 

\begin{figure}
   \centering
  \includegraphics[width=0.45\textwidth]{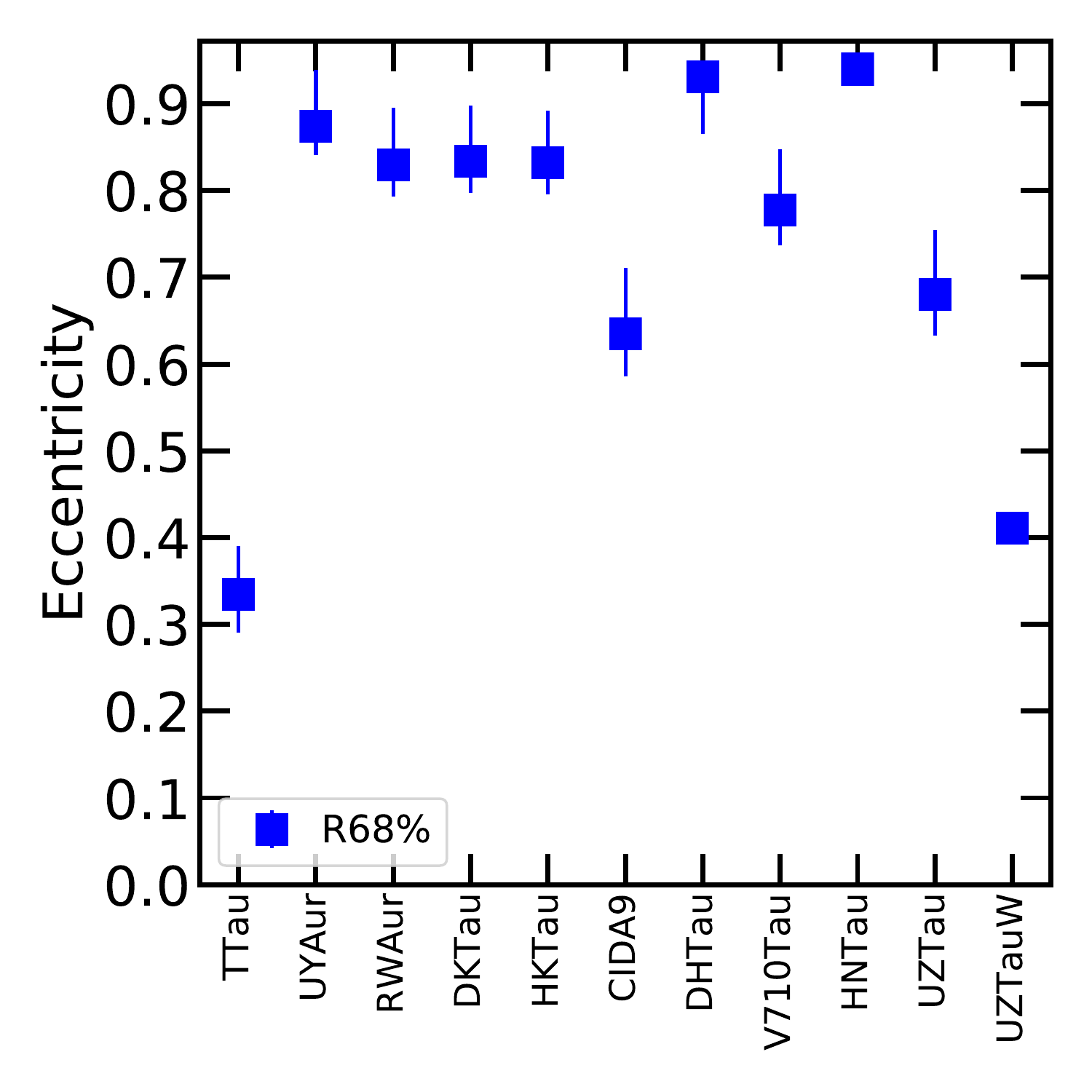}
  \includegraphics[width=0.45\textwidth]{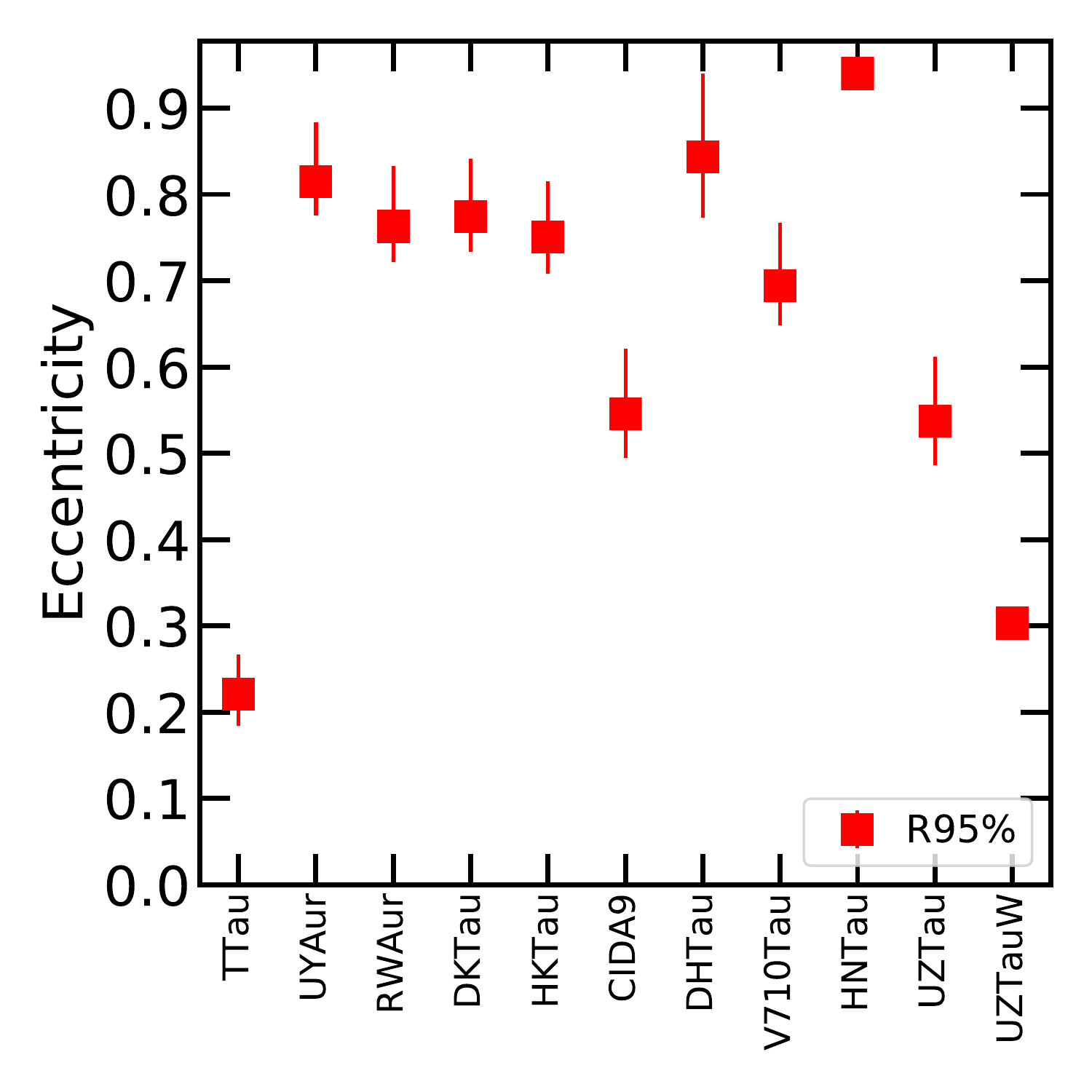}
   \caption{Minimum eccentricities obtained comparing the measured dust disk radii with the theoretical prediction of truncation assuming orbital planes co-planar with the primary disk. The error bars are dominated by the difference in the models with different Reynolds numbers. The upper panel refers to the definition of dust radii as \reff, while the lower panel refers to the definition of dust radii as \rdust. }
              \label{fig::ecc_trunc_mod_incl}%
    \end{figure}

\begin{figure}
   \centering
  \includegraphics[width=0.45\textwidth]{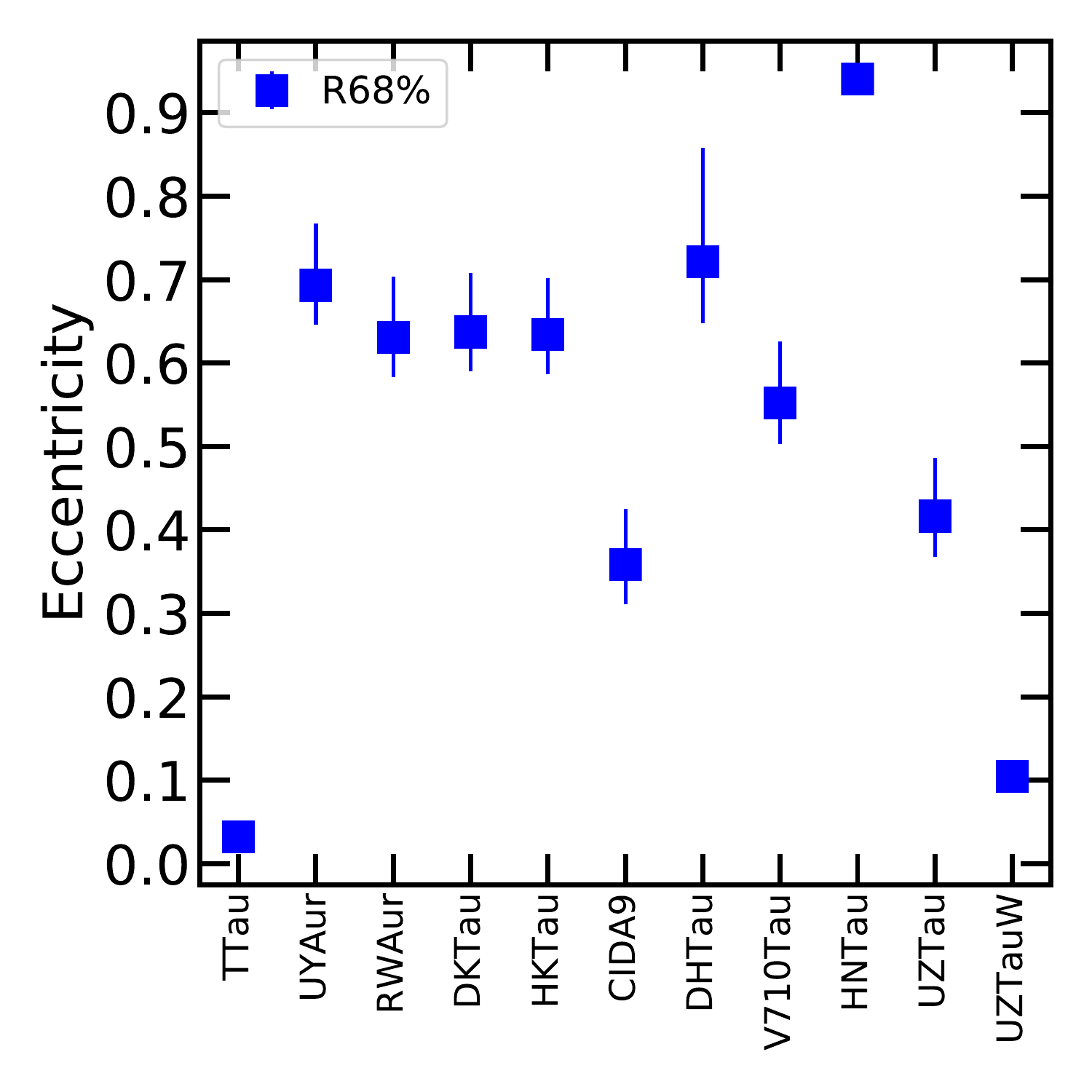}
  \includegraphics[width=0.45\textwidth]{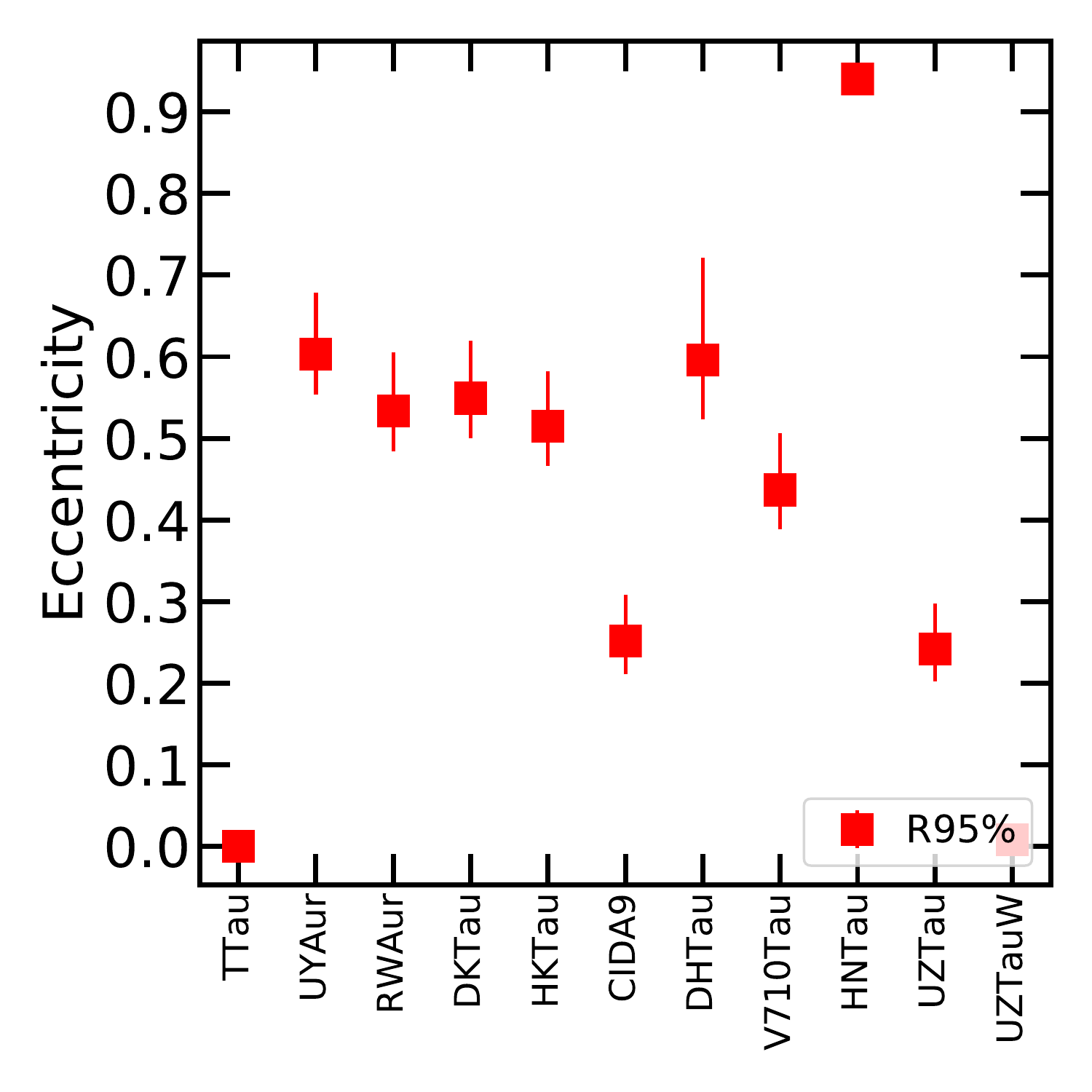}
   \caption{Minimum eccentricities obtained assuming truncation radii equal to twice the measured disk radii and comparing them with the theoretical prediction of truncation assuming orbital planes co-planar with the primary disk. The error bars are dominated by the difference in the models with different Reynolds numbers. The upper panel refers to the definition of dust radii as \reff, while the lower panel refers to the definition of dust radii as \rdust. }
              \label{fig::ecc_trunc_mod_gas_incl}%
    \end{figure}

{\bf Case 2: Binary coplanar with circumprimary disk.}  This assumption is probably not representative of the reality since in many cases these two planes are not aligned in hydrodynamical simulations of star formation \citep[e.g.,][]{bate18} or in observations where disks are resolved and the orbit is constrained \citep[e.g.,][]{rodriguez18,vdP19}. In any case, also under this assumption, the derived orbital eccentricities are very high  assuming either $R_{\rm trunc}$=\reff~($e>0.6$ in 9/11 cases, see Fig. \ref{fig::ecc_trunc_mod_incl}) or two times \reff~ ($e>0.5$ in 7/11 cases, see Fig. \ref{fig::ecc_trunc_mod_gas_incl}).

%

%

%

%



\section{Discussion}\label{sect::discussion}

The analysis of the dust continuum emission in our sample of disks around stars in multiple stellar systems in Taurus shows that these disks are smaller in size than disks around single stars, that their outer edges present a more abrupt truncation than disks around single stars, and that very high orbital eccentricities are expected if we assume that the observed values of \rdust/$a_{\rm p}$ correspond to the tidal truncation due to the binary. 

The dust continuum emission probed by our observation is a tracer of the (sub)millimeter dust grains in the disks, and these grains do not directly respond to the gas disk dynamics. The location and emission profile of dust grains in disks depends on the details of how dust grains grow and drift in the disk \citep[e.g.,][]{testi14} and by their opacity profile \citep[e.g.,][]{rosotti19}. 

Nevertheless, our results are already showing that the disks around multiple stars are different from isolated disks. The dust disk radii are smaller, implying that either the disks are intrinsically smaller, possibly due to tidal truncation, or that the drift of dust grains is more efficient in these disks in multiple systems. The latter could again be due to the effect of truncation on the outer edge of the gaseous disk, which could make the timescales of dust growth and drift shorter. The sharper profile of the dust emission in disks around multiple stars could also be similarly interpreted. Dust radial drift has been shown to imply a sharp outer edge in disks \citep{BA14}, although this observed sharp outer edge could be an effect of the dust opacity profile \citep{rosotti19}. Work needs to be done to verify whether a sharp truncation in the gas disk implies smaller and more sharply truncated dust disks, as we observe here. 

Another possibility is that the smaller observed sizes of disks in multiple systems is not an effect of disk evolution; instead, it is the result of a smaller disk size in multiple systems at formation. Wide protostar binaries are found in simulations of star formation \citep[e.g.,][]{bate18} and in observations. In the latter case the circumbinary disk is found to be large \citep[$\sim$300 au,][]{takakauwa17,delavillarmois18}.
Current work is thus not yet ready to constrain whether disks in multiple systems are small at the time of their formation. 

The comparison of the dust disk sizes with the predictions from theoretical models of tidal truncation implies very high orbital eccentricities (contrary to the expected eccentricity distribution in such young disks), suggesting that either predictions are to be revised or, more plausibly, that the dust disk sizes are smaller than the gas disk sizes even  in disks in multiple stellar systems, as suggested also by \citet{cox17}. 
As mentioned, it is known that gas disk radii in singles are larger than dust disk radii with differences from a factor of $\sim$2 \citep[e.g.,][]{ansdell18} to 5 or more \citep{facchini19}. Under the assumption that the analytic predictions of tidal truncation are correct, we can conclude that a factor of $\sim$2 is also needed  in binary systems to obtain values of orbital eccentricities more in line with expectations.

%

\section{Conclusions}\label{sect::concl}

We have presented here the analysis of our sample of ten multiple systems in the Taurus star-forming region observed with ALMA in the 1.3 mm continuum emission at spatial resolution $\sim$0.12\arcsec. The sample, comprising eight binaries and two triples, is part of a larger sample of disks in Taurus observed by our group \citep{long18,long19}. This allowed us to make a comparison between the properties of disks in multiple systems and disks around single stars. 

We   derived the brightness profile of the disks performing a fit of the data in the visibility plane. Assuming that the dust disk radii are traced by the radius at which a given fraction (68\% or 95\%) of the total emission is enclosed, we demonstrated that the disks around stars in multiple systems are smaller than in single systems in a statistically significant way. We also showed that the inferred brightness profiles for disks in multiple systems present a steeper outer edge than disks around single stars. This is  clear evidence that the disks in multiple systems are different from those in single systems, most plausibly due to the effect of tidal truncation. 
Our data also shows that the relative inclination of the disks in a system and their sizes do not have a strong dependence on the observed separation.

Finally, we compared our measurements with theoretical predictions for the effect of tidal truncation in binary systems on the disk sizes. In general, the measured dust disk radii are $\lesssim$0.1 $a_{\rm p}$. When comparing these values with the expectations from theoretical predictions, this would imply that 9 out of 11 of the disk pairs are in orbit with $e>0.5$, which is highly implausible. However, when assuming that the gas disk radii, which directly trace the dynamical truncation, are twice as large as the dust disk radii, the values of the eccentricity for the orbits are more reasonable. Our data, missing the information on the gas emission in these disks, do not allow us to constrain the ratio between the gas and dust disk radii in multiple systems. Future ALMA observations targeting the gas emission in multiple systems at similar spatial resolutions to those used in  the data presented here are needed to constrain the theory of tidal truncation and, in turn, the ability of protoplanetary disks in multiple systems to form planets.

\begin{acknowledgements}
We acknowledge insightful discussions with and contributions from I. Pascucci.
CFM acknowledges an ESO fellowship.
This paper makes use of the following ALMA data: ADS/JAO.ALMA\#2016.1.00164.S. ALMA is a partnership of ESO (representing its member states), NSF (USA) and NINS (Japan), together with NRC (Canada), MOST and ASIAA (Taiwan), and KASI (Republic of Korea), in cooperation with the Republic of Chile. The Joint ALMA Observatory is operated by ESO, AUI/NRAO and NAOJ.
This project has received funding from the European Union’s Horizon 2020 research and innovation program under the Marie Sklodowska-Curie grant agreement No 823823 (RISE DUSTBUSTERS).
This work was partly supported by the Deutsche Forschungs-Gemeinschaft (DFG, German Research Foundation) - Ref no. FOR 2634/1 TE 1024/1-1. 
MT has been supported by the DISCSIM project, grant agreement 341137 funded by the European Research Council under ERC-2013-ADG and by the UK Science and Technology research Council (STFC).
DJ is supported by the National Research Council Canada and by an NSERC Discovery Grant. 
ER acknowledges financial support from the European Research Council (ERC) under the European Union’s Horizon 2020 research and innovation program (grant agreement No 681601).
GP, FM, and YB acknowledge funding from ANR of France under contract number ANR-16-CE31-0013 (Planet-Forming Disks).
This work has made use of data from the European Space Agency (ESA) mission
{\it Gaia} (\url{https://www.cosmos.esa.int/gaia}), processed by the {\it Gaia}
Data Processing and Analysis Consortium (DPAC,
\url{https://www.cosmos.esa.int/web/gaia/dpac/consortium}). Funding for the DPAC
has been provided by national institutions, in particular the institutions
participating in the {\it Gaia} Multilateral Agreement.
This research has made use of the SIMBAD database,
operated at CDS, Strasbourg, France. This research made use of APLpy, an open-source plotting package for Python \citep{aplpy}.

\end{acknowledgements}



\appendix

\section{Discussion on individual targets}

\subsection{T Tau}\label{app::ttau}

\begin{figure}
   \centering
  \includegraphics[width=0.5\textwidth]{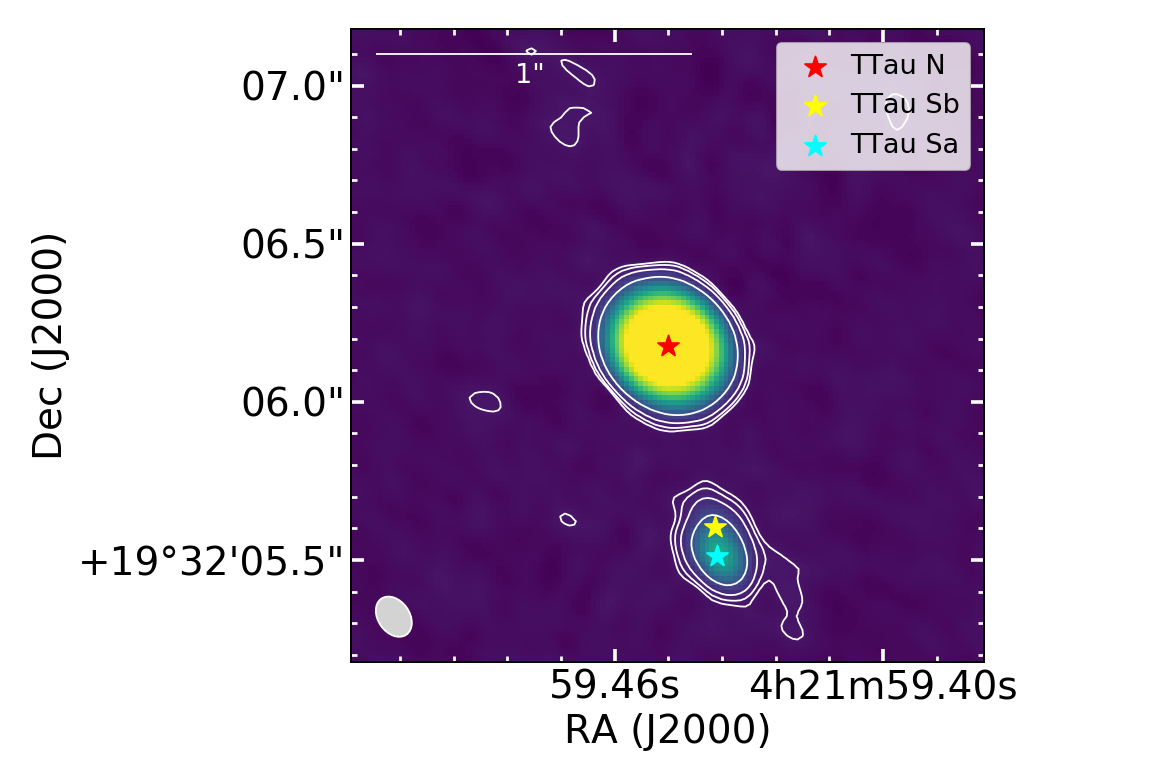}
     \includegraphics[width=0.5\textwidth]{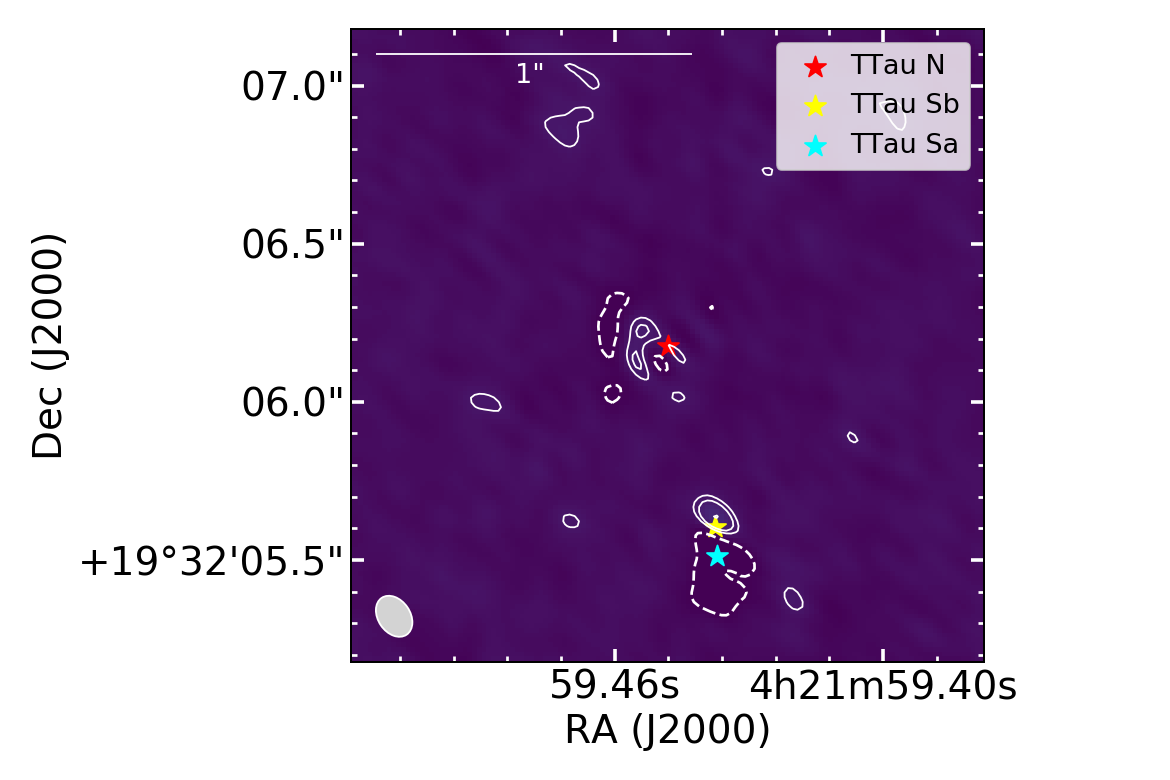}
\caption{Plot of the ALMA data on T~Tau (top panel) and the residuals of our fit (bottom panel) with the position of the three components of the T~Tau system at the time of the observation. Contours show the 3, 5, 10, and 30 $\sigma$ of the rms of the data. In the bottom panel, the dashed contours represents the -3$\sigma$ of the rms of the data.  }
              \label{fig::ttau_pos}%
    \end{figure}

The position of T Tau N, T Tau Sa, and T Tau Sb based on the orbital parameters derived by \citet{kohler16} are shown in Fig.~\ref{fig::ttau_pos} for the data and the residuals of the fit. This shows that the southern disk is centered on TTau Sa, but we cannot resolve whether this southern disk is a circumbinary disk, a disk around only T Tau Sa or Sb, or two unresolved disks. 

Large residuals are found after the fit, possibly because the disk around the southern components is composed of two disks or highly structured, as hinted by the 3$\sigma$ tail to the southwest of the southern component. The asymmetric residuals in the primary disk also point to the presence of other structures in this disk, not fitted by our smooth and axisymmetric functional form.

\subsection{CIDA 9}
The primary component of this system is a transition disk, meaning that it shows a large cavity and a ring-like emission structure \citep[see also][]{long18}. The formation mechanism of a transition disk is not yet understood, and the plausible candidates include photoevaporation, planet formation, or dead zones \citep[e.g.,][]{pinilla16,EP17}. 
\citet{RC18} discuss that X-ray photoevaporation can lead to the formation of cavities in binary systems only if the tidal truncation radius is larger than $\sim$10 au, depending on the mass of the targets and its X-ray luminosity. Following \citet{PP77} and \citet{RC18}, the truncation radius for these two targets is expected to be $\sim 0.3-0.4 \cdot a$, where $a$ is the separation of the two components. Given the observed separation in the system  of $\sim$400 au, the truncation radius for the disk of CIDA~9A is much larger than those at which \citet{RC18} would predict an outside-in clearing due to photoevaporation. Thus, it is not possible to use this system  to test their theory.

\subsection{V710 Tau}\label{app::v710tau}
This binary system is composed of two stars, the northern component (04:31:57.79, +18:21:37.95) as an M1-M2 young stellar object and the southern component (04:31:57.797, +18:21:35.06) as an M3-M3.5 object \citep[e.g.,][]{reipurth93,leinert93}. \citet{leinert93} refers to the northern component as the A component of the binary and the southern component as the B component (see also \citealt{KH09}, who also discuss a very distant C component). Similarly, \citet{WG01} and \citet{AJ19} refer to the northern component as the A component. However, the southern component was referred to as ``A'' by \citet{HH14}.

The properties of the two stars are consistent throughout the literature.  The northern component is the earlier spectral type and consistently has emission detected from a protoplanetary disk, including active accretion, mid-IR excess, and submillimeter dust continuum \citep[e.g.,][]{WG01,mccabe06,HH14,AJ19}.  The southern component has properties that are consistent with the absence of a disk.

In this paper, we have adopted the nomenclature of \citet{leinert93} and others that the northern component is V710 Tau A and the southern diskless component is V710 Tau B.

   \begin{figure}
   \centering
  \includegraphics[width=0.5\textwidth]{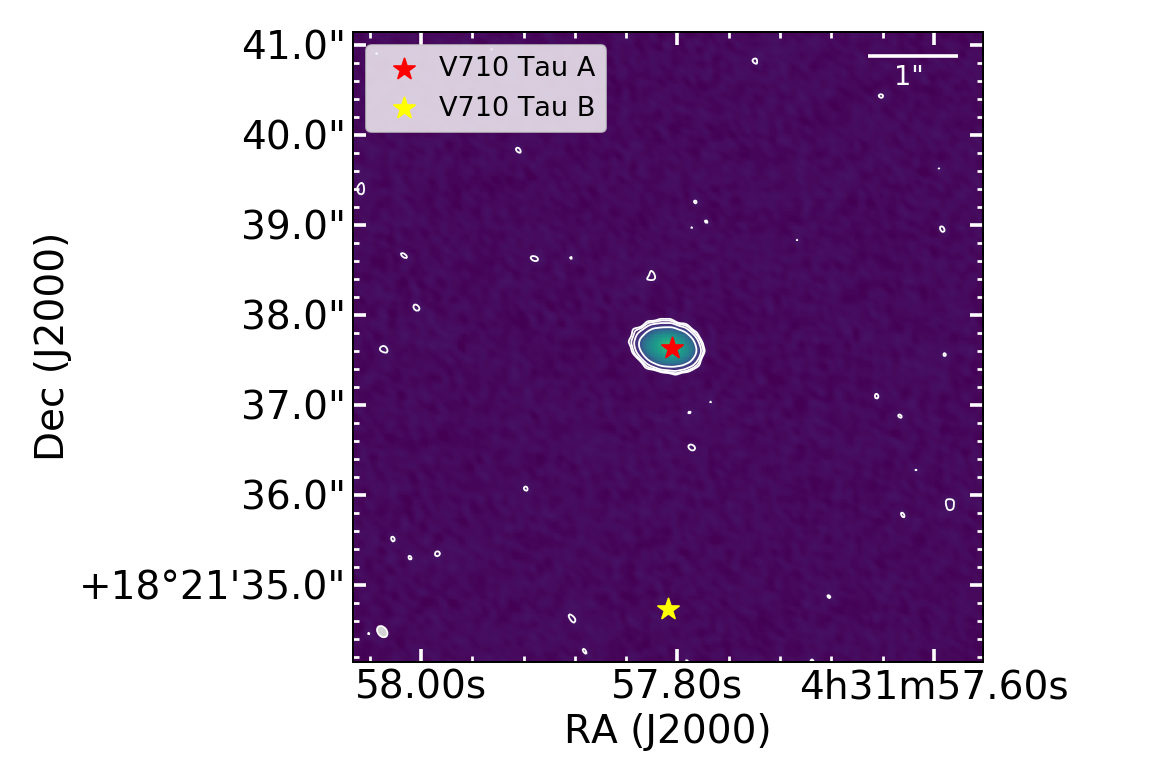}
\caption{Plot of the ALMA data on V710~Tau with the position of the components A and B at the time of the observation, considering their proper motion using the values from Gaia DR2 \citep{gaiadr2}. Contours show the 3, 5, 10, and 30 $\sigma$ of the rms of the data. }
              \label{fig::v710tau_map}%
    \end{figure}

\section{Best fit of the uv-data of the multiple systems}\label{app::bestfit}
 
As discussed in Sect.~\ref{sect::analysis}, the disks analyzed here are usually described by a power law with exponential cutoff (same as Eq.~\ref{eq::powexp}) with the following functional form:
\begin{equation}
I(r) = I_0 ~ r^{-\gamma1}  \exp\left(-\frac{r}{R_{\rm c}}\right)^{\gamma2}, 
\end{equation}
with $I_0$ defined such that
\begin{equation}
I_0  = F_{\rm tot} / \int_0^\infty 2\pi ~ r ~ r^{-\gamma1} ~ \exp\left(-\frac{r}{R_{\rm c}}\right)^{\gamma2} ~ dr.
\end{equation}
This functional form is chosen for all the disks not showing any clear substructure in their emission, i.e., smooth. The adopted final parameters and the uncertainties for these targets are reported in Table~\ref{tab::best_fit_pars}. While many parameters are well constrained, the uncertainty on $\gamma2$ is usually large, and the parameters for HN~Tau~B are very uncertain, since this disk is probably not resolved with our data. All the parameters are left unconstrained in the fit, with the exception of HK~Tau~B, whose inclination is constrained to be $>$80\degree.

For CIDA~9, a system composed of a transition disk showing a clear ring-like emission around the primary and an unresolved disk around the secondary component, we describe the disk around the primary with the  functional form
\begin{equation}
I(r) = f_0 ~ \exp\left(-0.5~\left(\frac{(r-R_p)}{\sigma}\right)^2\right)
,\end{equation}
where $R_p$ and $\sigma$ describe the location and width of a Gaussian ring. 
Finally, the intensity profile of the disk around the primary (eastern) component of the UZ~Tau system is described as a set of three concentric Gaussian rings, with the following functional form:
\begin{multline}
I(r) = f_01 ~ \exp\left(-0.5~\left(\frac{(r-R_p1)}{\sigma1}\right)^2\right) + \\ + f_02 ~ \exp\left(-0.5~\left(\frac{(r-R_p2)}{\sigma2}\right)^2\right) +\\ + f_03 ~ \exp\left(-0.5~\left(\frac{(r-R_p3)}{\sigma3}\right)^2\right).
\end{multline}
The two disks around the two western components of the system are instead described as in Eq.~\ref{eq::powexp}. The adopted final parameters for these two systems are reported in Table~\ref{tab::best_fit_pars2} with their uncertainties.

 \begin{table*} 
\caption{\label{tab::best_fit_pars} Best fit parameters for the multiple systems in this study: targets with no clear substructures }
\renewcommand{\arraystretch}{1.5}
\centering 
\begin{tabular}{l | ccccccccc  }

Name    & & & & \multicolumn{3}{c}{Parameters} & &\\

\hline  
& \multicolumn{5}{c}{Single power law with exponential cutoff} & \\     
& log$F_\nu$  & $R_c$ &  $\gamma1$  &  $\gamma2$ &  inc &  PA   & $\Delta$RA  &  $\Delta$Dec  \\
\hline

V710 Tau         &      -1.076$^{+0.003}_{-0.003}$ & 
0.320$^{+0.001}_{-0.001}$ & 
0.482$^{+0.009}_{-0.010}$ & 
8.819$^{+0.606}_{-0.574}$ & 
48.912$^{+0.298}_{-0.305}$ & 
84.345$^{+0.363}_{-0.364}$ & 
-0.022$^{+0.001}_{-0.001}$ & 
2.685$^{+0.000}_{-0.000}$ &  \\

DH Tau  & -1.555$^{+0.005}_{-0.005}$ & 
0.139$^{+0.003}_{-0.004}$ & 
0.381$^{+0.069}_{-0.075}$ & 
5.734$^{+1.367}_{-1.113}$ & 
16.946$^{+2.038}_{-2.217}$ & 
18.870$^{+7.372}_{-7.268}$ & 
-0.108$^{+0.000}_{-0.000}$ & 
-0.113$^{+0.001}_{-0.001}$ & \\

\hline
\hline

& \multicolumn{5}{c}{Single power law with exponential cutoff (both components)} & \\    
& log$F_\nu$  & $R_c$  & $\gamma1$ &  $\gamma2$  &  inc  &  PA &  $\Delta$RA  & $\Delta$Dec  &\\
\hline

T Tau N &       -0.690$^{+0.001}_{-0.001}$ & 
0.150$^{+0.000}_{-0.000}$ & 
0.680$^{+0.002}_{-0.002}$ & 
49.623$^{+0.535}_{-1.149}$ & 
28.251$^{+0.170}_{-0.181}$ & 
87.493$^{+0.343}_{-0.338}$ & 
-0.085$^{+0.000}_{-0.000}$ & 
-0.000$^{+0.000}_{-0.000}$ &  \\
T Tau S &
-1.690$^{+0.039}_{-0.038}$ & 
0.040$^{+0.302}_{-0.046}$ & 
1.501$^{+0.186}_{-0.342}$ & 
0.297$^{+0.173}_{-0.128}$ & 
61.558$^{+8.761}_{-4.831}$ & 
7.864$^{+3.742}_{-3.533}$ & 
-0.249$^{+0.000}_{-0.000}$ & 
-0.656$^{+0.001}_{-0.000}$ & \\

RW Aur A        &       -1.206$^{+0.004}_{-0.004}$ & 
0.139$^{+0.001}_{-0.001}$ & 
0.700$^{+0.019}_{-0.023}$ & 
26.711$^{+14.780}_{-13.308}$ & 
55.048$^{+0.500}_{-0.400}$ & 
41.132$^{+0.571}_{-0.552}$ & 
-0.080$^{+0.000}_{-0.000}$ & 
-0.123$^{+0.000}_{-0.000}$ &   \\
RW Aur B &
-1.812$^{+0.239}_{-0.198}$ & 
0.090$^{+0.015}_{-0.018}$ & 
0.074$^{+0.610}_{-1.363}$ & 
11.945$^{+9.652}_{-11.477}$ & 
74.556$^{+3.821}_{-8.245}$ & 
40.994$^{+3.562}_{-3.654}$ & 
-1.515$^{+0.001}_{-0.001}$ & 
-0.532$^{+0.002}_{-0.002}$ & \\

DK Tau A        &       -1.511$^{+0.005}_{-0.005}$ & 
0.122$^{+0.001}_{-0.001}$ & 
0.598$^{+0.025}_{-0.026}$ & 
38.642$^{+14.840}_{-19.740}$ & 
12.859$^{+2.520}_{-2.757}$ & 
4.513$^{+9.950}_{-9.714}$ & 
-0.034$^{+0.000}_{-0.000}$ & 
0.164$^{+0.000}_{-0.000}$ &  \\
DK Tau B & -1.929$^{+0.464}_{-0.219}$ & 
0.069$^{+0.026}_{-0.022}$ & 
-0.122$^{+1.285}_{-2.065}$ & 
26.849$^{+29.890}_{-30.743}$ & 
77.966$^{+6.119}_{-11.044}$ & 
28.050$^{+5.158}_{-5.440}$ & 
2.047$^{+0.002}_{-0.002}$ & -0.995$^{+0.002}_{-0.002}$ &\\

HN Tau A        &       -1.449$^{+0.024}_{-0.021}$ & 
0.142$^{+0.003}_{-0.003}$ & 
0.651$^{+0.048}_{-0.053}$ & 
16.148$^{+5.049}_{-7.522}$ & 
69.768$^{+1.402}_{-1.255}$ & 
85.299$^{+0.708}_{-0.694}$ & 
0.217$^{+0.001}_{-0.001}$ & 
-0.120$^{+0.000}_{-0.000}$ & 
\\
HN Tau B &
-3.447$^{+0.135}_{-0.067}$ & 
0.203$^{+0.374}_{-0.283}$ & 
2.467$^{+0.681}_{-0.885}$ & 
10.303$^{+12.343}_{-13.143}$ & 
49.461$^{+52.297}_{-62.540}$ & 
-2.690$^{+120.697}_{-112.721}$ & 
-1.820$^{+0.007}_{-0.007}$ & 
-2.535$^{+0.007}_{-0.007}$ &   \\

UY Aur A        &       -1.659$^{+0.032}_{-0.023}$ & 
0.043$^{+0.013}_{-0.009}$ & 
0.235$^{+1.063}_{-2.093}$ & 
8.616$^{+12.937}_{-7.997}$ & 
23.502$^{+8.591}_{-9.384}$ & 
-53.601$^{+10.073}_{-10.681}$ & 
0.161$^{+0.000}_{-0.000}$ & 
0.022$^{+0.000}_{-0.000}$ & 
 \\
UY Aur B &
-2.193$^{+0.106}_{-0.069}$ & 
0.055$^{+0.193}_{-0.047}$ & 
1.768$^{+0.501}_{-1.130}$ & 
10.444$^{+12.057}_{-11.951}$ & 
25.571$^{+35.315}_{-30.646}$ & 
-35.043$^{+186.769}_{-73.124}$ & 
-0.520$^{+0.001}_{-0.001}$ & 
-0.546$^{+0.001}_{-0.001}$ &  \\

HK Tau A        &       -1.217$^{+0.006}_{-0.006}$ & 
0.229$^{+0.003}_{-0.003}$ & 
0.919$^{+0.011}_{-0.013}$ & 
21.220$^{+18.845}_{-10.600}$ & 
56.882$^{+0.476}_{-0.481}$ & 
-5.117$^{+0.498}_{-0.494}$ & 
0.174$^{+0.000}_{-0.000}$ & 
-0.702$^{+0.001}_{-0.000}$ & 
 \\
HK Tau B &
-0.871$^{+0.014}_{-0.015}$ & 
0.499$^{+0.001}_{-0.003}$ & 
-1.182$^{+0.102}_{-0.091}$ & 
11.356$^{+3.819}_{-2.474}$ & 
83.242$^{+0.235}_{-0.236}$ & 
41.236$^{+0.205}_{-0.183}$ & 
0.599$^{+0.002}_{-0.002}$ & 
-2.982$^{+0.002}_{-0.002}$ &  \\

\hline
\end{tabular}
\tablefoot{Results of the fitting of the visibilities of the data. The reported uncertainties on the best fit paramters include the correction factor $\sqrt{3.5}$ discussed in Sect.~\ref{sect::analysis}.}
\end{table*}

\begin{table*}
\caption{\label{tab::best_fit_pars2} Best fit parameters for the multiple systems in this study: targets with clear substructures }
\renewcommand{\arraystretch}{1.5}
\centering 
\begin{tabular}{l | ccccccccc }

Name &   \multicolumn{3}{c}{Parameters}  \\
\hline
& \multicolumn{3}{c}{Gaussian ring (primary) and point source (secondary)} & \\    
& $f_0$ &   $\sigma$ &   $R_p$  &  inc &   PA  &  $\Delta$RAA  &  $\Delta$DecA &  \\
\hline
CIDA9A &         9.962$^{+0.006}_{-0.006}$ & 
0.067$^{+0.001}_{-0.001}$ & 
0.233$^{+0.001}_{-0.001}$ & 
46.393$^{+0.478}_{-0.439}$ & 
-76.454$^{+0.565}_{-0.591}$ & 
-0.507$^{+0.001}_{-0.001}$ & 
-0.734$^{+0.001}_{-0.001}$ & \\
\hline
& log$F_\nu$ &   $\Delta$RAB &   $\Delta$DecB  &\\
\hline
CIDA 9B &
-3.488$^{+0.079}_{-0.089}$ & 
1.514$^{+0.011}_{-0.012}$ & 
0.470$^{+0.013}_{-0.013}$ & \\

\hline
\hline

& \multicolumn{3}{c}{Gaussian rings (eastern component)} & \\      

&        $f_01$  &  $\sigma1$  &  $R_p1$  &  $f_02$ &   $\sigma2$  &  $R_p2$ &  \\
\hline
UZ Tau E  & 10.153$^{+0.023}_{-0.023}$ & 
0.227$^{+0.010}_{-0.011}$ & 
0.105$^{+0.023}_{-0.023}$ & 
10.275$^{+0.029}_{-0.026}$ & 
0.030$^{+0.004}_{-0.004}$ & 
0.091$^{+0.002}_{-0.002}$ & \\
\hline
& $f_03$ &   $\sigma3$ &   $R_p3$  &  incE  &  PAE  &   $\Delta$RAE & $\Delta$DecE  &\\
 \hline
 &  9.460$^{+0.025}_{-0.024}$ & 
0.045$^{+0.004}_{-0.005}$ & 
0.613$^{+0.003}_{-0.003}$ & 
55.211$^{+0.176}_{-0.175}$ & 
89.394$^{+0.208}_{-0.206}$ &   0.774$^{+0.001}_{-0.001}$ & -0.269$^{+0.000}_{-0.000}$ &\\
\hline
 & \multicolumn{3}{c}{Single power law with exponential cutoff (western components)} & \\    
 & log$F_\nu$  &  $R_c$  &  $\gamma_1$  &  $\gamma_2$  &  inc  &  PA &   $\Delta$RA &   $\Delta$Dec  & \\
\hline
UZ Tau Wa   & -1.523$^{+0.015}_{-0.013}$ &  0.132$^{+0.003}_{-0.004}$ & 
0.514$^{+0.072}_{-0.114}$ & 
18.259$^{+27.631}_{-17.951}$ & 
61.245$^{+1.089}_{-0.997}$ & 
91.532$^{+0.846}_{-0.875}$ & 
-2.697$^{+0.001}_{-0.001}$ & 
0.331$^{+0.001}_{-0.001}$ & 
 \\
UZ Tau Wb &
-1.508$^{+0.012}_{-0.011}$ & 
0.129$^{+0.003}_{-0.003}$ & 
0.436$^{+0.068}_{-0.088}$ & 
20.999$^{+22.019}_{-16.925}$ & 
59.896$^{+0.932}_{-0.876}$ & 
92.913$^{+0.809}_{-0.828}$ & 
-2.793$^{+0.001}_{-0.001}$ & 
-0.032$^{+0.000}_{-0.000}$ & \\

\hline

\end{tabular}
\tablefoot{Results of the fitting of the visibilities of the data. The reported uncertainties on the best fit parameters include the correction factor $\sqrt{3.5}$ discussed in Sect.~\ref{sect::analysis}.}
\end{table*}

\begin{figure*}
  \centering
 \includegraphics[width=\textwidth,page=1]{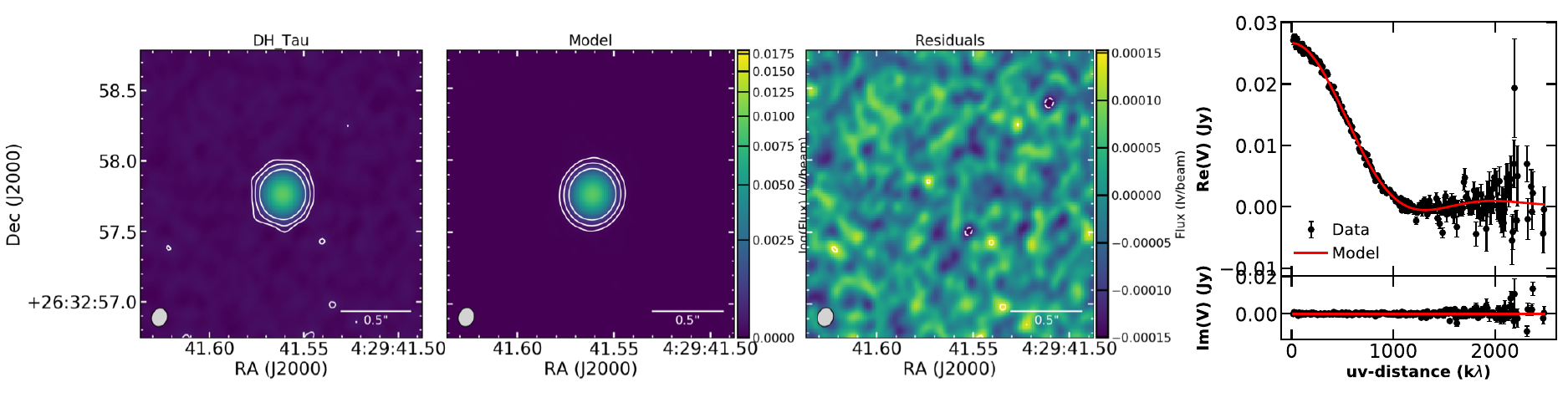}
  \caption{Fit for DH Tau with a power law with exponential cutoff profile. From left to right we show the data, the model, and the residuals. Contours show 3$\sigma$, 10$\sigma$, and 30$\sigma$, while the dashed contours in the residuals plot shows  -3$\sigma$. The beam size is shown in the bottom right. The rightmost panel shows the best fit of the visibilities.  }
             \label{fig::DHTau}%
    \end{figure*}

\begin{figure*}
  \centering
 \includegraphics[width=\textwidth,page=2]{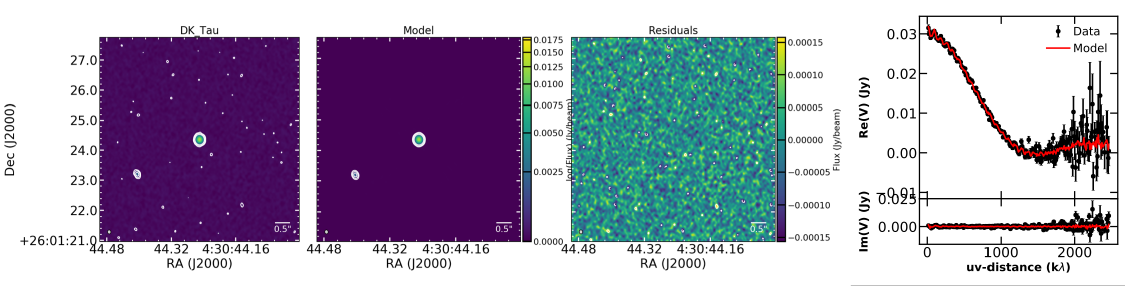}
  \caption{Fit for DK Tau with a power law with exponential cutoff profile. Same panels and symbols as in Fig.~\ref{fig::DHTau}. }
             \label{fig::DKTau}%
    \end{figure*}

\begin{figure*}
  \centering
 \includegraphics[width=\textwidth,page=3]{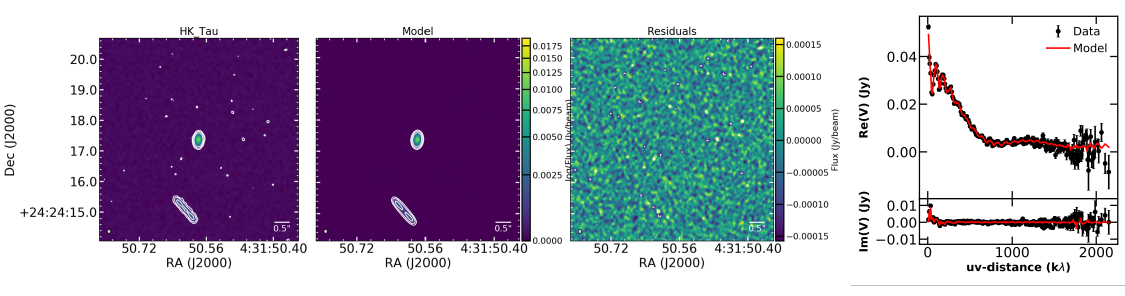}
  \caption{Fit for HK Tau with a power law with exponential cutoff profile. Same panels and symbols as in Fig.~\ref{fig::DHTau}. }
             \label{fig::HKTau}%
    \end{figure*}

\begin{figure*}
  \centering
 \includegraphics[width=\textwidth,page=4]{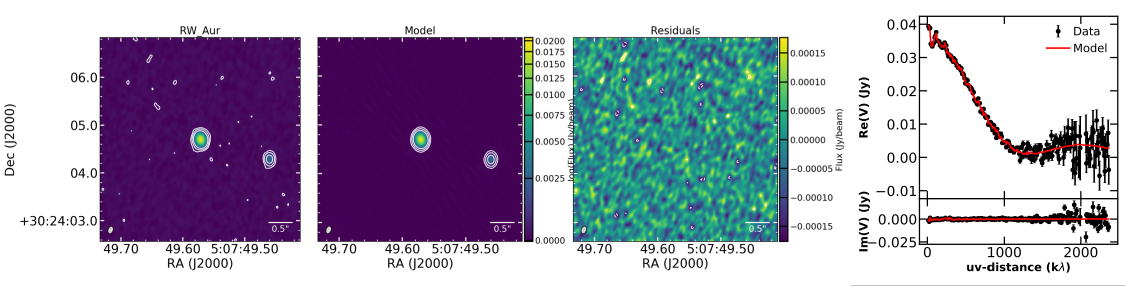}
  \caption{Fit for RW~Aur with a power law with exponential cutoff profile. Same panels and symbols as in Fig.~\ref{fig::DHTau}.}
             \label{fig::RWAur}%
    \end{figure*}

\begin{figure*}
  \centering
 \includegraphics[width=\textwidth,page=5]{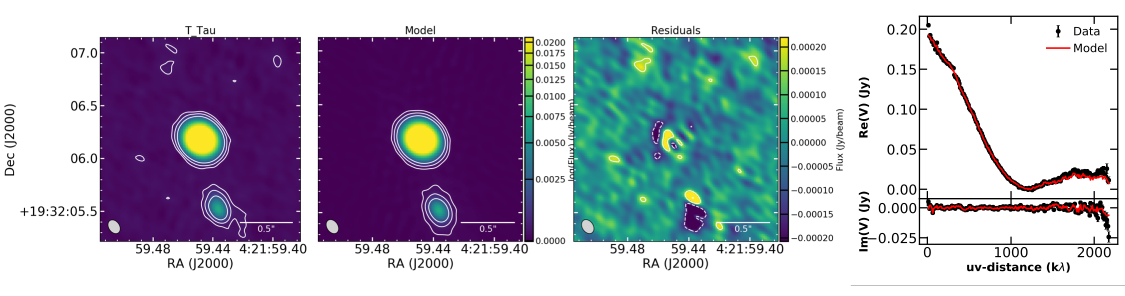}
  \caption{Fit for T Tau with a power law with exponential cutoff profile. Same panels and symbols as in Fig.~\ref{fig::DHTau}. }
             \label{fig::TTau}%
    \end{figure*}

\begin{figure*}
  \centering
 \includegraphics[width=\textwidth,page=6]{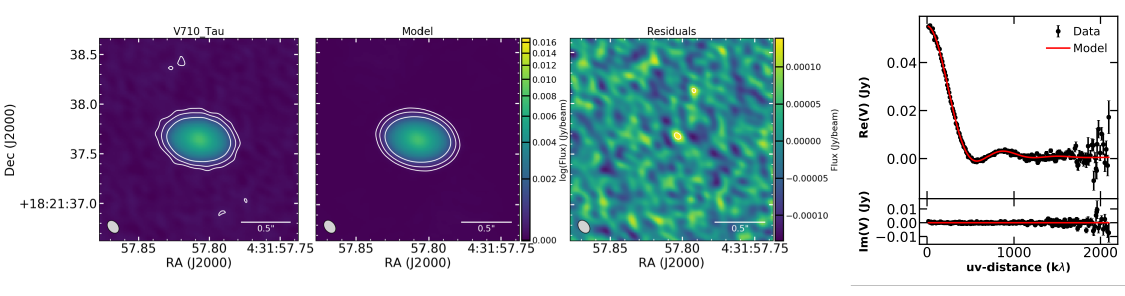}
  \caption{Fit for V710 Tau with a power law with exponential cutoff profile. Same panels and symbols as in Fig.~\ref{fig::DHTau}.  }
             \label{fig::V710Tau}%
    \end{figure*}

\begin{figure*}
  \centering
 \includegraphics[width=\textwidth,page=7]{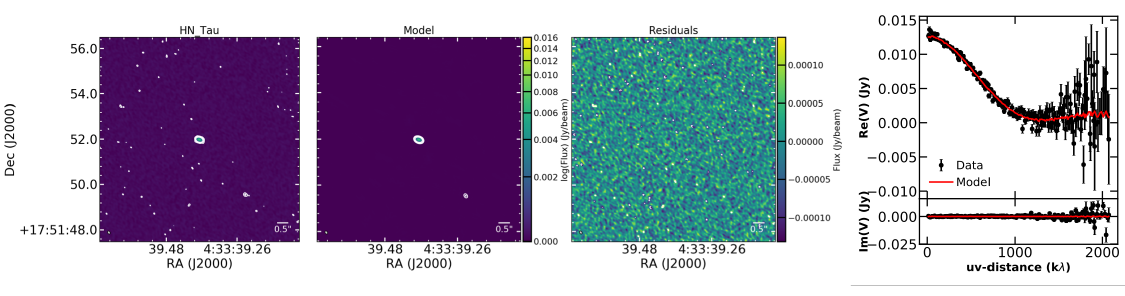}
  \caption{Fit for HN Tau with a power law with exponential cutoff profile. Same panels and symbols as in Fig.~\ref{fig::DHTau}.  }
             \label{fig::HNTau}%
    \end{figure*}

\begin{figure*}
  \centering
 \includegraphics[width=\textwidth,page=8]{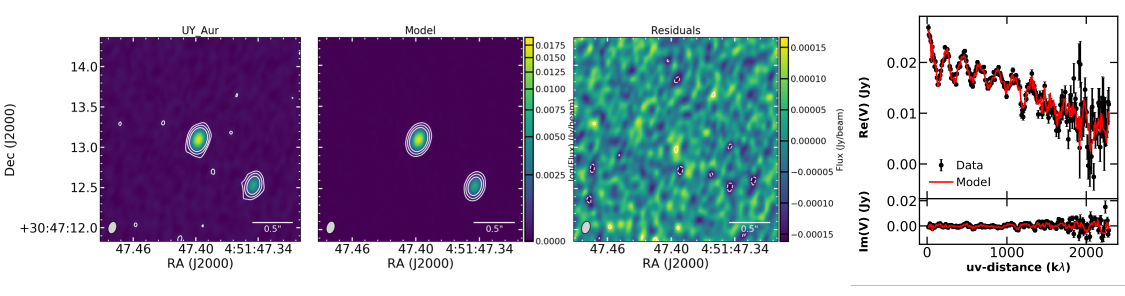}
  \caption{Fit for UY~Aur with a power law with exponential cutoff profile. Same panels and symbols as in Fig.~\ref{fig::DHTau}. }
             \label{fig::UYAur}%
    \end{figure*}

\begin{figure*}
  \centering
 \includegraphics[width=\textwidth,page=9]{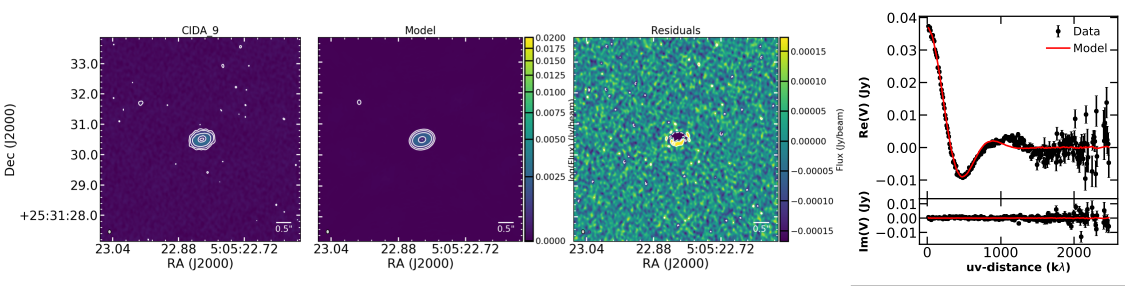}
  \caption{Fit for CIDA~9 with a Gaussian ring for the primary and a point source for the secondary. Same panels and symbols as in Fig.~\ref{fig::DHTau}.  }
             \label{fig::CIDA9}%
    \end{figure*}

\begin{figure*}
  \centering
 \includegraphics[width=\textwidth,page=10]{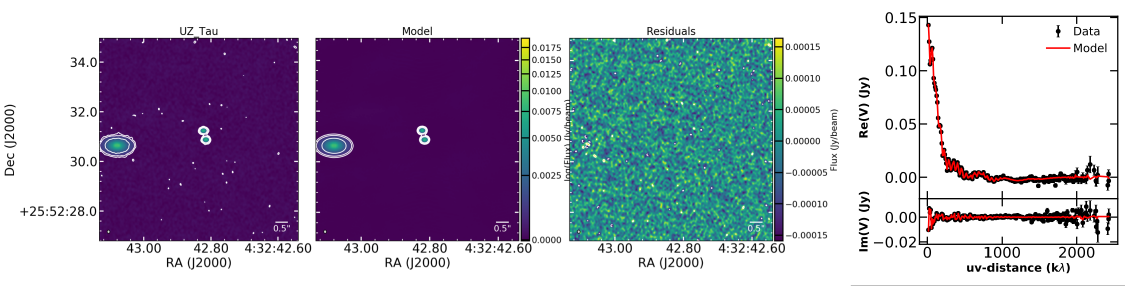}
  \caption{Fit for UZ Tau with a power law with exponential cutoff profile for the two western component, and a multi-ring for the eastern component. Same panels and symbols as in Fig.~\ref{fig::DHTau}.  }
             \label{fig::UZTau}%
    \end{figure*}

The final parameters are then used to produce synthetic images of the models and to compare the data with the model. These images are shown in Figures~\ref{fig::DHTau}-\ref{fig::UZTau} together with the comparison between the observed and model visibilities.

%

%

%

%

%

\section{Comparison of disk sizes to analytic predictions of tidal truncation}
Here we describe the analytic solutions to the models of tidal truncation and we then show the comparison of the observed dust disk radii to separation ratio with the analytic solutions. 

\subsection{Theoretical models of tidal truncations}\label{app::models}

\citet{AL94} have described tidal truncation in circumstellar and circumbinary disks both analytically (in terms of resonant tidal interaction) and numerically. The location of disk truncation is set by balancing the resonant torques with the disk viscous torques and hence depends on the mass ratio, the orbital eccentricity, and the Reynolds number ($\rey=\alpha_\nu^{-1}(r/H)^2$) in the disk.  In the case of zero eccentricity, disk truncation is actually due mostly to nonresonant interaction \citep{PP77}, in which case the truncation radius does not depend on viscosity and is just a function of the mass ratio, that we can express in terms of the mass parameter $\mu=M_2 / (M_1+M_2)$. 

It is useful for the analysis carried out in Sect.~\ref{sect::models_analytic} to obtain an analytical function $R_{\rm t}(M_1,M_2,e,a), $ which for a given set of binary parameters returns the value of the truncation radius of the circumstellar disks predicted by the theory of \citet{AL94}. 

Following the approach of \citet{pichardo} we fit the results in both $e$ and $\mu$ with an exponential function multiplied by the Roche Lobe radius ($R_{i,\rm Egg}$) of the appropriate star. The fitting function will therefore be
\begin{equation}\label{eqfittingfunc}
R_{\rm t}(M_1,M_2,e,a)=R_{i,\rm Egg}\cdot(b\cdot e^c+h\cdot\mu^k),
\end{equation}
where \textit{b}, \textit{c}, \textit{h}, and \textit{k} are the fitting parameters and
\begin{equation}
\frac{R_{i,\rm Egg}}{a}=\frac{0.49\cdot q_{i}^{2/3}}{0.6\cdot q_{i}^{2/3} +ln(1+q_{i}^{1/3})},
\end{equation}
where
$q_{1}={M_{1}}/{M_{2}}$  and  $q_{2}={M_{2}}/{M_{1}}$.

\begin{figure}
\begin{center}
\includegraphics[width=0.48\textwidth]{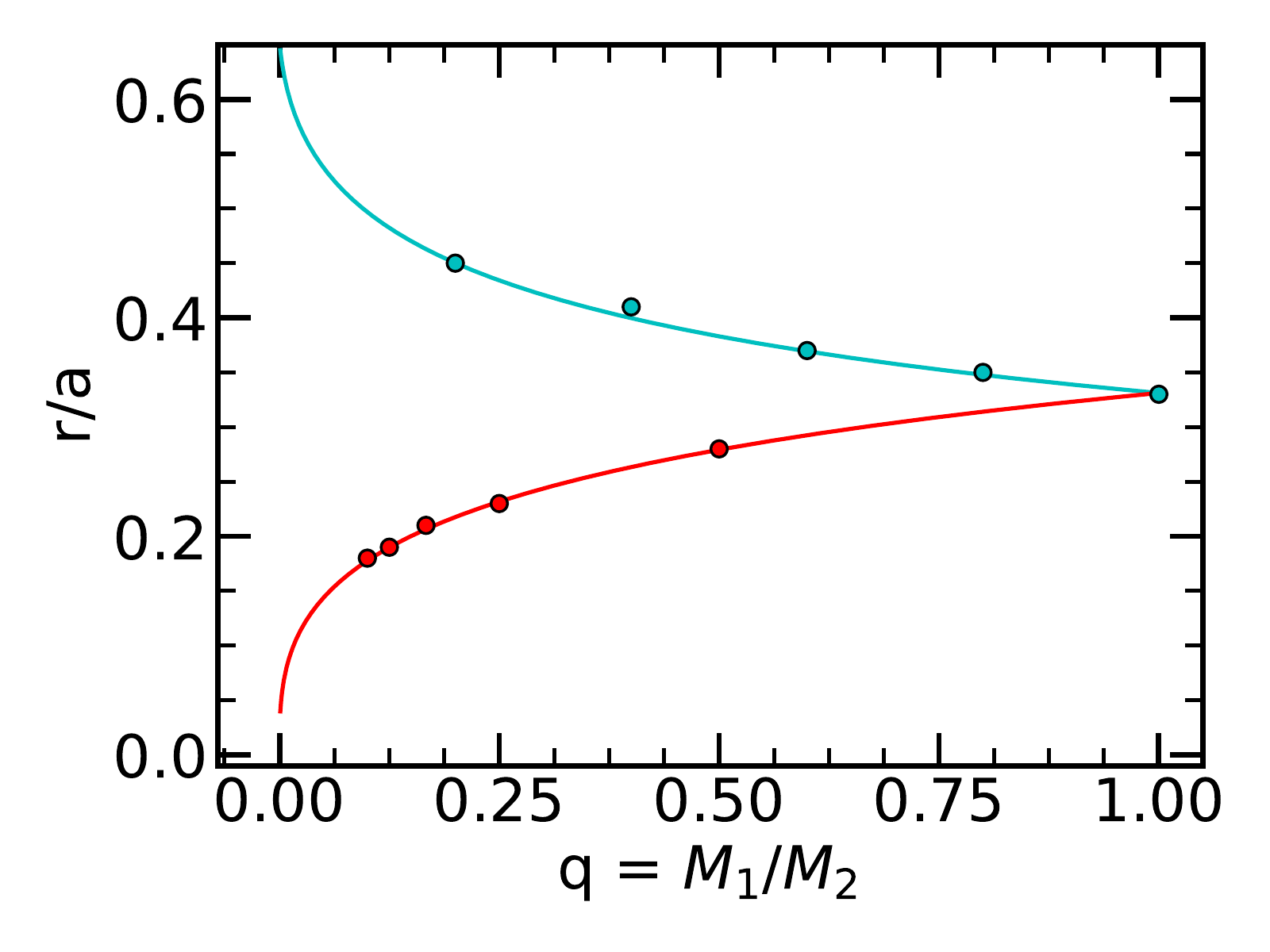}
\caption{Tidal truncation radius predictions for a zero-eccentricity binary as a function of the mass ratio q. The solid lines represent the fitting function $R_{i,\rm Egg}0.88\cdot \mu^{0.01}$. In particular the blue line refers to the circumprimary disk, and the red line to the circumsecondary. Similarly, the blue dots are the theoretical predictions for the circumprimary and the red dots for the circumsecondary (data from \citealt{PP77}). The yellow square refers to the equal mass case.} \label{figexpofitpp} 
\end{center}
\end{figure}

We note that Equation \ref{eqfittingfunc} is composed of two terms. The first  contains all the dependence on the eccentricity \textit{e} ($R_{i,\rm Egg}b\cdot e^c$); the second  ($R_{i,\rm Egg}h\cdot \mu^k$) describes how the zero-eccentricity truncation radius varies with varying $(M_1,M_2,a)$. As explained above, the truncation in zero-eccentricity binaries is determined by the \citet{PP77} mechanism: we can therefore obtain the value of \textit{h} and \textit{k} simply by fitting the results obtained by them. The fitting function (line) is overplotted to the data from \citet{PP77} (dots) in Figure \ref{figexpofitpp}, where the fitted parameters are $h=0.88$ and $k=0.01$. The exponent of $\mu$ is very small, the dependence on the masses is only inside $R_{i,\rm Egg}$, and in general the truncation occurs at $0.85-0.9$ times the size of the Roche Lobe.

Now we calculate \textit{b} and \textit{c} by fitting the numerical results from \citet{AL94} with
\begin{equation}\label{eqfuncfit}
R_{\rm t}(M_1,M_2,e,a)=R_{i,\rm Egg}(b\cdot e^c+0.88\cdot\mu^{0.01}).
\end{equation}

Table~\ref{tabfitvalues} reports the fitting parameters for some mass ratios and Reynolds numbers for both the circumprimary and the circumsecondary disk. The fitting parameters do not depend much on $\mu$. For a general choice of $\mu$ we simply interpolate the fitting parameters reported in Table ~\ref{tabfitvalues}.

\begin{table}[h]
\begin{center}
\begin{tabular}{ccc|ccc}
\multicolumn{3}{c}{Circumprimary}&
\multicolumn{3}{c}{Circumsecondary}\\
\hline
\hline

$\rey$ & \textit{b} & \textit{c} & $\rey$ & \textit{b} & \textit{c} \\
\hline
        \multicolumn{6}{c}{\textbf{$\mu=0.1$}}\\
\hline
$10^4$  &       -0.66   &       0.84    & $10^4$        &       -0.81   &       0.98    \\
$10^5$  &       -0.75   &       0.68    & $10^5$        &       -0.81   &       0.80    \\
$10^6$  &       -0.78   &       0.56    & $10^6$        &       -0.83   &       0.69    \\
\hline
        \multicolumn{6}{c}{\textbf{$\mu=0.2$}}\\
\hline
$10^4$  &       -0.72   &       0.88  & $10^4$  &       -0.81   &       0.99    \\
$10^5$  &       -0.78   &       0.72  & $10^5$  &       -0.82   &       0.82    \\
$10^6$  &       -0.80   &       0.60  & $10^6$  &       -0.83   &       0.70    \\
\hline
        \multicolumn{6}{c}{\textbf{$\mu=0.3$}}\\
\hline
$10^4$  &       -0.76   &       0.92  & $10^4$  &       -0.79   &       0.97    \\
$10^5$  &       -0.80   &       0.75  & $10^5$  &       -0.82   &       0.81    \\
$10^6$  &       -0.81   &       0.63  & $10^6$  &       -0.83   &       0.69    \\
\hline
        \multicolumn{6}{c}{\textbf{$\mu=0.4$}}\\
\hline
$10^4$  &       -0.77   &       0.95  & $10^4$  &       -0.80   &       0.98    \\
$10^5$  &       -0.81   &       0.78  & $10^5$  &       -0.82   &       0.80    \\
$10^6$  &       -0.82   &       0.66  & $10^6$  &       -0.83   &       0.68    \\
\hline
        \multicolumn{6}{c}{\textbf{$\mu=0.5$}}\\
\hline
$10^4$  &       -0.78   &       0.94  & $10^4$  &       -0.79   &       0.95    \\
$10^5$  &       -0.81   &       0.78  & $10^5$  &       -0.81   &       0.78    \\
$10^6$  &       -0.82   &       0.66  & $10^6$  &       -0.82   &       0.66    \\
\hline
\hline
\end{tabular}
\caption{Best fit parameters for Equation \ref{eqfuncfit} for different values of $\mu$ and $\rey$, both for circumprimary and circumsecondary disks.\label{tabfitvalues}}
\end{center}
\end{table}

\subsection{Comparison with observations}\label{app::models_comp}
We report here the plots of the comparison between the measured ratio of the dust disk radii and the projected separation with the expectations from analytic models, as described in Sect.~\ref{sect::models_analytic}.

   \begin{figure}
   \centering
  \includegraphics[width=0.45\textwidth]{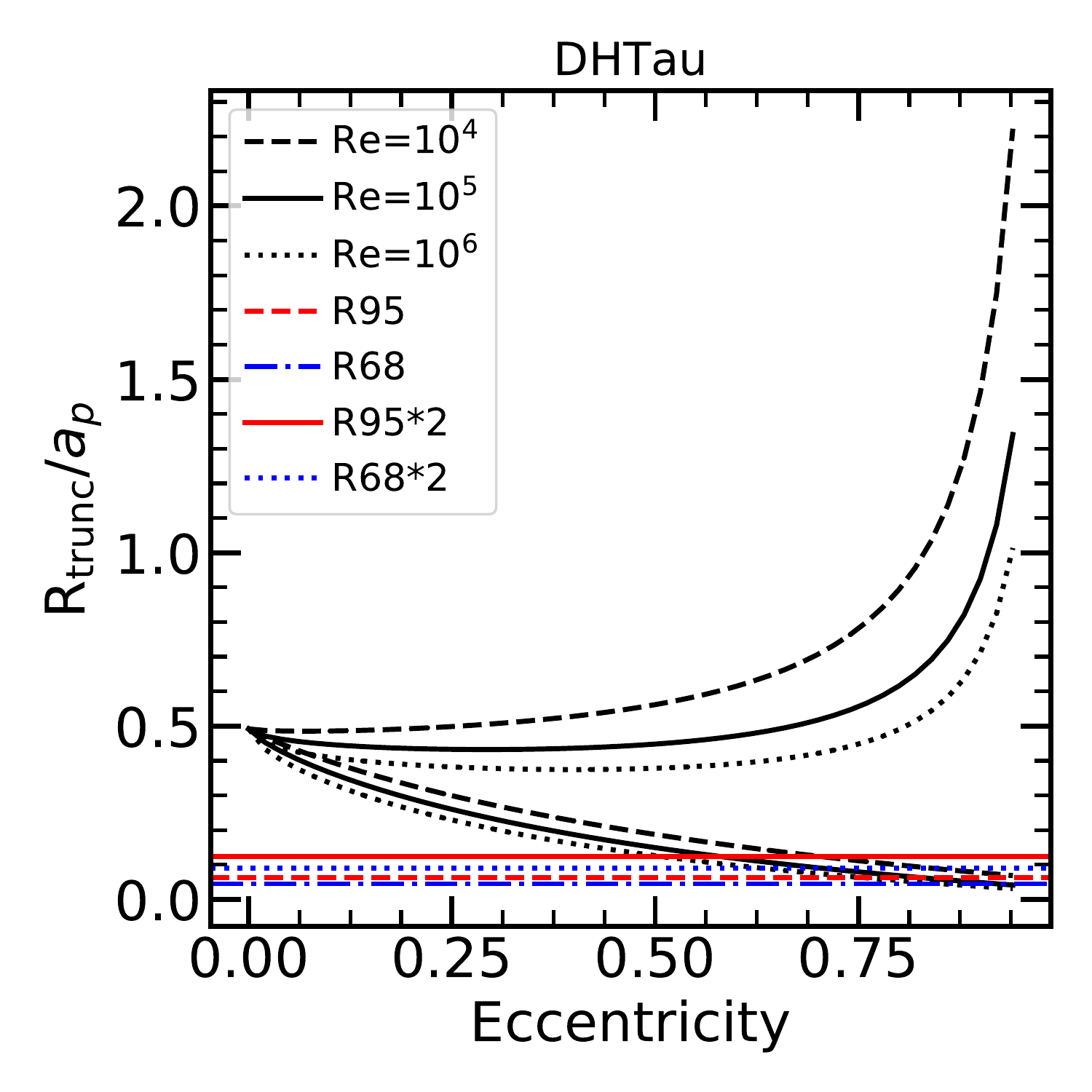}
   \caption{Same as Fig.~\ref{fig::RWAur_trunc_mod_gas_example}, but for DH~Tau. }
              \label{fig::DHTau_trunc_mod_gas}%
    \end{figure}

%

   \begin{figure}
   \centering
  \includegraphics[width=0.45\textwidth]{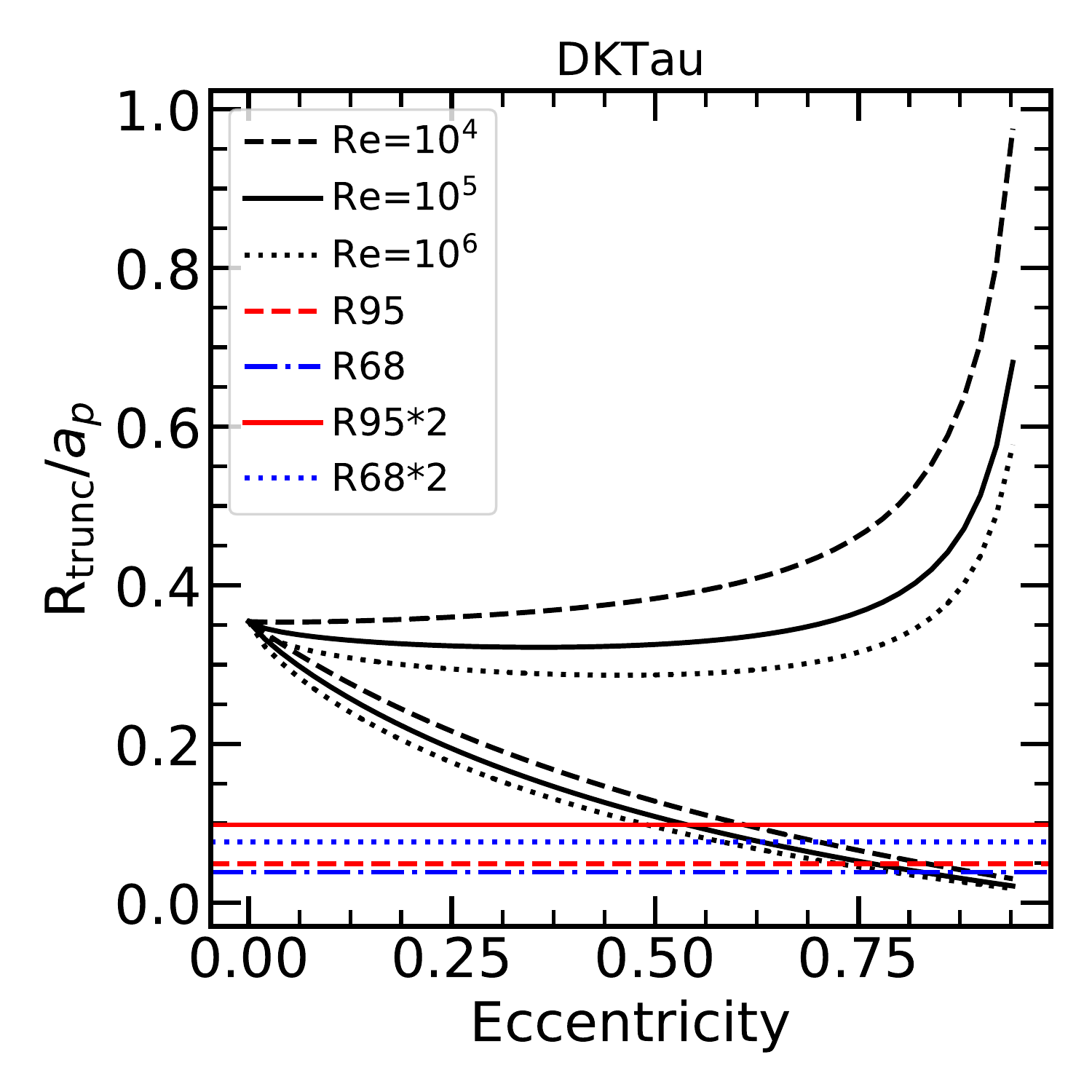}
   \caption{Same as Fig.~\ref{fig::RWAur_trunc_mod_gas_example}, but for DK Tau. }
              \label{fig::DKTau_trunc_mod_gas}%
    \end{figure}

%

   \begin{figure}
   \centering
  \includegraphics[width=0.45\textwidth]{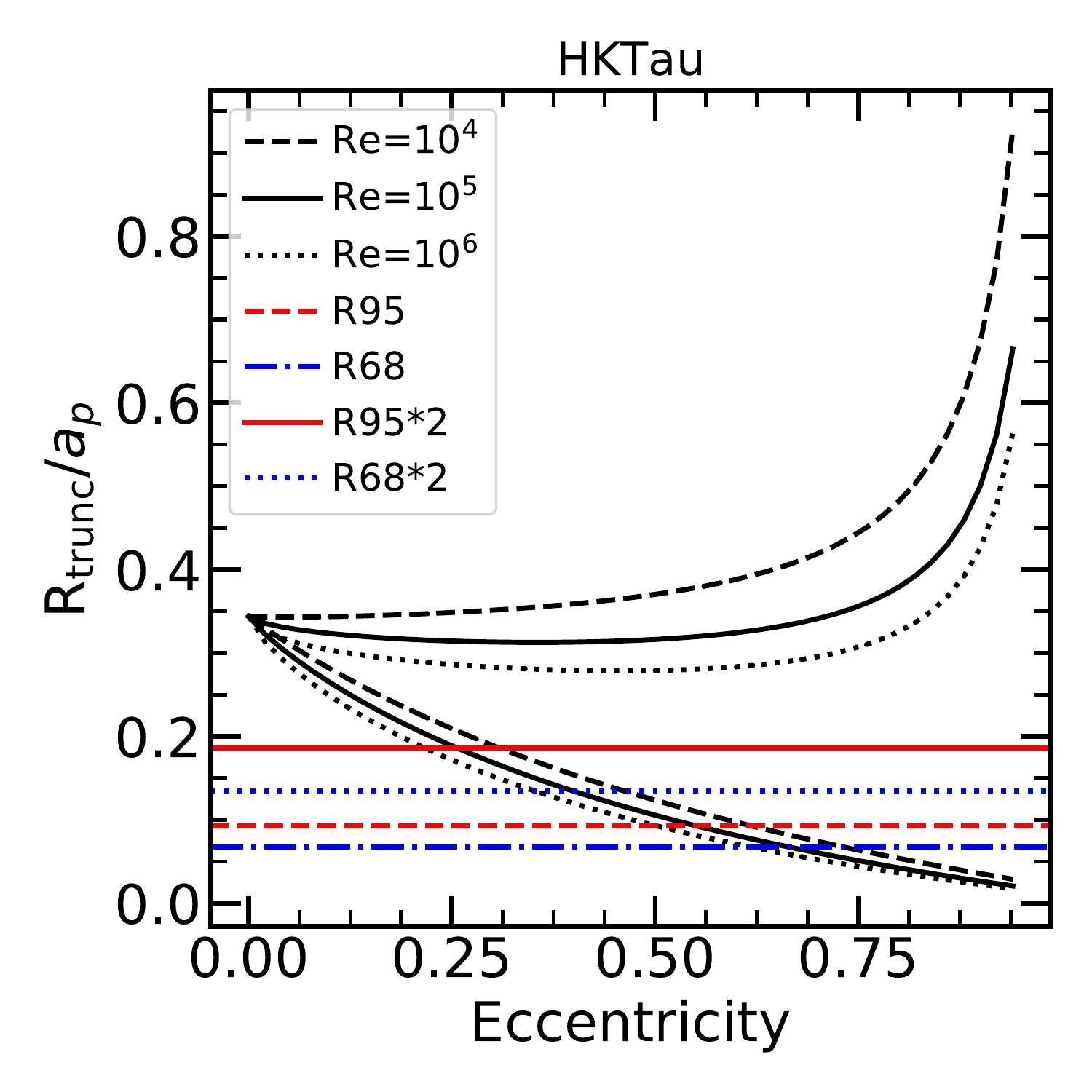}
   \caption{Same as Fig.~\ref{fig::RWAur_trunc_mod_gas_example},  but for HK Tau. }
              \label{fig::HKTau_trunc_mod_gas}%
    \end{figure}

%

   \begin{figure}
   \centering
  \includegraphics[width=0.45\textwidth]{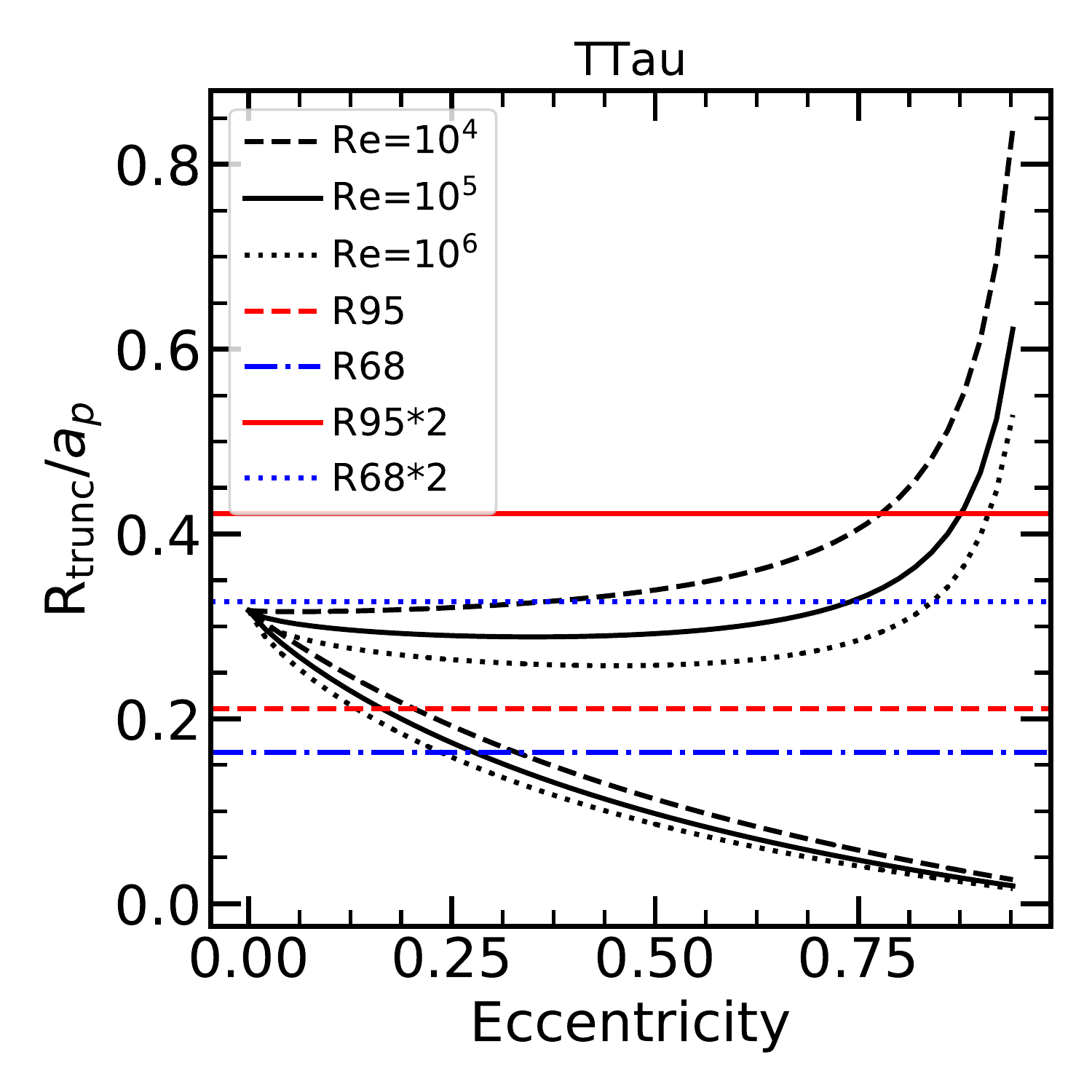}
   \caption{Same as Fig.~\ref{fig::RWAur_trunc_mod_gas_example}, but for T Tau. }
              \label{fig::TTau_trunc_mod_gas}%
    \end{figure}

   \begin{figure}
   \centering
  \includegraphics[width=0.45\textwidth]{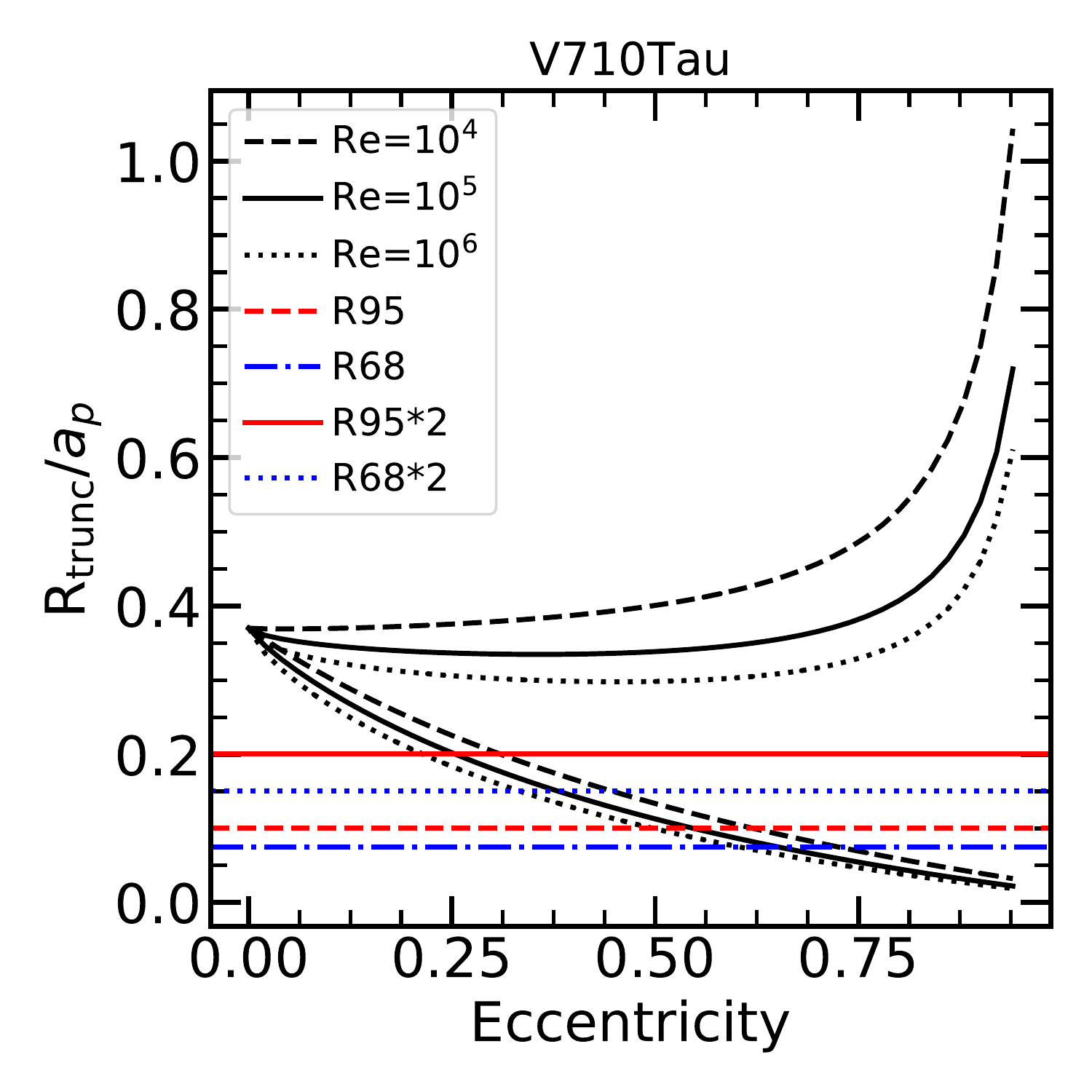}
   \caption{Same as Fig.~\ref{fig::RWAur_trunc_mod_gas_example}, but for V710 Tau. }
              \label{fig::V710Tau_trunc_mod_gas}%
    \end{figure}

   \begin{figure}
   \centering
  \includegraphics[width=0.45\textwidth]{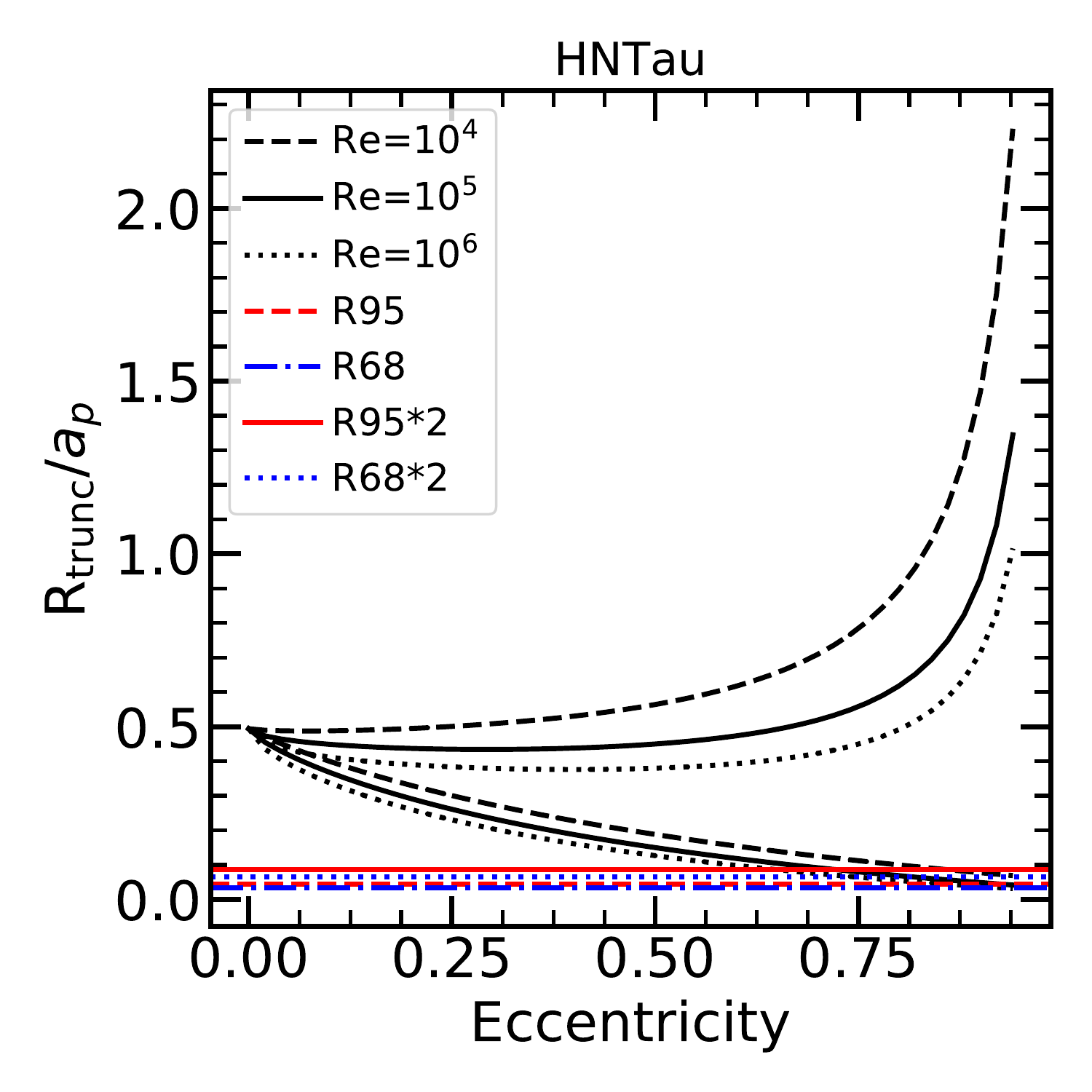}
   \caption{Same as Fig.~\ref{fig::RWAur_trunc_mod_gas_example}, but for HN Tau. }
              \label{fig::HNTau_trunc_mod_gas}%
    \end{figure}

   \begin{figure}
   \centering
  \includegraphics[width=0.45\textwidth]{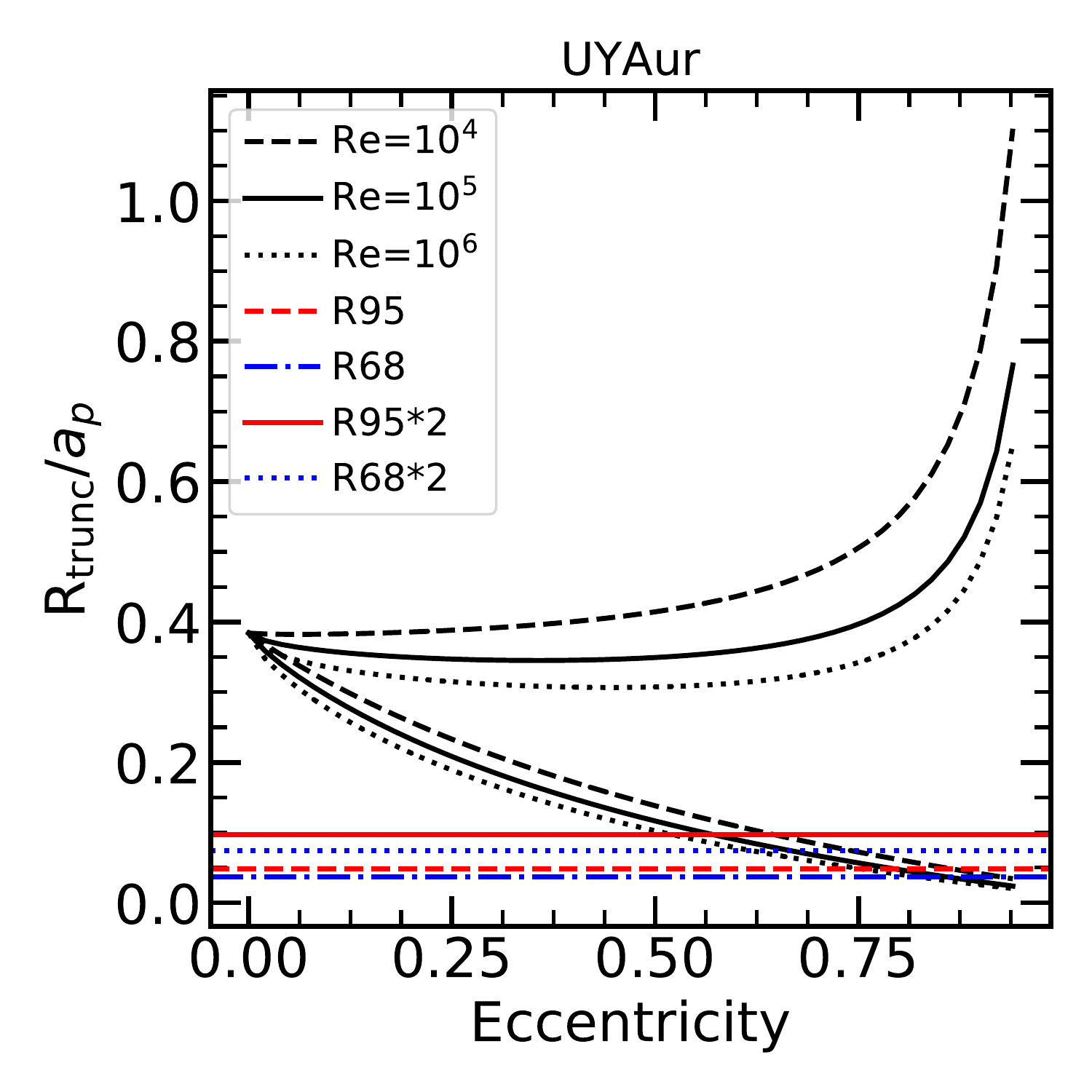}
   \caption{Same as Fig.~\ref{fig::RWAur_trunc_mod_gas_example}, but for UY Aur. }
              \label{fig::UYAur_trunc_mod_gas}%
    \end{figure}
 
%

   \begin{figure}
   \centering
  \includegraphics[width=0.45\textwidth]{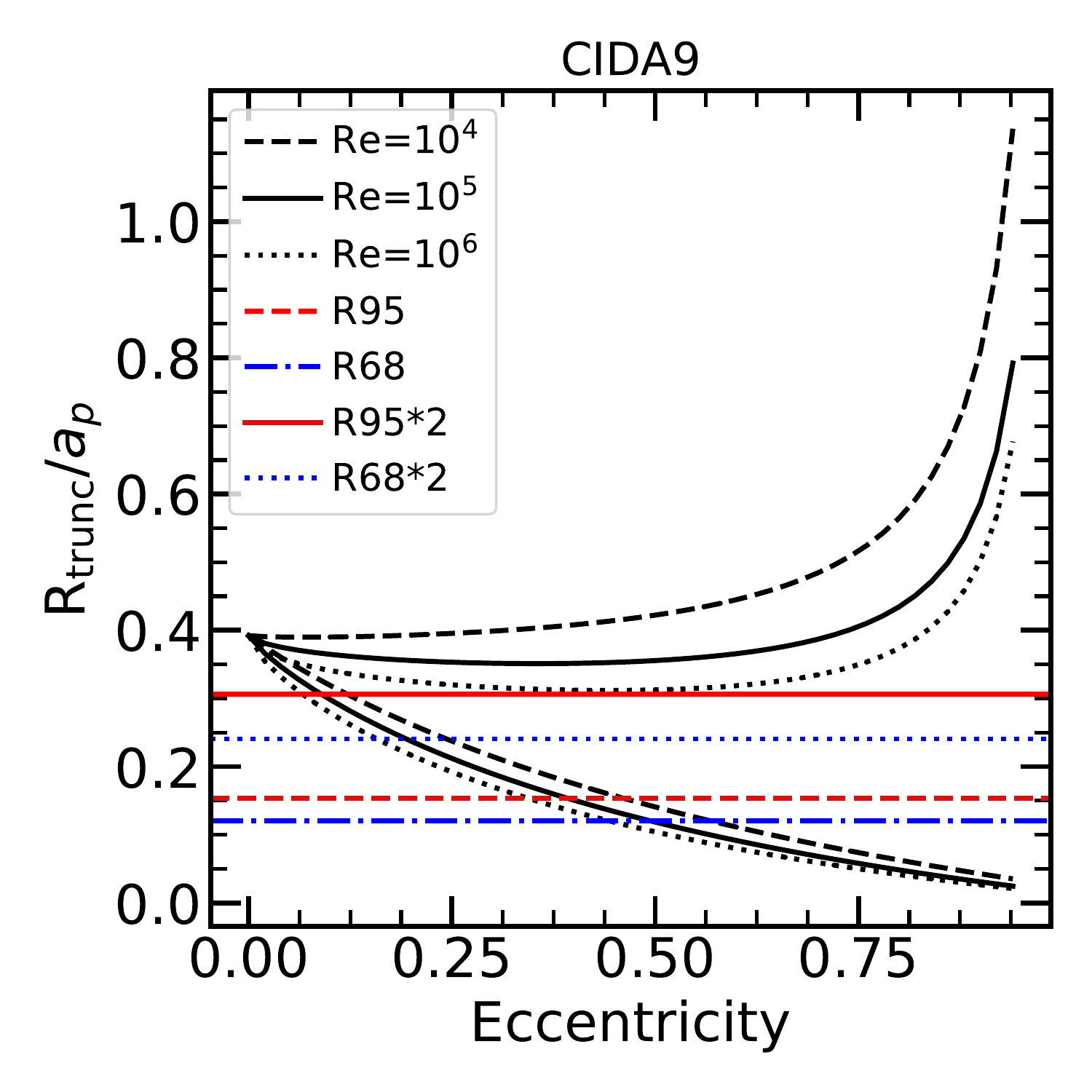}
   \caption{Same as Fig.~\ref{fig::RWAur_trunc_mod_gas_example}, but for CIDA 9. }
              \label{fig::CIDA9_trunc_mod_gas}%
    \end{figure}
 
%

   \begin{figure}
   \centering
  \includegraphics[width=0.45\textwidth]{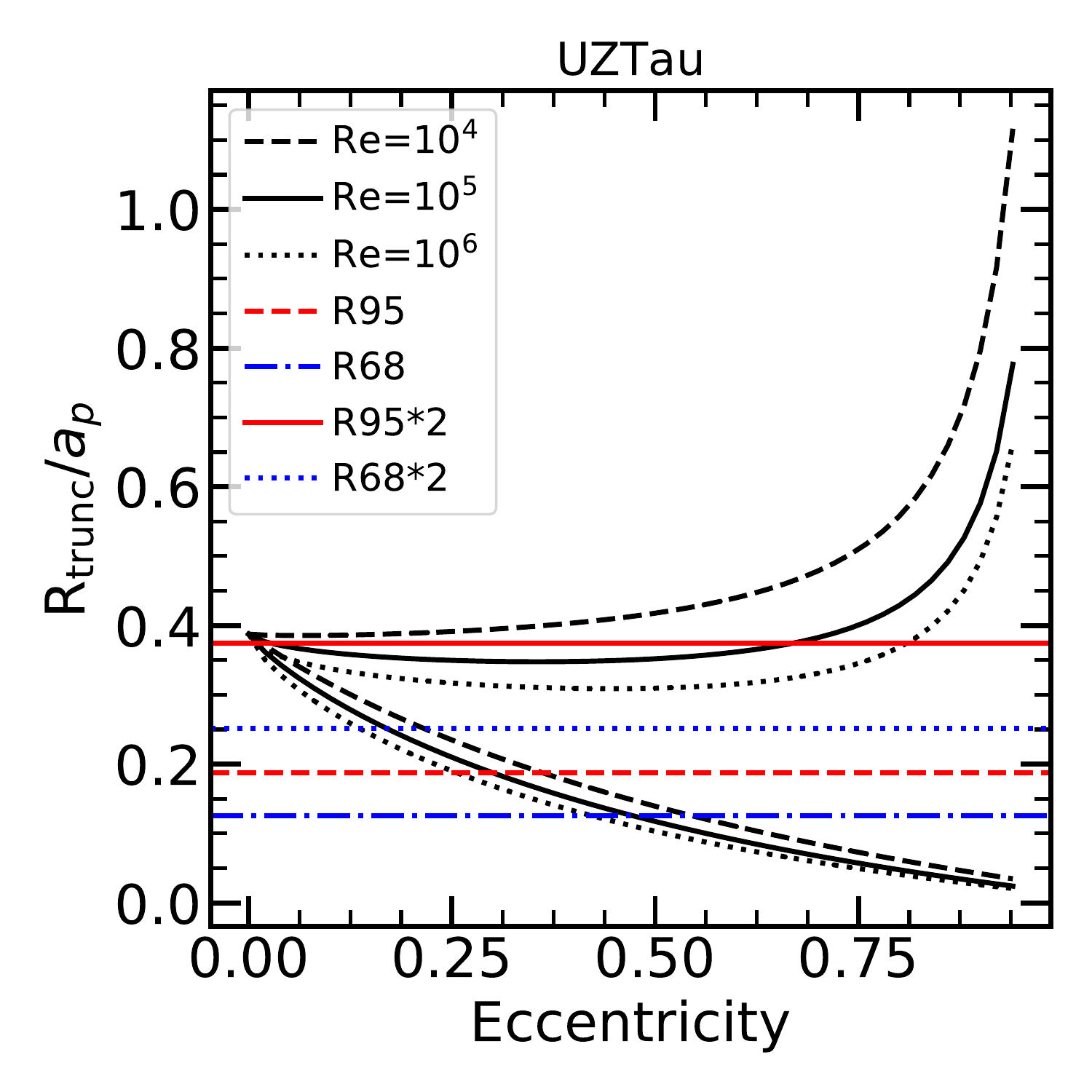}
   \caption{Same as Fig.~\ref{fig::RWAur_trunc_mod_gas_example}, but for UZ Tau. }
              \label{fig::UZTau_trunc_mod_gas}%
    \end{figure}

\end{document}